\documentclass[fleqn,usenatbib]{mnras}

\usepackage[T1]{fontenc}
\usepackage{ae,aecompl}

\usepackage{graphicx,color}   
\usepackage{amsmath}    
\usepackage{amssymb}    
\usepackage{ascmac}

\newcommand{\bfv}{\mbox{\boldmath$v$}}

\newcommand{\bfx}{\mbox{\boldmath$x$}}
\newcommand{\bfk}{\mbox{\boldmath$k$}}

\newcommand{\rmv}{{u}}
\newcommand{\Qchat}{\widehat{Q}_{\rm c}}
\newcommand{\Qc}{Q_{\rm c}}
\newcommand{\tauc}{\tau_{\rm c}}
\newcommand{\tauchat}{\widehat{\tau}_{\rm c}}

\newcommand{\omegam}{\Omega_{\rm m,0}}

\title[Post-collapse perturbation theory in 1D]{Post-collapse perturbation theory in 1D cosmology -- beyond shell-crossing}

\author[A. Taruya and S. Colombi]{
Atsushi Taruya$^{1,2}$
and
St\'ephane Colombi$^{1,3}$
\\
$^{1}$Center for Gravitational Physics, Yukawa Institute for Theoretical Physics, Kyoto University, Kyoto 606-8502, Japan\\
$^{2}$Kavli Institute for the Physics and Mathematics of the Universe, Todai Institutes for Advanced Study, The University of Tokyo, \\
~~Kashiwa, Chiba 277-8583, Japan (Kavli IPMU, WPI)\\
$^{3}$Institut d'Astrophysique de Paris, CNRS UMR 7095 and UPMC, 98bis bd Arago, F-75014 Paris, France
}

\date{Accepted XXX. Received YYY; in original form ZZZ}

\pubyear{2015}

\begin{document}
\label{firstpage}
\pagerange{\pageref{firstpage}--\pageref{lastpage}}
\maketitle

\begin{abstract}
We develop a new perturbation theory (PT) treatment that can describe gravitational dynamics of large-scale structure after shell-crossing in the one-dimensional cosmological case. Starting with cold initial conditions, the motion of matter distribution follows at early stages the single-stream regime, which can, in one dimension, be described exactly by the first-order Lagrangian perturbation, i.e. the Zel'dovich solution. However, the single-stream flow no longer holds after shell-crossing and a proper account of the multi-stream flow is essential for post-collapse dynamics. In this paper, extending previous work by \citet[][MNRAS 446, 2902]{Colombi:2014lda}, we present a perturbative description for the multi-stream flow after shell-crossing in a cosmological setup. In addition, we introduce an adaptive smoothing scheme to deal with the bulk properties of phase-space structures. The filtering scales in this scheme are linked to the next-crossing time in the post-collapse region, estimated from our PT calculations. Our PT treatment combined with adaptive smoothing is illustrated in several cases. Predictions are compared to simulations and we find that post-collapse PT with adaptive smoothing reproduces the power spectrum and phase-space structures remarkably well even at small scales, where Zel'dovich solution substantially deviates from simulations. 
\end{abstract}

%


\section{Introduction}
\label{sec:Introduction}


It is currently admitted that processes of structure formation in the Universe are mainly dominated at large scale by an invisible component called dark matter.  Although the microscopic origin of dark matter is still unclear, it is macroscopically described as a self-gravitating collisionless fluid following the collisionless Boltzmann or Vlasov equation in a cosmological background,
\begin{align}
\Biggl[\frac{\partial }{\partial t} + \frac{\bfv}{a}\cdot
\frac{\partial }{\partial \bfx} -
\frac{1}{a} \nabla \Phi\cdot
\frac{\partial }{\partial \bfv}\Biggr]
f(\bfx,\bfv,t) =0,
\label{eq:vp_3d}
\end{align}
supplemented with the Poisson equation for the Newton potential $\Phi$,
\begin{align}
\frac{1}{a^2}\nabla^2\Phi=4\pi\,G\,\Biggl[
\frac{1}{a^3}\int {\rm d}^3\bfv\,f(\bfx,\bfv,t)-\overline{\rho}_{\rm m} 
\Biggr],
\label{eq:poisson_3d}
\end{align}
where $f(\bfx,\bfv,t)$ is the phase-space density at comoving position $\bfx$, peculiar velocity $\bfv$ and time $t$, $a$ is the expansion factor of the Universe, $\Phi$ is the gravitational potential and $\overline{\rho}_{\rm m}$ is the average dark matter density. 

In the standard picture of structure formation, dark matter was initially cold i.e. with a virtually null local velocity dispersion, so the six-dimensional phase-space distribution is effectively reduced to a three-dimensional hyper-surface and this remains true at all times thanks to the Hamiltonian nature of the system. At early times, dark matter thus follows the single-stream flow regime, with a velocity field ${\overline \bfv}$ uniquely determined as a function of position, and its evolution is that of a pressure-less fluid with a phase-space distribution function given by 
\begin{align}
f(\bfx,\bfv,t) =\overline{\rho}_m\,a^3\,
\Bigl\{\,1+\delta(\bfx,t)\,\Bigr\}\,\delta_{\rm D}
\Bigl[\bfv-\overline{\bfv}(\bfx,t)\Bigr],
\label{eq:DF_single-stream}
\end{align}
where $\delta$ is the density contrast of the dark matter distribution, initially of very small magnitude, as well as ${\overline \bfv}$. 

Substituting Eq.~(\ref{eq:DF_single-stream}) into Eqs.~(\ref{eq:vp_3d}) and (\ref{eq:poisson_3d}) and taking the zeroth and first velocity moments yields the Eulerian formulation of large-scale structure dynamics
\begin{align}
&\frac{\partial\delta}{\partial t}+\frac{1}{a}\,\nabla\Bigl[\Bigl\{1+\delta\Bigr\}\,{\overline \bfv}\Bigr]=0,
\label{eq:continuity}\\
&\frac{\partial{\overline \bfv}}{\partial t}+\frac{1}{a}\,\Bigl\{{\overline \bfv}\cdot\nabla\Bigr\}\overline\bfv=-\frac{1}{a}\,\nabla\Phi,
\label{eq:Euler}\\
&\frac{1}{a^2}\nabla^2\Phi=4\pi\,G\,\overline{\rho}_{\rm m}\,\delta.
\label{eq:Poisson_v2}
\end{align}
Although the single-stream flow is, strictly speaking, valid only during the early phase of structure formation, the above equations have been shown in practice to describe accurately nonlinear mode-coupling in the weakly nonlinear regime and provide a solid basis for perturbation theory (PT) calculations to predict statistical quantities of large-scale structure, such as the power spectrum or the two-point correlation function of the matter distribution \citep[see, e.g.][]{Bernardeau:2001qr}.

With the advent of cosmological observations aiming at precisely mapping galaxy clustering at large scales, techniques have been developed to improve on the slow convergence of the perturbative expansion as well as to accelerate higher-order calculations \citep[e.g.,][]{Crocce:2005xy,Crocce:2005xz,Crocce:2007dt,Valageas2007,Taruya:2007xy,Matsubara:2007wj,Bernardeau:2008fa,Pietroni:2008jx,2009PhRvD..79j3526H,Taruya:2009ir,Okamura:2011nu,2011PhRvD..83h3518M,Bernardeau:2011dp,Crocce:2012fa,Bernardeau:2012ux,Taruya:2012ut,Valageas:2013gba}. Improved calculations involving the next-to-next-to-leading order called two-loop give a remarkable agreement with cosmological $N$-body simulations at weakly nonlinear scales and have been applied to observations. However, this does not ensure the convergence of perturbative calculations including higher-order corrections, e.g., three-loop. In fact, a direct calculation at three-loop order suggests a very large UV contribution to the large-scale modes through nonlinear mode-coupling \citep[][]{Bernardeau:2012ux,Blas:2013aba} and indicates a break down of higher-order perturbative expansions even at large scales. 

Deficiency of PT calculations has been also highlighted in recent numerical analyses. \citet{Nishimichi:2014rra} directly measured the coupling between different scales in cosmological $N$-body simulations and found that the actual contribution from small scales to the large-scale modes is suppressed, as opposed to the prediction of PT based on the single-stream approximation. These facts imply that the validity of the single-stream treatment is questionable even at large scales, and higher-order perturbative corrections need to be remedied with a proper account of small-scale dynamics, where the multi-stream flow contribution is important. 

One way to account for small-scale dynamics consists of using some ansatz to summarize the main physical effects in the multi-stream regime, such as Burgers' equation \citep[e.g.,][]{Gurbatov89,2010PhRvD..81d3516B} and various alternatives \citep[e.g.,][]{Sahni95}, which roughly amount to adding a source term in the right hand side of equation (\ref{eq:Euler}). In particular, it was proposed recently to use effective fluid equations giving account of the non-vanishing stress tensor arising in Eq.~(\ref{eq:Euler}) when calculating the first velocity moment of Vlasov equation in the multi-stream regime. This {\it effective-field theory} approach has attracted a lot of interest and has been studied in detail \citep[e.g.,][]{2012JCAP...07..051B,2012JHEP...09..082C,2014PhRvD..89d3521H,Baldauf:2015aha}. The drawback of this approach is however that the parameters in the stress tensor characterizing the small-scale dynamics need to be calibrated with $N$-body simulations to keep predictions of perturbative calculations under control. Furthermore, these parameters generally vary with cosmology and with time and no prediction with PT is really possible in this framework independently of $N$-body simulations. 

An alternative approach that we consider in this paper consists in going back to a more fundamental description, i.e. the Vlasov-Poisson system, Eqs.~(\ref{eq:vp_3d}) and (\ref{eq:poisson_3d}), and trying to follow accurately the phase-space distribution function in the multi-stream regime, which is essential for describing the formation of dark matter halos. For this purpose, it is useful to employ Lagrangian PT \citep[e.g.,][and references therein]{1970A&A.....5...84Z,1989RvMP...61..185S,Bouchet92,Buchert92,Buchert93,Bouchet95,Bernardeau:2001qr}, where the small parameter is the displacement field.  We shall consider one-dimensional gravitational dynamics in standard (three-dimensional) cosmology. In this case, large-scale structure dynamics is described by the gravitational interaction of massive parallel infinite planes moving left and right along a fixed axis, while Hubble expansion is taking place as usual in all the directions following standard Friedman-Lema\^{\i}tre equations. Despite its simplicity, one dimensional dynamics defined as such still displays a rich physical content which somewhat shares the same features as 3D clustering. This is partly the reason why the 1D model has recently attracted much attention \citep[e.g.,][]{Benhaiem2013,McQuinn:2015tva,Vlah:2015zda,Baldauf:2015fbu}. In particular, the Zel'dovich solution provides an exact solution for the dynamics of massive sheets before shell-crossing \citep{1970A&A.....5...84Z,1989RvMP...61..185S}, and thus, starting with Zel'dovich solution, a tractable perturbative treatment of multi-stream flows is made possible based on a Lagrangian description. The analysis in the present paper is an extension of the method developed in \citet{Colombi:2014lda} to the cosmological setup. We will describe perturbatively post-collapse dynamics around the shell-crossing region and apply it to several cases including random initial conditions.

In addition, we shall present a novel regularization scheme reducing the impact of small-scale clustering and improving greatly our post-collapse PT predictions of large-scale structure statistics. The idea is to apply adaptive smoothing to initial density peaks and to better capture the bulk properties of phase-space structures in the post-collapse regions, where the interaction or merger of halos is supposed to be significant. The idea is similar to the peak-patch treatment proposed by \citet{1996ApJS..103....1B} and subsequent works \citep[see, e.g.][]{2002MNRAS.331..587M}, but we here implement it in the PT prescription in order to better describe the late-time post-collapse dynamics. Indeed, our perturbative approach will not allow us to follow post-collapse dynamics beyond next-crossing time, although an iterative prescription such as proposed by \citet{Colombi:2014lda} could make this possible but is out of the scope of the present work. Here, we will show instead that the predictions with adaptive smoothing reproduce simulations remarkably well even at nonlinear scales, where Zel'dovich solution significantly deviates from simulations. 

Note finally that an analytical study in 1D of course represents only a first step toward a proper description of 6D phase-space dynamics.  Apart from a few examples including self-similar solutions \citep{1984ApJ...281....1F,1984ApJ...281....9F,1985ApJS...58...39B,1993ApJ...418....4R,2011ApJ...734..100L,Alard2013}, little is indeed known analytically for the phase-space dynamics of Vlasov-Poisson systems, particularly in cosmology. Therefore, further development of analytical treatment in this framework seems indispensable, complementary to simulations, and even helpful to cross check simulation codes. This is all the more motivated by the fact that simulations in 6D phase space have now become available thanks to recent efforts on the development of pure Vlasov codes \citep{2013ApJ...762..116Y,Sousbie:2015uja,Hahn:2015sia}.

This paper is organized as follows.  In \S~\ref{sec:1Dcosmology}, we begin by describing the basic setup of our calculations in one-dimensional cosmology. We then discuss in \S~\ref{sec:postcollapse} the analytic treatment beyond shell-crossing and develop post-collapse PT. \S~\ref{sec:improvement} introduces a novel filtering scheme to improve the perturbative description of post-collapse dynamics, suited for a system with interacting clusters.   In \S~\ref{sec:comparison_nbody}, analytic calculations in the post-collapse perturbative framework are tested against controlled $N$-body experiments. Several cases are considered, including the single initial sine wave and random initial conditions. Finally, \S~\ref{sec:conclusion} is devoted to discussion and conclusion.

\section{1D cosmology}
\label{sec:1Dcosmology}

\subsection{Basic setup}
\label{subsec:basic_setup}

We consider an ensemble of massive parallel infinite planes moving along $x$ axis and interacting through gravitational force in the expanding Universe. 
The Lagrangian equations of motion of the planes are given by
\begin{align}
&\frac{{\rm d}x}{{\rm d}t}=\frac{v}{a}, 
\label{eq:eom0_x}
\\
&\frac{{\rm d}v}{{\rm d}t}+H\,v=-\frac{1}{a}\nabla_x\phi,
\label{eq:eom0_v}
\\ 
&\nabla^2_x\phi(x)=4\pi\,G\overline{\rho}_{\rm m}\,a^2\,\delta(x),
\label{eq:eom0_delta}
\end{align}
where $x(t)$ and $v(t)$ are respectively the comoving position and peculiar velocity of each plane, $\phi$ the gravitational potential, $\overline{\rho}_{\rm m}$ the average matter density, $\delta$ the density contrast and $a$ the expansion factor of the Universe. 

To simplify the equations, it is useful to introduce the {\it super-conformal time} $\tau$ defined by \citep[e.g.,][]{1973Ap......9..144D,MartelShapiro1998},
\begin{align}
{\rm d}\tau = \frac{{\rm d}t}{a^2},
\end{align}
the new velocity $\rmv$ and potential $\Phi$:
\begin{align}
\rmv \equiv a\,v,
\qquad
\Phi\equiv a^2\phi.
\end{align}
Then, Eqs.~(\ref{eq:eom0_x})--(\ref{eq:eom0_delta}) are simplified:
\begin{align}
&\frac{{\rm d}x}{{\rm d}\tau} = \rmv,
\label{eq:eom1_x}\\
&\frac{{\rm d}\rmv}{{\rm d}\tau} = -\nabla_x\Phi,
\label{eq:eom1_p}\\
&\nabla_x^2\Phi = 4\pi G\overline{\rho}_{\rm m}\,a^4\,\delta = \frac{3}{2}\,\omegam H_0^2 \,a\,\delta,
\label{eq:eom1_delta}
\end{align}
where $\Omega_{{\rm m},0}$ is the matter density parameter of the Universe and $H_0$ is the Hubble constant. 

To deal with this system, especially for the dynamics after shell-crossing, it is useful to introduce in the cold case considered here the Lagrangian coordinate $q$ defining the initial position of the planes in the absence of perturbation
\begin{align}
q \equiv x(\tau\to0),
\end{align}
and to express the subsequent position and velocity of each mass element as $x(q,t)$ and $\rmv(q,t)$. In particular, mass conservation implies
\begin{align}
{\rm d}q =[1+\delta(x)]\,{\rm d}x \,\,\Longrightarrow\,\,
\delta(x)=\left(\frac{\partial x}{\partial q}\right)^{-1}-1.
\label{eq:jacobian_delta}
\end{align}
This equation is valid in the single-stream regime, i.e. as long as $x(q,\tau)$ remains monotonic as a function of $q$. 

With the new expressions above, the solution of the equations of motion can be formally written as:
\begin{align}
&x(q;\tau)=x(q;\tau_{\rm ini})+\int_{\tau_{\rm ini}}^\tau {\rm  d}\tau'\,\rmv(q;\tau'),
\label{eq:sol1_x}\\
&\rmv(q;\tau)=\rmv(q;\tau_{\rm ini})-\int_{\tau_{\rm ini}}^\tau {\rm d}\tau'\,\nabla_x\Phi[x(q;\tau');\tau'],
\label{eq:sol1_p}
\end{align}
where $x(q;\tau_{\rm ini})$ and $\rmv(q;\tau_{\rm ini})$ are the initial positions and velocities given at a starting time $\tau_{\rm ini}$, which will be specified below. 

In what follows, we consider the dynamics of the cosmological system in a finite-size box with periodic boundaries, $0\leq x\leq L$. The solution of Poisson equation, Eq.~(\ref{eq:eom1_delta}), can be expressed in an integral form as: 
\begin{align}
&\Phi(x)=\frac{3}{2}\,\omegam H_0^2\,a\,
\nonumber\\
&~~\quad\times
\int_0^L{\rm d}x'\,
\left[-\frac{L}{2}\left\{\left(\frac{|x-x'|}{L}-\frac{1}{2}\right)^2-\frac{1}{12}\right\}\right]\,\delta(x').
\label{eq:Phi_Poisson}
\end{align}
The derivation of this expression is presented in Appendix \ref{sec:Green_func}. Then, the force exerted on a mass element at position $x$ is given by: 
\begin{align}
F(x)&\equiv -\nabla_x\Phi(x)
\nonumber\\
&=
 -\frac{3}{2}\,\omegam H_0^2\,a\, \nonumber \\
& \times \,\Bigl[ \int_0^L{\rm d}x'\,
\frac{\delta(x')}{2}\,\left\{\Theta(x-x')-\Theta(x'-x)\right\}
 \nonumber\\
& \quad
+\frac{1}{L}\int_0^L {\rm d}x'\,x'\,\delta(x')\,\Bigr],
\label{eq:Force_Poisson}
\end{align}
where function $\Theta(x)$ represents the Heaviside step function. 
In the above, we used the fact that fluctuations averaged over space vanish, $\int_0^L\,dx'\,\delta(x')=0$. 
In the limit $L\to\infty$, equation (\ref{eq:Force_Poisson}) of course converges to the well-known result in the infinite space.

\subsection{Pre-collapse dynamics}
\label{sec:Initial}

In one-dimension and in the cold case, Zel'dovich approximation \citep[][]{1970A&A.....5...84Z} is known to provide the exact solution for the dynamics of mass elements before shell-crossing. It can be explicitly written as
\begin{align}
x(q;\tau) = q+\psi(q)\,D_+(\tau),
\qquad
\rmv(q;\tau) = \psi(q)\,\frac{{\rm d}D_+(\tau)}{{\rm d}\tau}.
\label{eq:Zeldovich_sol}
\end{align}
Here, function $D_+$ corresponds to the linear growth factor satisfying the following equation:
\begin{align}
\left[\frac{{\rm d}^2}{{\rm d}\tau^2}-\frac{3}{2}\omegam H_0^2\,a(\tau)\right]\,D_+(\tau)=0.
\label{eq:evolv_lin}
\end{align}
Note that in terms of cosmic time $t$, Eq.~(\ref{eq:evolv_lin}) reduces to the standard form of linear evolution equation:
\begin{align}
\left[\frac{{\rm d}^2}{{\rm d}t^2}+2H(t)\frac{{\rm d}}{{\rm d}t}-\frac{3}{2}\frac{\omegam H_0^2}{a^3(t)}\right]\,D_+(t)=0.
\end{align}
The Zel'dovich solution in Eq.~(\ref{eq:Zeldovich_sol}) contains an arbitrary function $\psi(q)$ that we call displacement field. It is related to the linear density 
field $\delta_{\rm L}(q)$ given at a very early time ($\tau_{\rm ini}\to-\infty$ or 
$t_{\rm ini}\to 0$) through
\begin{align}
\frac{{\rm d}\psi(q)}{{\rm d}q}\,D_+(\tau_{\rm ini})=-\delta_{\rm L}(q;\,\tau_{\rm ini})=-\delta_{\rm L}(q)\,D_+(\tau_{\rm ini}).
\label{eq:def_deltaL}
\end{align}
Since Zel'dovich solution is exact until shell-crossing, we do not necessarily assume that the evolved density field $\delta(x)$ is small. One may thus consider the situation where in a region around a Lagrangian coordinate $q_0$, the density field becomes large and the region will undergo shell-crossing at time $\tau_0$. Recalling the fact that in 1D, a shell-crossing point corresponds to an inflection point of the mapping from Lagrangian to Eulerian space, the actual conditions for shell-crossing are given by 
\begin{align}
\left.\frac{\partial x}{\partial q}\right|_{q_0}=0,\quad
\left.\frac{\partial^2 x}{\partial q^2}\right|_{q_0}=0,\quad
\left.\frac{\partial^3 x}{\partial q^3}\right|_{q_0}>0.
\label{eq:condition_shellcrossing}
\end{align}
At the time of shell-crossing, $\tau_0$, the solution (\ref{eq:Zeldovich_sol}) around the shell-crossing region can be expanded as follows: 
\begin{align}
x(q;\tau_0) \simeq &\,\, q_0+\psi(q_0)D_+(\tau_0) \nonumber \\
 & + \left\{1+\frac{{\rm d}\psi(q_0)}{{\rm d}q_0}D_+(\tau_0)\right\}(q-q_0)
\nonumber\\
 &+
\sum_{n=2}\frac{1}{n!}\,\frac{{\rm d}^n\psi(q_0)}{{\rm d}q_0^n}D_+(\tau_0)\,(q-q_0)^n.
\end{align}
Using Eq.~(\ref{eq:def_deltaL}), the conditions (\ref{eq:condition_shellcrossing}) for shell-crossing can be rewritten as
\begin{align}
\delta_{\rm L}(q_0)=\frac{1}{D_+(\tau_0)},\quad
\left.\frac{{\rm d}\delta_{\rm L}(q)}{{\rm d}q}\right|_{q_0}=0,\quad
\left.\frac{{\rm d}^2\delta_{\rm L}(q)}{{\rm d}q^2}\right|_{q_0}<0.
\label{eq:shellcrossing_peak}
\end{align}
In other words, shell-crossing takes place at local density peaks. Hence, in the 1D case, the conditions for shell-crossing are equivalent to peak constraints and collapse take places exactly when the Eulerian linear density contrast at the peak positions becomes equal to unity. 
\section{Perturbative treatment of post-collapse dynamics}
\label{sec:postcollapse}

\begin{figure}
\includegraphics[width=8.5cm]{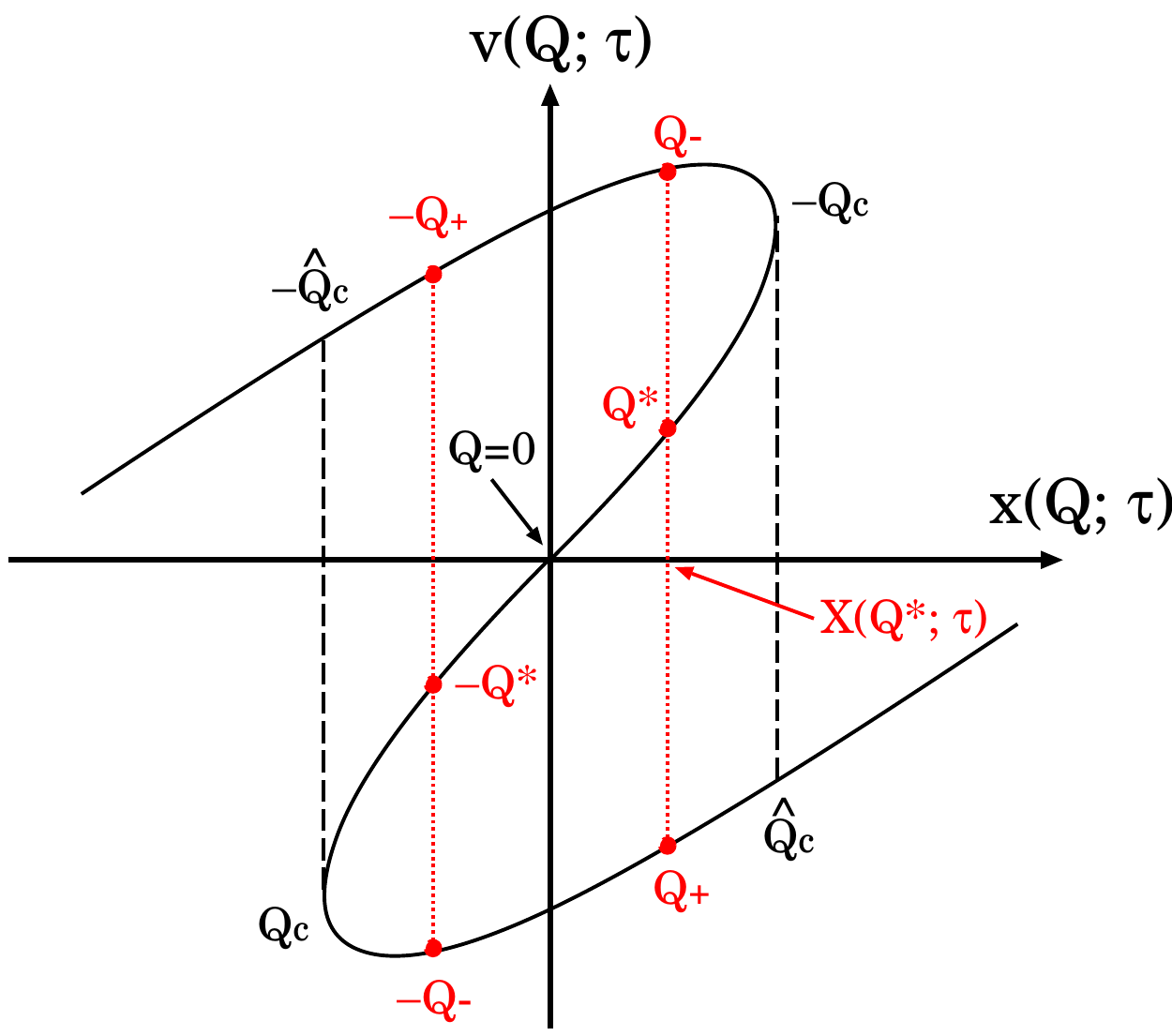}
\caption{Geometrical configuration of phase space around a local density peak after shell-crossing, extrapolating the Zel'dovich solution (see Eq.~\ref{eq:ballistic_shellcrossing}). The structure of the system around the density peak is symmetric with respect to the Lagrangian coordinate $Q\equiv q-q_0$, where $q_0$ is the Lagrangian position of the shell-crossing point
\citep[see also][]{Colombi:2014lda}.  
\label{fig:shellcrossing}}
\end{figure}

We are interested in the dynamics of mass elements after shell-crossing, when the Zel'dovich solution is no longer valid. In this section, extending the work of \cite{Colombi:2014lda},  we develop perturbative calculations to deal with the multi-stream motion around the shell-crossing region. 

\subsection{Post-collapse PT}
\label{subsec:formalism_pcpt}

The basic formalism to treat post-collapse dynamics can be described as follows. Starting with the cold initial conditions in Sec.~\ref{sec:Initial}, we first follow pre-collapse dynamics with the exact Zel'dovich solution. Then, in regions undergoing shell-crossing, we switch to a perturbative treatment and compute the backreaction to the Zel'dovich flow based on an explicit functional form of the displacement field in the shell-crossing region. To be precise, we compute the force from Eq.~(\ref{eq:Force_Poisson}) exerted at each position using the extrapolation of the Zel'dovich flow.  Integrating the force over time, we obtain the correction for the velocity to the Zel'dovich motion from Eq.~(\ref{eq:sol1_p}). Further integrating the corrected velocity over time, a correction for the position is obtained from Eq.~(\ref{eq:sol1_x}). Throughout these calculations, we assume that the collapsing region is small, which allows us to express the phase-space configuration in terms of polynomial forms of low order of the Lagrangian coordinate. 

Let us focus on a collapse point centered on Lagrangian position $q_0$ and perform a perturbative description of the multi-stream flow around $q=q_0$. As we mentioned earlier, the conditions for shell-crossing are given by Eq.~(\ref{eq:condition_shellcrossing}). Shortly after shell-crossing time $\tau_0$, the multi-stream flow has just started growing in a small Lagrangian region around $q_0$ and the displacement field $\psi(q)$ can be approximately described by a third-order polynomial in $q-q_0$. Since the deviation from Zel'dovich flow is small, the motion of a mass element around $q=q_0$  may be expanded at third-order in $q$ for $\tau > \tau_0$  as
\begin{align}
x(q;\tau)&\simeq A(q_0;\tau)\,-\,B(q_0;\tau)\,(q-q_0)\,
\nonumber\\
&\qquad\qquad\qquad\qquad+\,C(q_0;\tau)\,(q-q_0)^3+\cdots
\label{eq:ballistic_shellcrossing}
\end{align}
with the time-dependent coefficients $A$, $B$ and $C$ defined by  
\begin{align}
A(q_0;\tau)& \equiv x(q_0;\tau)=q_0+D_+(\tau)\,\psi(q_0),
\label{eq:def_A}
\\
B(q_0;\tau)&\equiv -\left.\frac{\partial x}{\partial q}\right|_{q_0}=
-1-D_+(\tau)\psi'(q_0) 
\nonumber\\
&= \left\{D_+(\tau)-D_+(\tau_0)\right\}\delta_{\rm L}(q_0),
\label{eq:def_B}
\\
C(q_0;\tau)&\equiv 
\frac{1}{6}\left.\frac{\partial^3 x}{\partial q^3}\right|_{q_0}=
\frac{1}{6}\,D_+(\tau)\,\psi'''(q_0)
\nonumber\\
&=-\frac{1}{6}\,D_+(\tau)\,\delta_{\rm L}''(q_0),
\label{eq:def_C}
\end{align}
where the prime denotes the derivative with respect to $q$. 
Here, we used Eqs.~(\ref{eq:def_deltaL}) and (\ref{eq:shellcrossing_peak}). The above expressions imply that the shell-crossing point slightly moves from 
$x=x(q_0;\tau_0)$ to $x=A(q_0;\tau)$ and the shell-crossing structure develops as an ``$\mathcal{S}$'' shape, as shown in Fig.~\ref{fig:shellcrossing}. 

We now introduce the new Lagrangian coordinate, $Q\equiv q-q_0$, for which the structure of the local density peak is symmetric. Since we expand the displacement field at third-order, the equation $x(q,\tau)=x_0$ inside the shell-crossing region has three solutions. An example is shown in Fig.~\ref{fig:shellcrossing} for the equation $x(Q)=x(Q_*)$ with solutions $Q_-<Q_*<Q_+$. In general, the three ordered roots of the equation $x(Q)={\tilde x}$, when they exist, are related to each other through the following expressions:
\begin{align}
Q_*&=\frac{1}{2}\,\left\{-Q_\pm\pm\sqrt{3(\widehat{Q}_{\rm c}^2-Q_\pm^2)}\right\},
\label{eq:relation_three-values_1}
\\
Q_\pm&=\frac{1}{2}\,\left\{-Q_*\mp\sqrt{3(\widehat{Q}_{\rm c}^2-Q_*^2)}\right\}
\nonumber\\
    &=\frac{1}{2}\,\left\{-Q_\mp\pm\sqrt{3(\widehat{Q}_{\rm c}^2-Q_\mp^2)}\right\}. 
\label{eq:relation_three-values_2}
\end{align}
Here, the Lagrangian extent $\Qchat$ of the boundary of the multi-valued region, also related to the Lagrangian position $\Qc$ of the left caustic in Fig.~\ref{fig:shellcrossing} through $x(\Qchat)=x(\Qc)$, is expressed as 
\begin{align}
\Qchat &=2\Qc=\sqrt{\frac{4B}{3C}}.
\label{eq:Qchat_Qc}
\end{align}
Because of the time dependence of the coefficients (see Eqs.~\ref{eq:def_B} and \ref{eq:def_C}), the position of the boundary gradually changes in time. The leading-order expression in powers of time $(\tau-\tau_0)$ reads
\begin{align}
\Qchat\simeq\left\{\frac{8}{\kappa(q_0,\tau_0)}\right\}^{1/2}(\tau-\tau_0)^{1/2}
+\mathcal{O}\left((\tau-\tau_0)^{3/2}\right),
\label{eq:Qchat_approx}
\end{align}
with 
\begin{align}
\kappa(q_0,\tau_0)\equiv\frac{-\delta_{\rm L}''(q_0)\,D_+(\tau_0)}{\displaystyle \delta_{\rm L}(q_0)\,\frac{{\rm d}D_+(\tau_0)}{{\rm d}\tau_0}}.
\label{eq:def_kappa}
\end{align}
In other words, it also means that a given element of fluid of Lagrangian position $Q$ will enter the multi-valued region at some time $\widehat{\tau}_{\rm c}$ defined by $\Qchat(\widehat{\tau}_{\rm c}) \equiv |Q|$ and then subsequently coincide exactly with the Lagrangian position of the caustic at a time $\tau_{\rm c} > \widehat{\tau}_{\rm c}$ with $Q_{\rm c}(\tau_{\rm c}) \equiv |Q|$. By inverting the relations $|Q|=Q_{\rm c}(\tau_{\rm c})$ and $|Q|=\Qchat(\widehat{\tau}_{\rm c})$, we obtain
\begin{align}
&\widehat{\tau}_{\rm c}(Q)-\tau_0 
\simeq \frac{1}{8}\kappa(q_0,\tau_0)\,Q^2 
\label{eq:hattau_c}
\end{align}
and
\begin{align}
&\tau_{\rm c}(Q)-\tau_0 = \widehat{\tau}_{\rm c}(2Q)-\tau_0
\simeq \frac{1}{2}\kappa(q_0,\tau_0)\,Q^2.
\label{eq:tau_c}
\end{align}
The quantities defined above thus play a crucial role to disentangle the single-stream regime from post-collapse dynamics, as well as the inner part of the ``${\cal S}$'' shape of Fig.~\ref{fig:shellcrossing} from its outer part, the Lagrangian position of the caustic acting as a separator in the latter case. The expression of the force will indeed be different between the Zel'dovich single-flow regime $|Q| > \Qchat$ and the multi-valued regime composed of an outer caustic region, $ \Qc <|Q| < \Qchat$, and an inner caustic region, $|Q| < \Qc$. 

One important remark for the subsequent calculations is that the boundaries coordinates $\Qchat$ and $Q_{\rm c}$ are assumed to be small and that the backreaction to the position and velocity inside the multi-valued region can be described perturbatively in a polynomial form of $Q=q-q_0$ [partly allowing the fractional power of $(\Qchat^2-Q^2)$]. Thus, the resultant expressions are, rigorously speaking, only valid for a short period after collapse time, but we shall see in practice, as also shown by \citet{Colombi:2014lda} in the non-cosmological case, that they remain impressively accurate even up to next-crossing time (see Sec.~\ref{sec:comparison_nbody}), which will allow us to set up the framework for a powerful self-adaptive scheme.

\subsection{Computing the force in the multi-valued region}
\label{subsec:computing_force}

To derive the corrections to the motion, we first compute the force exerted on a mass element inside the multi-valued region, $-\Qchat\leq Q\leq \Qchat$ (see Fig.~\ref{fig:shellcrossing}). Note again that the outer regions $x<x(-\Qchat)$ and $x>x(\Qchat)$ are described by the Zel'dovich solution. 

The force in the multi-valued region is computed using Eq.~(\ref{eq:Force_Poisson}), dividing each integral of the right-hand-side into three contributions: 
\begin{align}
\int_0^L\,{\rm d}x\quad \longrightarrow
\Bigl(\int_0^{x(-\Qchat)}+\int_{x(-\Qchat)}^{x(\Qchat)}+\int_{x(\Qchat)}^{L}\Bigr)\,{\rm d}x.
\label{eq:part_by_part_integrals}
\end{align}
Assuming that the collapse region, $|Q|\leq\Qchat$, is small enough, the contributions to the integrals from each domain can be computed analytically, based on the geometrical setup in Fig.~\ref{fig:shellcrossing}.  The detailed calculations are presented in Appendix \ref{sec:derivation}.  Summing up all the contributions given in Eqs.~(\ref{eq:integral_inner1}), (\ref{eq:integral_inner2}), (\ref{eq:integral_outer1}) and (\ref{eq:integral_outer2}), the force exerted on the mass element at $x=x(Q)$ inside the multi-valued region becomes 
\begin{align}
&F(x(Q;\,\tau))
=-\frac{3}{2}H_0^2\Omega_{\rm m,0}\,a(\tau)\,
\Bigl[ \mathcal{J}(Q;q_0,\tau) +\mathcal{F}(q_0,\tau)\Bigl]
\label{eq:force_inner}
\end{align}
with the functions $\mathcal{J}$ and $\mathcal{F}$ respectively defined by
\begin{align}
& \mathcal{J}(Q;q_0,\tau) =\left\{
\begin{array}{lc}
\Bigl\{1+B(q_0;\tau)\Bigr\}Q-C(q_0;\tau)\,Q^3 
\\
-\mbox{sgn}(Q)
\sqrt{3(\hat{Q}_{\rm c}^2-Q^2)} 
\\ \qquad\qquad; Q_{\rm c}<|Q|<\Qchat,
\\
\\
\Bigl\{-2+B(q_0;\tau)\Bigr\}Q-C(q_0;\tau)\,Q^3 
\\\qquad\qquad; |Q|<Q_{\rm c},
\end{array}
\right.
\label{eq:force_inner_Q}
\end{align}
and
\begin{align}
\mathcal{F}(q_0,\tau)=-\psi(q_0) \,D_+(\tau),
\label{eq:force_inner_const}
\end{align}
where the quantities $A$, $B$, and $C$ are defined by Eqs.~(\ref{eq:def_A})--(\ref{eq:def_C}). Note that in deriving Eq.~(\ref{eq:force_inner}), we have assumed that the system follows Zel'dovich solution if $|Q|>\Qchat$. Since the resultant expressions are written in terms of the local quantities characterizing the density peak at position $q_0$ and the shell-crossing time $\tau_0$, Eq.~(\ref{eq:force_inner}) is in fact still applicable to other shell-crossing regions possibly appearing elsewhere in the region  $|Q|>\Qchat$, allowing one to generalize the result to a smooth random initial field with multiple peaks. One issue, discussed later, is then to treat mergers, i.e. the case when for instance two multi-stream regions start overlapping.

\subsection{Corrections to the Zel'dovich flow: basic post-collapse PT results}
\label{subsec:corrections_pcpt}

Given the explicit expression for the force in the multi-stream region and using the formal solution given by Eqs.~(\ref{eq:sol1_x}) and (\ref{eq:sol1_p}), we now compute corrections to the Zel'dovich flow that we write as follows:
\begin{align}
&\Delta\rmv(Q;\tau,\,\widehat{\tau}_{\rm c})=\int_{\widehat{\tau}_{\rm c}}^\tau {\rm d}\tau'
\, F(x(Q,\tau')),
\label{eq:Delta_p_formal}\\
&\Delta x(Q;\tau,\,\widehat{\tau}_{\rm c})=\int_{\widehat{\tau}_{\rm c}}^\tau {\rm d}\tau'
\,\Delta \rmv(Q;\tau',\widehat{\tau}_{\rm c}).
\label{eq:Delta_x_formal}
\end{align}
We noticed in previous section that, depending on the Lagrangian position of interest, the expression for the force is different and we have to divide the domain of the integrals in Eqs.~(\ref{eq:Delta_p_formal}) and (\ref{eq:Delta_x_formal}) 
into several pieces:
\begin{enumerate}
\item $\tau_0\leq\tau<\widehat{\tau}_{\rm c}(Q)$:  the position $Q$ is located in the single-valued region (i.e., $|Q|>Q_{\rm c}$) and the motion is still described by the Zel'dovich solution. We have
\begin{align}
&x(Q;\tau)=x_{\rm Zel}(Q;\tau)\equiv q+\psi(q)D_+(\tau),
\label{eq:x_pre-collapse}\\
&\rmv (Q;\tau)=\rmv_{\rm Zel}(Q;\tau)\equiv \psi(q)\frac{{\rm d}D_+(\tau)}{{\rm d}\tau}.
\label{eq:v_pre-collapse}
\end{align}
\end{enumerate}

\smallskip

\begin{enumerate}
\setcounter{enumi}{1}
\item $\widehat{\tau}_{\rm c}(Q)\leq\tau<\tau_{\rm c}(Q)$:  the position $Q$ is in the multi-valued region and satisfies $Q_{\rm c} <|Q|\leq\Qchat$, i.e. lies in the outer part of the caustic. In addition to the Zel'dovich flow, the corrections arising from the multi-stream flow need to be added and we note them as follows:
\begin{align}
&x(Q;\tau)=x_{\rm Zel}(Q;\,\widehat{\tau}_{\rm c}(Q))+\Delta x_{\rm out}(Q;\tau,\,\widehat{\tau}_{\rm c}(Q)),
\label{eq:x_post-collapse_out}\\
&\rmv(Q;\tau)=\rmv_{\rm Zel}(Q;\,\widehat{\tau}_{\rm c}(Q))+\Delta\rmv_{\rm out}(Q;\tau,\,\widehat{\tau}_{\rm c}(Q)).
\label{eq:v_post-collapse_out}
\end{align}
In this region, which correspond to the tails of the ${\cal S}$ shape in Fig.~\ref{fig:shellcrossing}, the system is globally expanding in phase space and gaining energy, at variance with the central part, which on the contrary, contracts. This can be understood from computing the variation of energy of a typical test particle during a fraction of orbit. This phenomenon was studied in the non-cosmological case by \cite{Colombi:2014lda} and the results should not be fundamentally different in the cosmological case studied here. 
\end{enumerate}

\smallskip

\begin{enumerate}
\setcounter{enumi}{2}
\item $\tau_{\rm c}(Q)\leq\tau$:  this corresponds to $|Q|\leq Q_{\rm c}$, i.e. the position $Q$ now lies in the inner part of the multi-stream region with respect to the caustic. Similarly to the above case, the backreaction to Zel'dovich flow needs to be computed, including at present both the multi-stream dynamics of the inner part and the incoming flow from the outer part. We may write
\begin{align}
&x(Q;\tau)=x_{\rm Zel}(Q;\,\widehat{\tau}_{\rm c}(Q))+\Delta x_{\rm in}(Q;\tau,\,\widehat{\tau}_{\rm c}(Q)),
\label{eq:x_post-collapse_inner}\\
&\rmv(Q;\tau)=\rmv_{\rm Zel}(Q;\,\widehat{\tau}_{\rm c}(Q))+\Delta\rmv_{\rm in}(Q;\tau,\,\widehat{\tau}_{\rm c}(Q)). 
\label{eq:v_post-collapse_inner}
\end{align}
Note that the corrections $\Delta x_{\rm in}$ and $\Delta \rmv_{\rm in}$ partly come from outer part contributions, $\Delta x_{\rm out}(Q;\tauc,\tauchat)$ and $\Delta \rmv_{\rm out}(Q;\tauc,\tauchat)$. As mentioned in the previous point (ii), in this region, which corresponds to the central part of the ${\cal S}$ on Fig.~\ref{fig:shellcrossing}, the system is globally contracting and losing energy in favor of the tails of the ${\cal S}$. 
\end{enumerate}

\smallskip

In what follows, we compute the backreaction to Zel'dovich flow and derive the expressions for $\Delta x$ and $\Delta \rmv$ in each domain. The calculation for the corrected motion is rather straightforward but needs several steps. Here, we give a brief sketch of the calculation, deferring details to Appendix \ref{sec:pcpt_basic}. The final results are given in Eqs.~(\ref{eq:p_for_tau_c>tau>hattau_c}) and (\ref{eq:x_for_tau_c>tau>hattau_c}) for the outer part,  Eqs.~(\ref{eq:p_for_tau>tau_c}) and (\ref{eq:x_for_tau>tau_c}) for the inner part, together with the coefficients in Tables~\ref{tab:xv_pcpt1} and \ref{tab:xv_pcpt2}. Note that although our expansion is rigorously valid only at third-order in $Q$, higher-order contributions will appear in the corrections to provide a continuous solution, up to fifth- and seventh-order for the velocity and the position, respectively. 

\begin{table*}
 \caption{Coefficients for the corrections to Zel'dovich flow in the outer part of the collapse region, $\Qc<|Q|\leq\Qchat$.}
 \label{tab:xv_pcpt1}
 \begin{tabular}{cll}
  \hline
  Coefficients & velocity $\Delta \rmv_{\rm out}$ [Eq.~(\ref{eq:p_for_tau_c>tau>hattau_c})] & position $\Delta x_{\rm out}$ [Eq.~(\ref{eq:x_for_tau_c>tau>hattau_c})]\\
\\
  \hline
  $\widetilde{\alpha}$  & $T\,(\equiv\tau-\tau_0)$ & ${\displaystyle \frac{T^2}{2}}$\\
\\
  $\widetilde{\beta}$   & ${\displaystyle T\,\frac{\delta_{\rm L}''(q_0)}{6}\,D_+(\tau_0)- 
\frac{\kappa}{8}}$ & ${\displaystyle -\frac{\kappa}{8}\,T+\frac{\delta_{\rm L}''(q_0)}{12}D_+(\tau_0) \,T^2}$ \\
\\
  $\widetilde{\gamma}$  & ${\displaystyle -\mbox{sgn}(Q)\frac{\kappa}{4\sqrt{3}} }$ & ${\displaystyle -\mbox{sgn}(Q)\,\frac{\kappa^2}{80\sqrt{3}}}$\\
\\
  $\widetilde{\delta}$  & ${\displaystyle -\frac{\delta_{\rm L}''(q_0)}{48}\kappa\,D_+(\tau_0)}$ & ${\displaystyle \frac{1}{2}\left(\frac{\kappa}{8}\right)^2-\left(\frac{\kappa}{8}\right)\,\frac{\delta_{\rm L}''(q_0)}{6}D_+(\tau_0)
\,T}$ \\
\\
  $\widetilde{\zeta}$  & \rule{1.5cm}{0.1pt} & ${\displaystyle \frac{1}{2}\left(\frac{\kappa}{8}\right)^2 \frac{\delta_{\rm L}''(q_0)}{6}D_+(\tau_0)}$ \\
\\
$\widetilde{\epsilon}$ & ${\displaystyle \psi(q_0)\Bigl[\frac{{\rm d} D_+(\tau')}{{\rm d}\tau'}\Bigr]^\tau_{\widehat{\tau}_{\rm c}(Q)}}$ & ${\displaystyle \psi(q_0)\,\Biggl\{D_+(\tau)-D_+(\widehat{\tau}_{\rm c}(Q))-\left.\frac{{\rm d}D_+}{{\rm d}\tau}\right|_{\widehat{\tau}_{\rm c}(Q)}\,(\tau-\widehat{\tau}_{\rm c}(Q))\Biggr\}}$ \\
  \hline
 \end{tabular}
\end{table*}

\subsubsection{Velocity and position in the outer part: $\Qc<|Q|\leq \Qchat$}
\label{subsubsec:v_and_x_at_Qc<|Q|<Qchat}

In the outer part of the multi-valued region, (ii), the correction to the velocity can be expressed as
\begin{align}
&\Delta \rmv_{\rm out} (Q;\tau,\,\widehat{\tau}_{\rm c})=-\frac{3}{2}H_0^2\,\Omega_{\rm m,0}\,
\int_{\widehat{\tau}_{\rm c}(Q)}^\tau {\rm d}\tau'\,a(\tau')\,
\nonumber\\
&\qquad\qquad\qquad\qquad\qquad\times\Bigl\{\mathcal{J}(Q;\,q_0,\tau')+ \mathcal{F}(q_0,\tau')\Bigr\}.
\label{eq:v_ii}
\end{align}
In the above, while the first integral is performed with the help of formulae in Appendix \ref{sec:integral}, the second integral is computed exactly. The whole derivation is provided in Appendix \ref{subsec:Delta_v_out}. The resultant expression can be summarized as
\begin{align}
& \Delta\rmv_{\rm out}(Q;\tau,\,\widehat{\tau}_{\rm c})=-\frac{3}{2}H_0^2\,\Omega_{\rm m,0}\,a(\tau_0)
\Bigl[\,\widetilde{\alpha}_1(\tau)\,Q+ 
\widetilde{\beta}_1(\tau)\,Q^3
\nonumber\\
&\quad\quad+
\widetilde{\gamma}_1(\tau_0)\,
\Bigl\{\Qchat^2(\tau)-Q^2\Bigr\}^{3/2}+
\widetilde{\delta}_1(\tau_0)
\,Q^5\Bigr] +\,
\widetilde{\epsilon}_1(\tau,\widehat{\tau}_{\rm c}),
\label{eq:p_for_tau_c>tau>hattau_c}
\end{align}
with the time-dependent coefficients given in Table~\ref{tab:xv_pcpt1}. Note that the coefficient $\widetilde{\epsilon}_1$ implicitly depends on the Lagrangian position $Q$ through $\widehat{\tau}_{\rm c}(Q)\simeq \tau_0+(\kappa/8)\,Q^2$ but is not Taylor expanded with respect to $Q$, for simplicity. A fully analytical theory, in particular to predict the power spectrum of the projected density field, would in principle require such a Taylor expansion. Performing it should not change significantly the performances of post-collapse PT as presented in this article.

Once $\Delta \rmv_{\rm out}$ is obtained, the expression for the correction $\Delta x_{\rm out}$ is evaluated by further integrating Eq.~(\ref{eq:p_for_tau_c>tau>hattau_c}) over time:  
\begin{align}
&\Delta x_{\rm out}(Q;\tau,\,\widehat{\tau}_{\rm c})=\int_{\widehat{\tau}_{\rm c}(Q)}^\tau {\rm d}\tau'
\,\Delta\rmv_{\rm out}(Q,\tau').
\label{eq:x_ii}
\end{align}
The resultant expression becomes (see Appendix \ref{subsec:Delta_x_out} for derivation): 
\begin{align}
& \Delta x_{\rm out}(Q;\tau,\widehat{\tau}_{\rm c})=-\frac{3}{2}H_0^2\,\Omega_{\rm m,0}\,a(\tau_0)
\Bigl[\,\widetilde{\alpha}_2(\tau)\,Q
+\widetilde{\beta}_2(\tau)\,Q^3
\nonumber\\
&\quad\quad +\widetilde{\gamma}_2(\tau_0)\,
\Bigl\{\Qchat^2(\tau)-Q^2\Bigr\}^{5/2}
+\widetilde{\delta}_2(\tau)\,Q^5+
\widetilde{\zeta}_2(\tau)\,Q^7\,
\Bigr]
\nonumber\\
&\quad\quad +\,\widetilde{\epsilon}_2(\tau,\widehat{\tau}_{\rm c}),
\label{eq:x_for_tau_c>tau>hattau_c}
\end{align}
with the time-dependent coefficients presented in Table~\ref{tab:xv_pcpt1}. Note again the dependence on $Q$ of parameter $\widetilde{\epsilon}_2(\tau,\widehat{\tau}_{\rm c})$ through $\widehat{\tau}_{\rm c}$ and the fact that we did not Taylor expand it in polynomials of $Q$, as it would be required for a fully analytical theory.

\subsubsection{Velocity and position in inner part: $|Q|\leq \Qc$}
\label{subsubsec:v_and_x_at_|Q|<Qc}

Let us consider now the inner part of the multi-valued region (iii). For the correction to the velocity, $\Delta\rmv_{\rm in}$, 
the expression to be evaluated is divided into three contributions: 
\begin{align}
&\Delta \rmv_{\rm in}(Q;\tau,\,\widehat{\tau}_{\rm c})=-\frac{3}{2}H_0^2\,\Omega_{\rm m,0}\,
\Biggl[
\int_{\widehat{\tau}_{\rm c}(Q)}^{\tau_{\rm c}(Q)} {\rm d}\tau'\,a(\tau')\,\mathcal{J}(Q;\,q_0,\tau')
\nonumber\\
& \quad \quad \quad \quad +
\int_{\tau_{\rm c}(Q)}^{\tau} {\rm d}\tau'\,a(\tau')\,\mathcal{J}(Q;\,q_0,\tau') \nonumber \\
& \quad \quad \quad \quad +
\int_{\widehat{\tau}_{\rm c}(Q)}^{\tau} {\rm d}\tau'\,a(\tau')\,\mathcal{F}(q_0,\tau')
\Biggr].
\label{eq:Delta_p_2} 
\end{align}
Evaluating each contribution in Appendix \ref{subsec:Delta_v_in}, we obtain
\begin{align}
&\Delta\rmv_{\rm in}(Q;\tau,\widehat{\tau}_{\rm c})=-\frac{3}{2}H_0^2\,\Omega_{\rm m,0}\,a(\tau_0)
\Biggl[\,\widetilde{\alpha}_3(\tau)\,Q 
+ \widetilde{\beta}_3(\tau)\,Q^3
\nonumber\\
&\quad\quad +
\widetilde{\delta}_3(\tau_0)\,Q^5\Biggr]\,+\,
\widetilde{\epsilon}_1(\tau, \widehat{\tau}_{\rm c}),
\label{eq:p_for_tau>tau_c}
\end{align}
with the time-dependent coefficients given in Table~\ref{tab:xv_pcpt2}. 
Note that the coefficient $\widetilde{\epsilon}_1$ in the above equation is the same one as in Eq.~(\ref{eq:p_for_tau_c>tau>hattau_c}).
\begin{table*}
 \caption{Coefficients for the corrections to Zel'dovich flow in the inner part of the collapse region, $|Q|\leq\Qc$.}
 \label{tab:xv_pcpt2}
 \begin{tabular}{cll}
  \hline
  Coefficients & velocity $\Delta\rmv_{\rm in}$ [Eq.~(\ref{eq:p_for_tau>tau_c})] & position $\Delta x_{\rm in}$ [Eq.~(\ref{eq:x_for_tau>tau_c})]\\
\\
  \hline
  $\widetilde{\alpha}$  & $2T$ & ${\displaystyle -T^2}$\\
\\
  $\widetilde{\beta}$   & ${\displaystyle T\,\frac{\delta_{\rm L}''(q_0)}{6}\,D_+(\tau_0)\,+
\frac{5}{8}\kappa }$ & ${\displaystyle \frac{5\kappa}{8}\,T+\frac{\delta_{\rm L}''(q_0)}{12}\,D_+(\tau_0)\,T^2}$ \\
\\
  $\widetilde{\delta}$  & ${\displaystyle 
-\left(\frac{\kappa}{8}\right)\,\frac{\delta_{\rm L}''(q_0)}{6}\,D_+(\tau_0)}$ & ${\displaystyle -\left(\frac{\kappa}{4}\right)^2\,\frac{67}{40}
-\left(\frac{\kappa}{8}\right)\frac{\delta_{\rm D}''(q_0)}{6}\,D_+(\tau_0)\,T}$ \\
\\
  $\widetilde{\zeta}$  & \rule{1.5cm}{0.1pt} & ${\displaystyle \frac{1}{2}
\left(\frac{\kappa}{8}\right)^2\frac{\delta_{\rm L}''(q_0)}{6}\,D_+(\tau_0)}$ \\
\\
$\widetilde{\epsilon}$ & ${\displaystyle \psi(q_0)\Bigl[\frac{{\rm d} D_+(\tau')}{{\rm d}\tau'}\Bigr]^\tau_{\widehat{\tau}_{\rm c}(Q)}}$ & ${\displaystyle \psi(q_0)\,\Biggl\{D_+(\tau)-D_+(\widehat{\tau}_{\rm c}(Q))-\left.\frac{{\rm d}D_+}{{\rm d}\tau}\right|_{\widehat{\tau}_{\rm c}(Q)}\,(\tau-\widehat{\tau}_{\rm c}(Q))\Biggr\}}$ \\
  \hline
 \end{tabular}
\end{table*}

Similarly, the correction to the position, $\Delta x_{\rm in}$, is divided into three pieces:
\begin{align}
\Delta x_{\rm in}(Q;\tau,\,\widehat{\tau}_{\rm c})&=\left(\int_{\widehat{\tau}_{\rm c}(Q)}^{\tau_{\rm c}(Q)} {\rm d}\tau'
+\int_{\tau_{\rm c}(Q)}^{\tau} {\rm d}\tau'\right)\,\Delta \widetilde{\rmv}(Q;q_0,\tau')
\nonumber\\
&+\widetilde{\epsilon}_2(\tau,\widehat{\tau}_{\rm c}),
\label{eq:x_iii}
\end{align}
where the function $\Delta \widetilde{\rmv}$ represents the correction of the velocity given in either Eq.~(\ref{eq:p_for_tau_c>tau>hattau_c}) or Eq.~(\ref{eq:p_for_tau>tau_c}), but without the contribution corresponding to the Zel'dovich  motion, i.e., the last term in their expressions. The term $\widetilde{\epsilon}_2$ is the same quantity as in Eq.~(\ref{eq:x_for_tau_c>tau>hattau_c}) (see Table~\ref{tab:xv_pcpt1}).  

Based on the calculation presented in Appendix \ref{subsec:Delta_x_in}, we obtain the expression for $\Delta x_{\rm in}$:  
\begin{align}
& \Delta x_{\rm in}(Q;\tau,\widehat{\tau}_{\rm c})=-\frac{3}{2}H_0^2\,\Omega_{\rm m,0}\,a(\tau_0)
\Biggl[\,\widetilde{\alpha}_4(\tau)\,Q
+ \widetilde{\beta}_4(\tau)\,Q^3
\nonumber\\
&\quad\quad+
\widetilde{\delta}_4(\tau)\,Q^5+
\widetilde{\zeta}_4(\tau)\,Q^7\,
\Biggr]\,
+\,\widetilde{\epsilon}_2(\tau,\widehat{\tau}_{\rm c}),
\label{eq:x_for_tau>tau_c}
\end{align}
with the time dependent coefficients given in Table~\ref{tab:xv_pcpt2}.

\subsection{Higher-order corrections}
\label{subsec:higher-order}

The perturbative description of post-collapse dynamics introduces two kinds of approximations around the singularities (shell-crossing points). Firstly, we assumed the position and velocity to be described by third-order polynomials in $Q$ in the multi-stream regions. To preserve continuity between the multi-stream and the single-stream regime, we added higher-order terms in $Q$ to the solutions for $x$ and $v$, but strictly speaking, these solutions are valid locally only at third-order in $Q$ around the singularities. Secondly, these solutions are correct only very shortly after collapse time $\tau_0$ up to second-order in $\tau-\tau_0$. Our post-collapse dynamics can thus be improved at a twofold level: 
\begin{description}
\item[(a)]{\em Improvement at the spatial level:} we could try, starting from initial conditions, to use a description of the ${\cal S}$ shape of Fig.~\ref{fig:shellcrossing} at higher-order in $Q$, e.g. fourth- or fifth- instead of third-order. In this latter case, deriving the solution of equation $x(Q)=x(Q_*)$ comes down to find the roots of a third or fourth-order polynomial instead of a second-order one, which makes, in addition to the symmetry breaking of the ${\cal S}$ shape, the calculations much more cumbersome, even though a series expansion of the roots at fourth or fifth-order in $Q$ remains tractable and might bring interesting improvements on the description of the tails of the ${\cal S}$ shape. 
\end{description}

\begin{description}
\item[(b)]{\em Higher-order in time:} we can introduce higher-order corrections in time to follow more accurately the evolution of the central part of the singularity after collapse. Note that after integration of the equations of motion, such corrections bring out higher-order terms in $Q$. Strictly speaking, the right way to proceed would consist of performing ``post-post-collapse'' dynamics, i.e. of computing a counter-term to the Zel'dovich+leading-order post-collapse solution. This procedure is cumbersome, and seems to make sense only if performed simultaneously with a description at higher-order in $Q$, in particular because the tails of the ${\cal S}$ structure in Fig.~\ref{fig:shellcrossing} also influence the dynamics of the central part of the system (but this remains to be proved). 
\end{description}

Going beyond leading-order in post-collapse dynamics thus seems rather challenging even in the simple one-dimensional case considered in this article. 

However, we can still try to investigate some potential improvements of the theory while maintaining the complexity of the calculations at an acceptable level. For this purpose, we consider here mainly two approaches.  The first one just consists in incorporating next-to-leading order terms into the expressions for $\Qchat$ and $\tauchat$ (and $\Qc$ and $\tau_{\rm c}$). These quantities, which determine the boundary between the single-stream  and the multi-stream regime, are indeed crucial because they correspond to the main scale/timescale of the system. The second approach consists, in addition, of improving the evaluation of the time integrals in Eqs.~(\ref{eq:Delta_p_formal}) and (\ref{eq:Delta_x_formal}). In Sec.~\ref{subsec:corrections_pcpt}, when evaluating these time integrals, a part of the integrands is Taylor-expanded and the results are presented at leading-order in time. It is thus possible to include higher-order corrections from the Taylor-expanded integrands. 

Again, one has to be aware that the two corrections we propose might not improve the results, since a correct higher-order time approach, as discussed in point (b) above, would actually require to compute counter terms to the standard post-collapse result, which we do not do here. In fact, our higher-order corrections might even {\em worsen} the results since the corrections brought by such additional terms might have opposite sign compared to the iterated counter-term approach. 

The derivation of the higher-order corrections we propose here and summarize just below is rather straightforward compared to its rigorous alternatives (a) and (b) above, but still cumbersome. The resultant expressions now include higher-order polynomials of $Q$ as well as new contributions, which are given in Appendix \ref{sec:pcpt_higher-order}. To be more specific, we consider the following cases:
\begin{enumerate}
\item Higher-order for the expressions of critical time ${\hat \tau}_{\rm c}$ and position ${\hat Q}_{\rm c}$ ({\em hc}): based on the basic results in Sec.~\ref{subsec:corrections_pcpt}, we include the contributions arising from the next-to-leading order expressions for $\Qchat$ and $\tauchat$ (as well as $\Qc$ and $\tauc$). The calculation to derive higher-order terms is almost the same as presented in Sec.~\ref{subsec:corrections_pcpt}, but we replace the leading-order expressions for $\Qchat$ and $\tauchat$ in Eqs.~(\ref{eq:Qchat_approx}) and (\ref{eq:hattau_c}) with those including the next-to-leading order, given in Eqs.~(\ref{eq:higher_order_Qhat}) and (\ref{eq:higher_order_hattau_c}).  The higher-order results, denoted by $\Delta x^{\rm(hc)}$ and $\Delta \rmv^{\rm(hc)}$, correspond to the expressions summarized in Appendix \ref{sec:pcpt_higher-order}, but where the terms involving the quantities $g(\tau_0)$ or $H(\tau_0)$ are all set to zero. 
\end{enumerate}

\begin{enumerate}
\setcounter{enumi}{1}
\item Higher-order in time ({\em ho}): in addition to the above treatment, we also include higher-order corrections to the time-integrals, based on the formulas given in Appendix \ref{sec:integral}. The resultant expressions, $\Delta x^{\rm(ho)}$ and $\Delta \rmv^{\rm(ho)}$, are summarized in Appendix \ref{sec:pcpt_higher-order}. 
\end{enumerate}

Finally, we also consider a semi-analytic approach in order to smoothly connect the inner part of the multi-stream region to the Zel'dovich solution:
\begin{enumerate}
\setcounter{enumi}{2}
\item Third-order spline ({\em spl}): basic post-collapse PT is applied to the inner part, $|Q|\leq \Qc$, but we take the higher-order expression for $\Qc$.  In the outer part, $\Qc<|Q|<\Qchat$, on the other hand, a third-order spline interpolation is used to smoothly connect the inner post-collapse region to the Zel'dovich solution, with again, the higher-order expression for $\Qchat$. 
\end{enumerate}
\section{Improved treatment with adaptive smoothing}
\label{sec:improvement}
Post-collapse PT as described in \S~\ref{sec:postcollapse} provides a way to describe the dynamics around an initial density peak shortly beyond collapse time. In practice, the quality of the description is expected to be good until next-crossing time, at least this is what is known in the non-cosmological case \citep{Colombi:2014lda}. Using the leading-order expression given in Eq.~(\ref{eq:x_for_tau>tau_c}), the duration of time spent between collapse and next shell-crossing can be estimated from Eq.~(\ref{eq:x_post-collapse_inner}),
\begin{align}
\tau_{\rm cross}(q_0)\equiv \frac{\delta_{\rm L}(q_0)\,\displaystyle \frac{{\rm d}D_+(\tau_0)}{{\rm d}\tau_0}}
{\frac{3}{2} H_0^2\Omega_{\rm m,0}\,a(\tau_0)}, 
\label{eq:tau_cross}
\end{align}
which is derived by solving the equation $(\partial x/\partial q)_{q=q_0}=0$. 

Beyond time $\tau=\tau_0+\tau_{\rm cross}$, post-collapse PT is expected to fail, similarly as Zel'dovich approximation fails beyond $\tau=\tau_0$.  The only way to fix this problem at the fine grain level would be to use an iterative procedure similar to what was proposed by \cite{Colombi:2014lda} to reconstruct the spiral structure built up during the course of the dynamics. In addition, post-collapse PT assumes the initial density distribution to be sufficiently smooth. While this is certainly expected at very small scales in the standard Cold Dark Matter scenario, the issue of mergers arises. Indeed, post-collapse PT is valid only if the forming halo is isolated.  If there are two adjacent collapsing regions, they might, at some point, overlap in Lagrangian space, hence merge with each other in Eulerian space. After the merger, a separate description of each halo becomes inadequate. In this case, the prediction of post-collapse PT, which is in principle able to handle multiple but non-overlapping structures in Lagrangian space, will break down much earlier than expected. 

One simple way to fix both limitation in time and mergers issues, already applied in the past to the Zel'dovich approximation \citep[see, e.g.][and references therein]{Coles1993,Sahni95}, consists in trying to describe the dynamics at the coarse-grained level by introducing a smoothing procedure in Lagrangian space, so that post-collapse PT remains applicable, in particular that the condition $\tau \la \tau_0+\tau_{\rm cross}$ remains valid for the system with coarse-grained initial conditions. This means that single or composite halos as well as mergers are locally summarized by a simplified ${\cal S}$ shape (Fig.~\ref{fig:shellcrossing}) corresponding to a single peak of the coarse-grained initial density field and such that $\tau \la \tau_0+\tau_{\rm cross}$, while retaining the correct large-scale dynamics. The hope is that this simplified structure still captures the most important features of the nonlinear system, e.g. its mass and its typical size. 

Of course, the coarsening scale should depend on environment, hence on Lagrangian position $Q$, so for best results, {\em adaptive smoothing} of initial conditions has to be introduced. In this sense our method is very similar to the ``peak-patch'' treatment by \citet{1996ApJS..103....1B} \citep[see also][]{2002MNRAS.331..587M}, but we implemented it in the post-collapse PT framework. To be precise, the practical set up of post-collapse dynamics with adaptive smoothing is summarized as follows:

\begin{enumerate}
\item The initial density field is smoothed at various scales, employing the sharp-$k$ filter function in Fourier space: 
\begin{align}
W(k;k_{\rm cut})=\Theta(k_{\rm cut}-k)
\label{eq:sharp-k}
\end{align}
with $\Theta(x)$ being the Heaviside step function. The cutoff wavenumber $k_{\rm cut}$ ranges from the fundamental mode $k_{\rm min}=2\pi/L$ to a certain value $k_{\rm max}=2\pi\, m_{\rm max}/L$, where $m_{\rm max}$ is an integer corresponding to the effective resolution of the simulation, as discussed more in details in \S~\ref{subsec:simulation}. In our investigations, we covered all the possible values of $k_{\rm cut}$, i.e. $k_{\rm cut}=2\pi\, m/L$ with $m=1,2,\cdots,m_{\rm max}$, but this is costly and probably not required, in practice.
\end{enumerate}

\begin{enumerate}
\setcounter{enumi}{1}
\item Zel'dovich solutions are constructed from the smoothed density fields. Then we identify the critical points $q_0$ in Lagrangian space that have undergone shell-crossing at time $\tau_0\,\, <\tau$, where $\tau$ is the time of interest for the analyses. For each critical point, various quantities needed for perturbative calculations, such as $\delta_{\rm L}(q_0)$ and $\delta_{\rm L}''(q_0)$, are also computed. 
\end{enumerate}

\begin{enumerate}
\setcounter{enumi}{2}
\item Starting from the largest smoothing scale (i.e. $k_{\rm min}$), we compute the next-crossing time $\tau_{\rm cross}$ associated to each critical point. For the points satisfying $\tau \geq \tau_0+f_{\rm cross}\,\tau_{\rm cross}$ with $f_{\rm cross} \leq 1$ a constant parameter, the post-collapse correction to the Zel'dovich flow is obtained in the interval $I_0=[q_0-\Qchat(q_0),\,q_0+\Qchat(q_0)]$, which is labeled. 
\end{enumerate}

\begin{enumerate}
\setcounter{enumi}{3}
\item The procedure (iii) is repeated, but with a smaller smoothing scale, i.e., larger $k_{\rm cut}$, and only for peaks lying in the unlabeled regions, until the cutoff  wavenumber reaches $k_{\rm max}$.  Note that if two Lagrangian regions overlap, e.g. $I_0(k_{{\rm cut}, 1})$ and $I_0(k_{{\rm cut},2})$,  corresponding to two different values of the smoothing scale,  $k_{{\rm cut},1} < k_{{\rm cut},2}$, the interval $I_0(k_{{\rm cut},2})$ corresponding to the smallest smoothing scale, i.e. largest wavenumber $k_{{\rm cut},2}$, prevails. This is the way we solve the cloud-in-cloud problem, but not the only possible choice. 
\end{enumerate}

\begin{enumerate}
\setcounter{enumi}{4}
\item At the end of the iterative process, post-collapse dynamics is applied as well for the rest of the critical points corresponding to the highest resolution scale  $2\pi/k_{\rm max}$ that still require treatment, i.e. for those with $\tau_0 < \tau < \tau_0+f_{\rm cross}\,\tau_{\rm cross}$. The procedure is the same as described in point (iv) above, i.e. only critical points which do not fall in a labeled region are treated, and if it is the case, then their Lagrangian interval $I_0$ prevails for applying post-collapse dynamics. For the rest of the unlabeled regions, standard Zel'dovich dynamics is applied and the mapping uses the initial density corresponding, of course, to the highest resolution scale.
\end{enumerate}

Note that the approximation of the phase-space density distribution constructed this way is generally discontinuous, because we collect perturbative solutions for the displacement field in phase space coming from different coarse-graining scales without imposing smoothness at the transition between these solutions. In this respect, the procedure given above is not entirely consistent and remains to be improved in the future. Nevertheless, the discontinuities only affect the results at small scales and in regions where the projected density is small. Indeed, the discontinuities correspond to transitions between tails of the ${\cal S}$ shape representing a halo and a non collapsed region or transitions between tails of two ${\cal S}$ shapes in the case of a merger. As illustrated below through numerical examples, these regions of phase space have small density contrast and thus contribute weakly to second- or higher-order statistics, hence to the power spectrum which is the statistics under focus in this work. With an appropriate choice of $f_{\rm cross}$, we shall show that the adaptive smoothing procedure provides a substantial improvement on the post-collapse prediction for the overall phase-space structure of the system and the power spectrum of the projected density.

\section{Comparison with controlled $N$-body experiments}
\label{sec:comparison_nbody}

We are now in position to compare the predictions of post-collapse PT with simulation results. In \S~\ref{subsec:simulation}, we briefly describe the one-dimensional cosmological $N$-body code we wrote to perform the numerical analyses of this article and discuss various important parameters settings. We then present the results of the comparison between theory and numerical experiments in various cases with cold initial conditions: formation of an isolated single halo (\S~\ref{subsec:single_cluster}), merging of two neighboring halos (\S~\ref{subsec:merging_clusters}), as well as random initial conditions with Cold Dark Matter-like (CDM) and power-law power spectra (\S~\ref{subsec:CDM_initial} and \ref{subsec:power-law_initial}). In the later case, we shall discuss more in details highly nonlinear dynamics, in particular predictions from stable clustering, the expected outcome on the power spectrum from halos with power-law density profiles, as well as the statistical properties of the phase-space sheet resulting from a caustic distribution.  
\subsection{Simulations}
\label{subsec:simulation}
To solve numerically Vlasov-Poisson equations in one dimension, we wrote a Fortran90 particle-mesh code, {\tt Vlafroid}. In this code, publicly available through the following web page, {\tt www.vlasix.org}, Poisson equation is solved on a periodic grid of fixed resolution $N_{\rm grid}$ using Fast Fourier Transform, after projecting the $N_{\rm particles}$ particles on the grid using simple Cloud-in-Cell interpolation \citep[see, e.g.][]{1988csup.book.....H}. The particles positions and velocities are updated at each time step using a standard predictor-corrector integration scheme. After the predictor step,
\begin{align}
x(\tau+\Delta \tau/2) =x(\tau)+v(\tau) \Delta \tau/2,
\end{align}
Poisson equation is solved and the force is estimated on the grid by direct derivation in Fourier space. The acceleration $\gamma(t+\Delta t/2)$ of each particle is computed by linear interpolation of the force. This is followed by the corrector step:
\begin{align}
v(\tau+\Delta \tau)&=v(\tau)+\gamma(\tau+\Delta \tau/2) \Delta \tau, \\
x(\tau+\Delta \tau)&=x(\tau+\Delta \tau/2) + v(\tau+\Delta \tau) \Delta \tau/2.
\end{align}
The value of the time step $\Delta \tau$ is slowly varying with time and is bounded by 3 constraints: 
\begin{enumerate}
\item[(i)] a limit on the relative variations of the expansion factor during a time step to have a good description of the linear growing modes at early stages of the simulation,
\begin{align}
\Delta \tau \leq C_{\rm dloga} \left( \frac{{\rm d} \log a}{{\rm d}\tau} \right)^{-1},
\end{align}
\item[(ii)] a dynamical condition which states that the time step should be a small fraction of the smallest harmonic oscillator time scale \citep[see, e.g.][]{2014MNRAS.441.2414C},
\begin{align}
\Delta \tau \leq \frac{C_{\rm dyn}}{\sqrt{\displaystyle \frac{3}{2}a \,\Omega_{{\rm m},0}\,\rho_{\rm max}\,H_0^2 }},
\end{align}
where $\rho_{\rm max}$ is the maximum value of the projected density (normalized to be unity in average), 
\item[(iii)] the standard Courant-Friedrichs-Lewy (CFL) condition
\begin{align}
\Delta \tau \leq C_{\rm CFL} \frac{L}{N_{\rm grid}v_{\rm max}},
\end{align}
with $v_{\rm max}$ the maximum magnitude of the velocity, stating that a particle should not travel more than a fraction of a grid cell size during a time step. 
\end{enumerate}
For all the runs performed in this article, our practical and safe choice of the time step parameters was:
\begin{align}
C_{\rm dloga}=0.1, \quad C_{\rm dyn}=0.01, \quad C_{\rm CFL}=0.25.
\end{align}
The evolution of single or composite objects as studied in \S~\ref{subsec:single_cluster} and \ref{subsec:merging_clusters} does not pose any problem from the numerical point of view. Indeed, we consider here only a small number of dynamical times, so we do not expect significant defects due to the discrete nature of the particle distribution  \citep[for instance, such as described in][]{Melott1997}. The case of a random field, as studied in \S~\ref{subsec:CDM_initial} for CDM and in \S~\ref{subsec:power-law_initial} for power-law  power spectra, is less trivial. Indeed, one needs accurate numerical modeling of the dynamics of each single structure, which has to be sampled with many particles and be resolved with a sufficient number of grid elements. In particular, to avoid contamination by the discrete nature of the particle distribution, we require, in average, many particles per grid element:
\begin{align}
N_{\rm particles} \gg N_{\rm grid}.
\end{align}
To avoid the formation of small structures that would be insufficiently resolved by the computational grid, there must be a coarse-graining scale below which the initial conditions are smooth. This also facilitates, of course, the calculation of (post-collapse) PT predictions. Here we impose a cut-off on the power spectrum of initial conditions using the sharp-$k$ filter introduced in Eq.~(\ref{eq:sharp-k}), with 
\begin{align}
k_{\rm max} \ll N_{\rm grid} \frac{2\pi}{L}.
\end{align}
In practice, a choice which works well and adopted in this paper is
\begin{align}
N_{\rm particles}=10\, N_{\rm grid}=100\, k_{\rm max} \frac{L}{2\pi},
\end{align}
which means that $k_{\rm max}$ is five time smaller than the Nyquist frequency of the grid. Since the cut-off scale on the power spectrum can affect the dynamics compared to the true, un-smooth system, we shall perform some convergence studies to determine the available dynamic range in which the results are not influenced by $k_{\rm max}$. 

\subsection{Formation of a single structure}
\label{subsec:single_cluster}
Let us first present the simplest case of the formation of a single structure seeded with the following sinusoidal density perturbation:
\begin{align}
& \delta_{L}(x) =-A\,\cos\left(\frac{2\pi}{L}x\right).
\label{eq:delta_init_single_cluster}
\end{align}
We adopt the Einstein-de Sitter cosmology ($\Omega_{\rm m,0}=1$, $\Omega_{\rm \Lambda}=0$ with $h \equiv H_0/100=0.7$) and the box size of the simulation is arbitrarily set to $L=1$. The simulation was started at expansion factor $a=0.01$ with initial amplitude $A=0.1$ and was run using $(N_{\rm particles},N_{\rm grid})=(10,000; 1,000)$. PT analyses were performed on the fly as part of routines of the code, when needed. 

\begin{figure*}
\includegraphics[width=14.4cm]{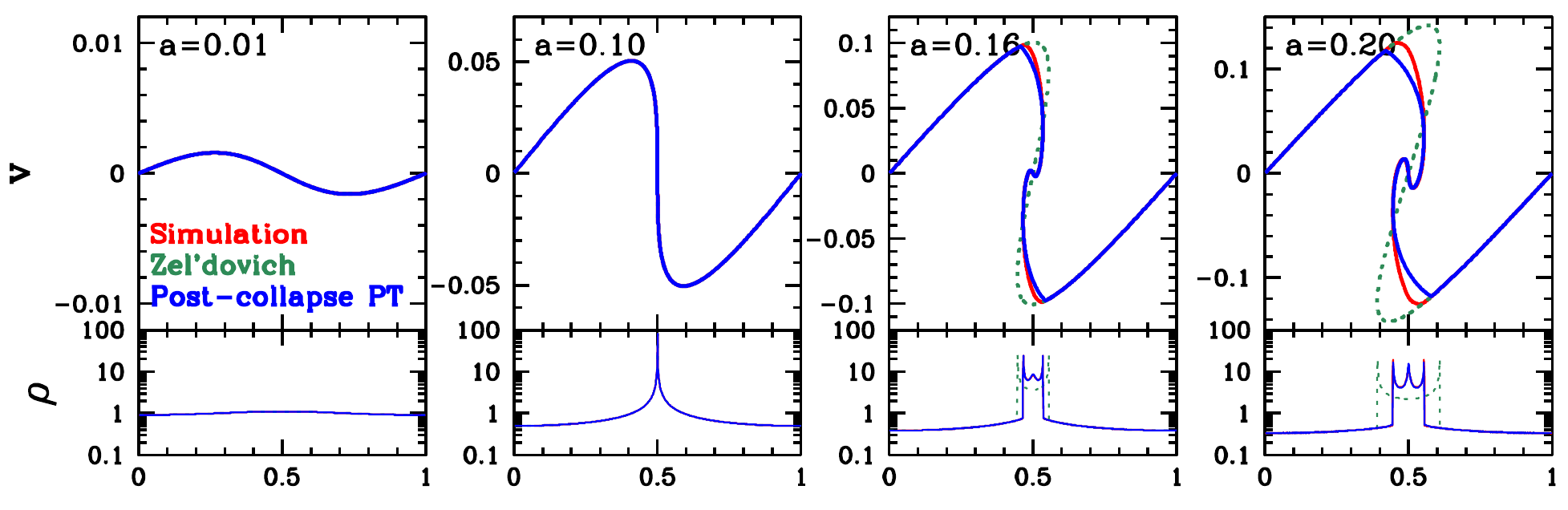}
\includegraphics[width=14.4cm]{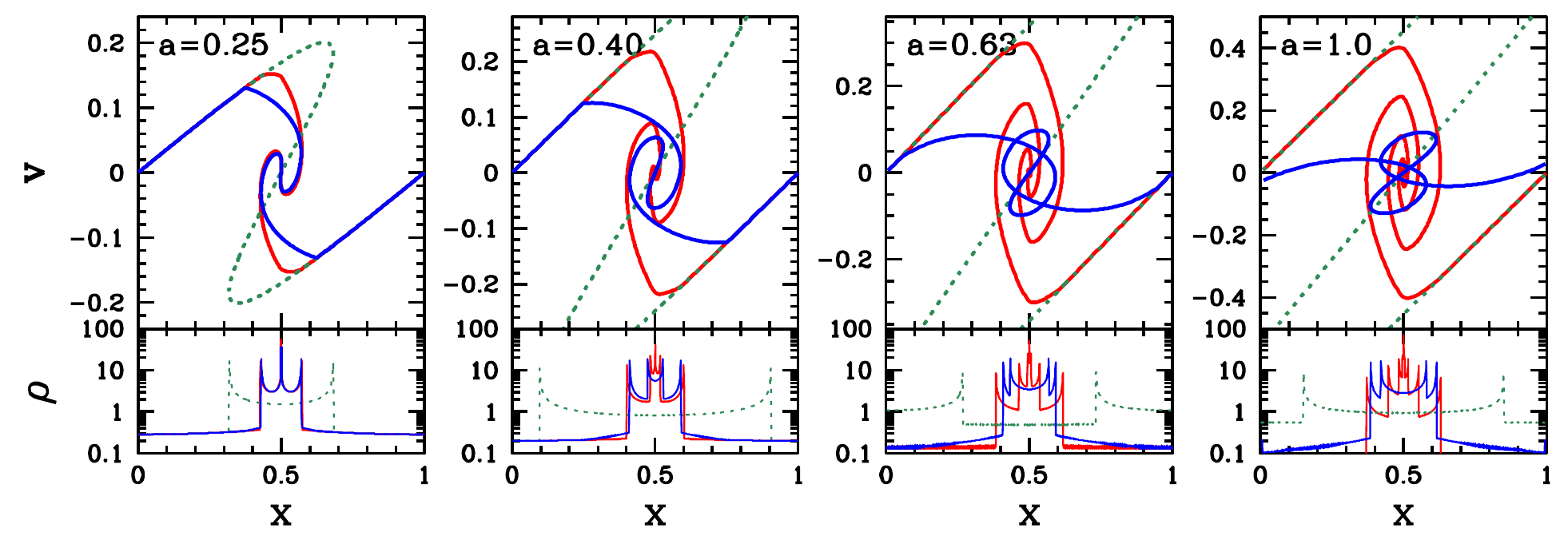}
\caption{Snapshots of the phase-space structure (upper insert of each panel) and projected density profile (lower insert of each panel) of a single halo at different times in an Einstein-de Sitter universe. The initial conditions represented in upper left panel correspond to the projected initial density contrast given by Eq.~(\ref{eq:delta_init_single_cluster}).  On each panel, results of $N$-body simulations are depicted as a red curve, the Zel'dovich solution  as a green dotted line and the basic post-collapse PT prediction as a blue solid line. 
\label{fig:single-cluster_xv_basic}}
\end{figure*}

Fig.~\ref{fig:single-cluster_xv_basic} shows snapshots of the phase-space structure (upper inserts) and projected density profiles (lower inserts) obtained from the simulation (red) and post-collapse PT (blue) based on the basic prescription described in Sec.~\ref{subsec:corrections_pcpt}. Also, for reference, the Zel'dovich solution is plotted as a green dashed line on each panel. Although the three curves are identical at early time, differences show up, as expected, after collapse, which occurs at $a\simeq0.1$. A closer look at the central part of the system reveals however that post-collapse PT captures the trends of the $N$-body simulation quite well and succeeds to reproduce the structure of the inner part of the system until next-crossing time, which occurs at $a\approx0.25$. This is in marked contrast with Zel'dovich solution, which, without the back reaction contribution present in the post-collapse correction, overshoots the displacement of the phase-space sheet after first crossing time, in particular of the caustics. After next-crossing time, $a \ga 0.25$, the phase-space structure given by post-collapse PT goes away from the $N$-body result. Nevertheless, the predicted projected density profile still provides a reasonably good approximation of the evolved density structure, which is clearly not the case for the Zel'dovich solution. 
\begin{figure*}
\includegraphics[width=14.4cm]{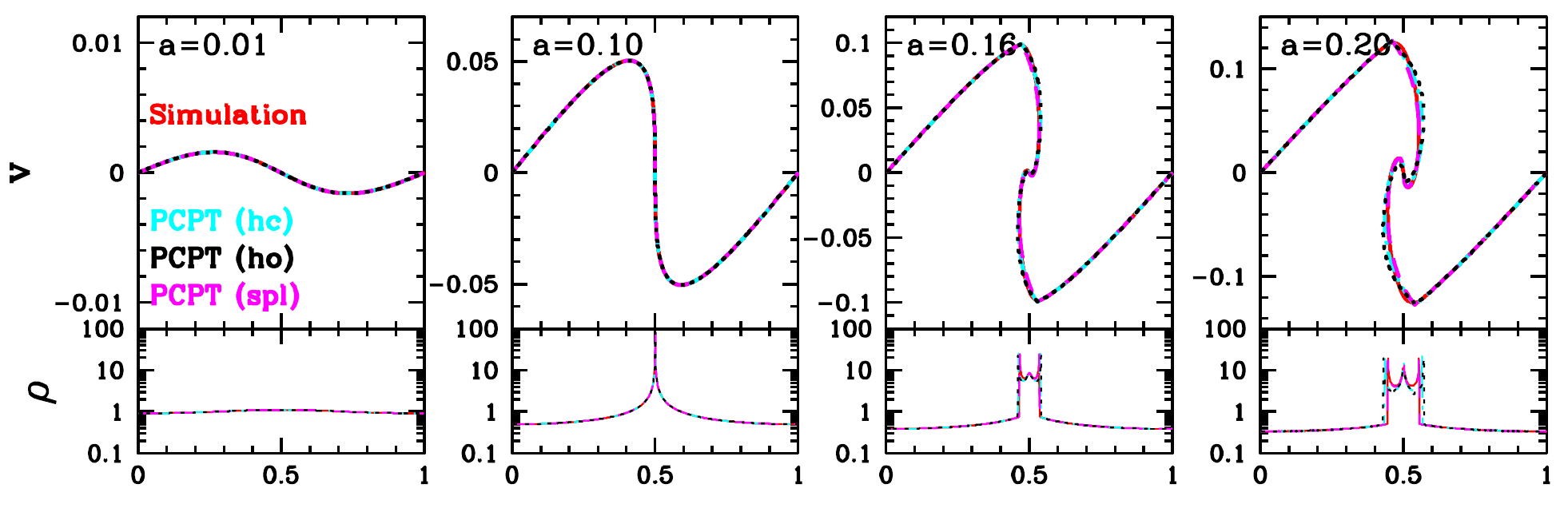}
\includegraphics[width=14.4cm]{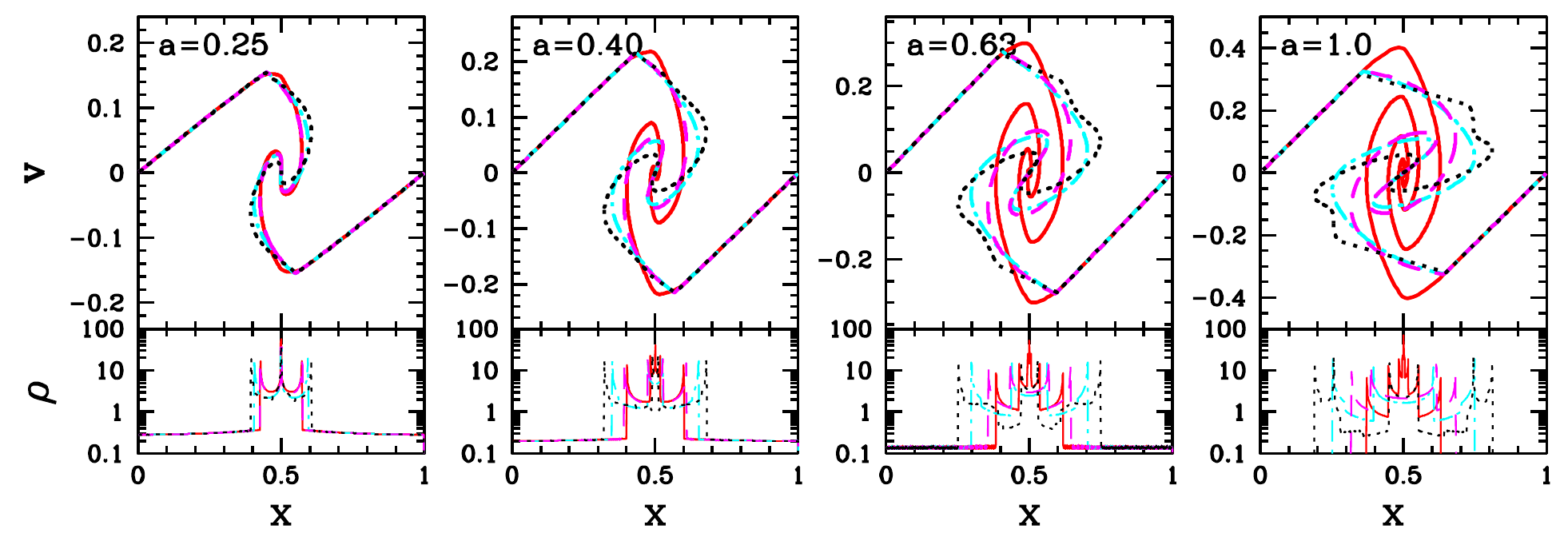}
\caption{Same as in Fig.~\ref{fig:single-cluster_xv_basic}, but variants of the post-collapse PT calculation that include higher-order corrections are compared with the $N$-body simulation, still in red: higher-order for critical times ({\em hc}, cyan dot-dashed), higher-order for critical times and integrands of the equations of motion ({\em ho}, black dotted), and higher-order for critical times combined with spline interpolation to connect inner part with Zel'dovich solution ({\em spl}, dashed magenta). 
\label{fig:single-cluster_xv_variant}}
\end{figure*}

Fig.~\ref{fig:single-cluster_xv_variant} presents results of post-collapse PT  when including the higher-order corrections discussed in \S~\ref{subsec:higher-order}. While the three prescriptions examined here, namely {\em hc}, {\em ho} and {\em spl}, account for higher-order terms in different manners, they are all based on the force expression in Eq.~(\ref{eq:force_inner}), which is valid only shortly after collapse. Indeed, as discussed in \S~\ref{subsec:higher-order}, our implementation of higher-order corrections is not performed in a fully rigorous way, that would require to compute a back reaction to post-collapse PT, similarly as post-collapse PT introduces a back reaction correction to Zel'dovich solution. It also does not propose an iteration procedure that would allow us, similarly as in \cite{Colombi:2014lda}, to follow the system behind next-crossing time. Hence, all of the treatments break down at the end of the simulation. Nevertheless, taking the higher-order corrections to $\Qchat$ and $\tauchat$ into account improves the prediction of the outer boundary of the multi-valued region, and the phase-space description becomes visually better than that of the basic PT prescription in Fig.~\ref{fig:single-cluster_xv_basic}. Even after next-crossing time ($a \ga 0.25$), the predicted phase-space structure still remains consistent, although the size of the halo is prone to be over-estimated, leading to an extended density profile, the best result being obtained with the {\em spl} prescription. Still, one can notice that if only the projected density is under consideration, basic post-collapse PT does better than any of its higher-order counter-part, except for {\em spl} which, not surprisingly, provides nearly equivalent results, since it is exactly the same as post-collapse PT in the center of the system. As far as statistics depending on the projected density are concerned, such as the power spectrum that will be studied later, we thus expect from the analyses of this single halo that the simplest prescription of post-collapse PT is going to perform best.

\begin{figure*}
\includegraphics[width=14.4cm]{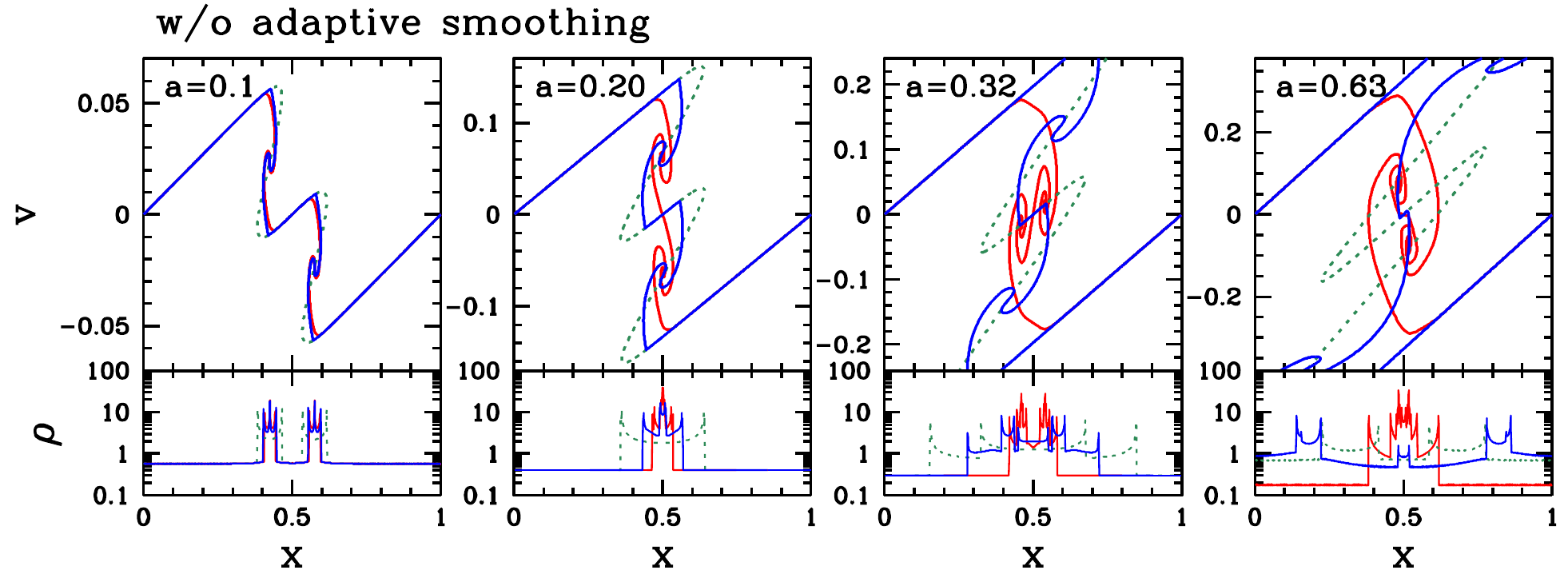}

\vspace*{0.0cm}

\includegraphics[width=14.4cm]{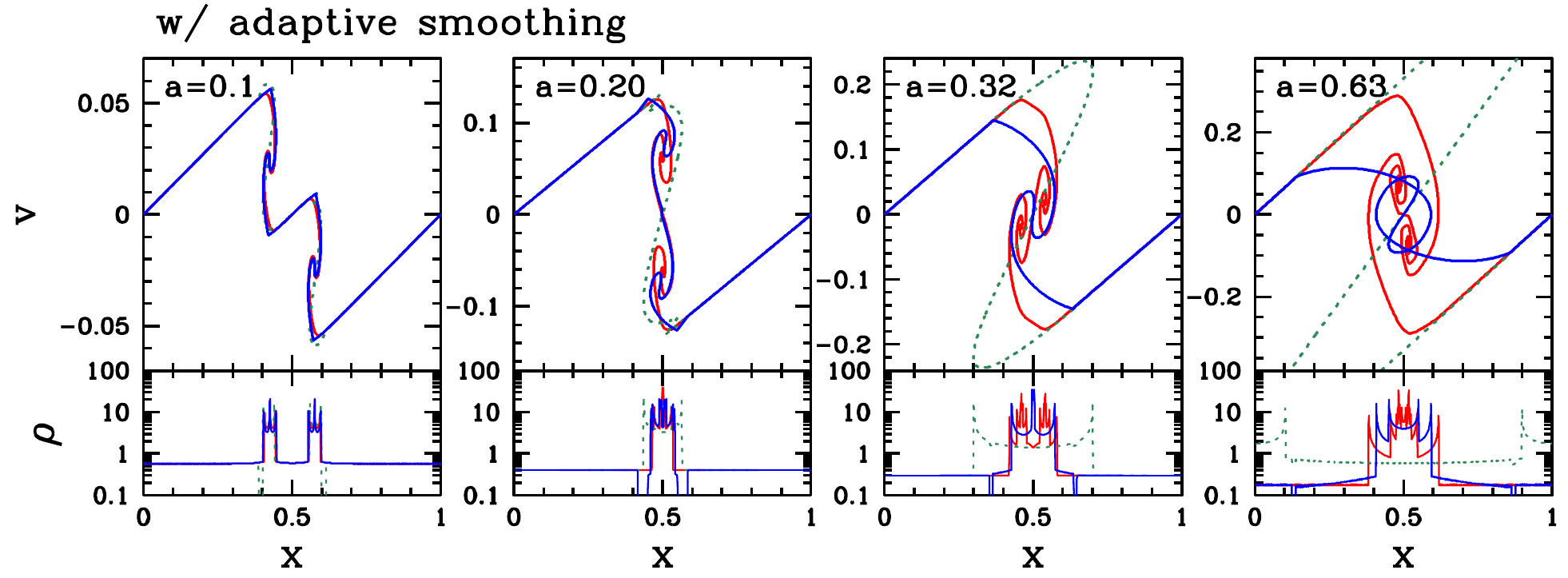}

\vspace*{-0.2cm}

\caption{Phase-space structure (upper inserts in each panel) and density profile (lower insert in each panel) of merging clusters in an Einstein-de Sitter universe. For the initial density contrast given in Eq.~(\ref{eq:delta_init_two_clusters}), results of $N$-body simulations are depicted as red lines, the Zel'dovich solution is shown as green dotted lines and the blue solid lines correspond to the prediction of post-collapse PT. Upper and lower panels respectively show the results without and with adaptive smoothing.
\label{fig:merging-cluster_xv_basic}}
\includegraphics[width=14.4cm]{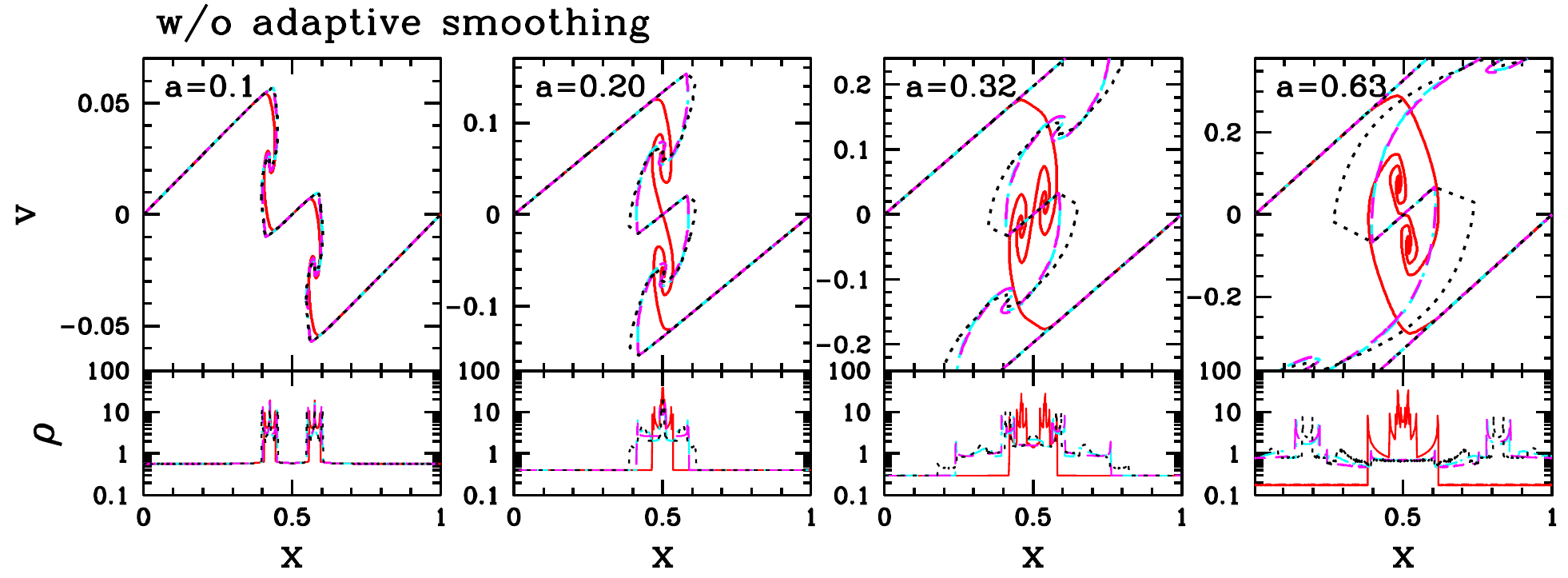}

\vspace*{0.0cm}

\includegraphics[width=14.4cm]{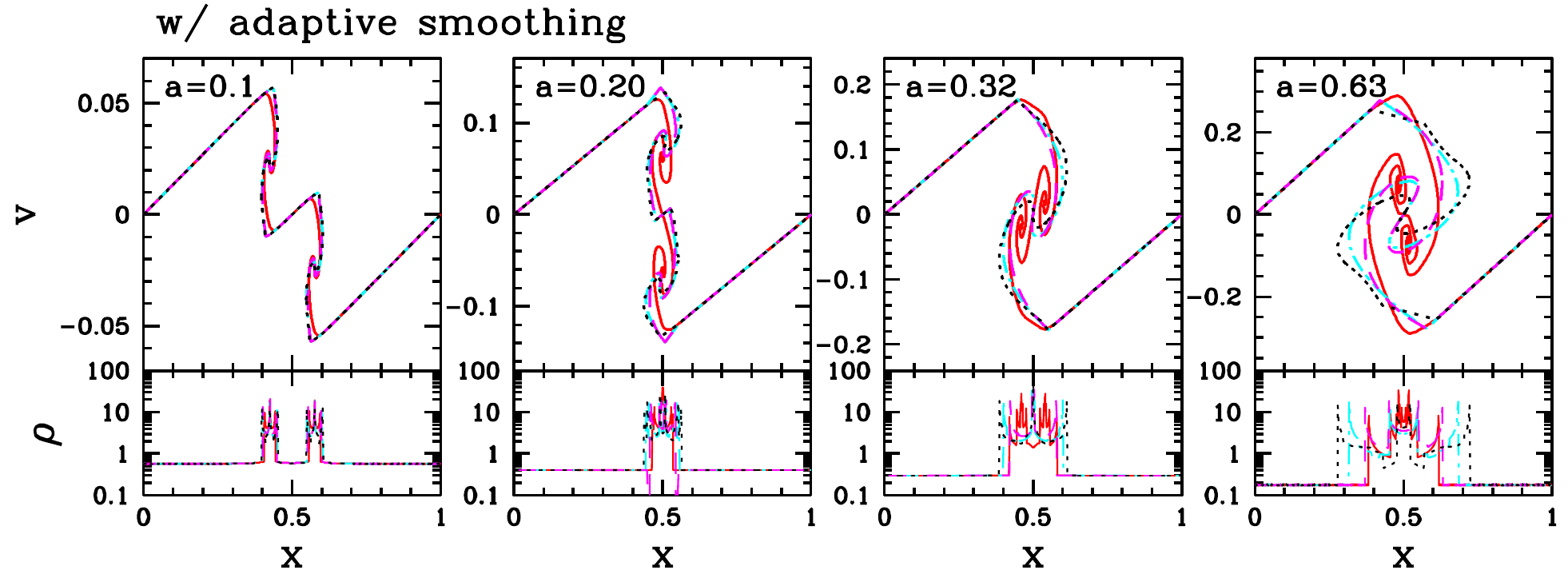}

\vspace*{-0.2cm}

\caption{Same as in Fig.~\ref{fig:merging-cluster_xv_basic}, but the simulation results are compared with improved treatments of post-collapse PT without (upper panels) and with (lower panels) adaptive smoothing. The coding for line types is the same as in Fig.~\ref{fig:single-cluster_xv_variant}.
\label{fig:merging-cluster_xv_variant}}
\end{figure*}

\subsection{Merger}
\label{subsec:merging_clusters}

We examine now the second simplest case of two merging structures in Einstein-de Sitter cosmology ($\Omega_{\rm m,0}=1$, $\Omega_{\Lambda}=0$ and $h \equiv H_0/100=0.7$). The initial density contrast is given by 
\begin{align}
& \delta_{L}(x)=A\Bigl[ \exp\left\{-\left(\frac{x-x_1}{\sigma}\right)^2\right\} \nonumber \\
& \quad \quad \quad \quad \quad \quad +\exp\left\{-\left(\frac{x-x_2}{\sigma}\right)^2\right\} -c\Bigr].
\label{eq:delta_init_two_clusters}
\end{align}
The constant $c$ is determined so as to satisfy the condition $\int_0^L {\rm d}x \,\delta_{L}(x)=0$. With $x_1=0.35L$, $x_2=0.65L$ and $\sigma=0.07L$, we have $c=0.248L$. The overall amplitude is set to $A=0.3$ and the simulation, of box size unity ($L=1$), was started at $a=0.01$ and was run using $(N_{\rm particles},N_{\rm grid})=(10,000; 1,000)$.

The results are shown in Fig.~\ref{fig:merging-cluster_xv_basic} for basic post-collapse PT and Fig.~\ref{fig:merging-cluster_xv_variant} for the higher-order variants of post-collapse PT. Here, we focus on the merger phase which takes place after collapse of each individual structure. Figs.~\ref{fig:merging-cluster_xv_basic} and \ref{fig:merging-cluster_xv_variant} demonstrate how the adaptive smoothing introduced in Sec.~\ref{sec:improvement} improves PT prescriptions to describe the overall phase-space structure after the merger. In each figure, the upper and lower panels show the results without and with adaptive smoothing, respectively. The free parameter controlling adaptive smoothing, $f_{\rm cross}$, is set here to $1$ for post-collapse PT and $0.5$ for the Zel'dovich solution.

Without adaptive smoothing, as soon as the two clusters cross each other ($a=0.2$), both post-collapse PT and Zel'dovich solution start failing to describe the real dynamics in the $N$-body simulation. Indeed, while post-collapse PT tries to provide a local perturbative correction for each single cluster, it is clearly unable to account for nonlinear couplings taking place at larger scales, where it simply follows Zel'dovich motion. As a result, the location of multi-valued regions predicted by post-collapse PT, as well as their shape, largely deviates from the correct one.

On the other hand, implementing adaptive smoothing in the PT calculations significantly improves the predicted phase-space structure, starting already from the beginning of the merger phase ($a=0.2$). After merging, dynamics of the central part of the system is now described by the displacement field derived from the smoothed initial density field. By construction, it summarizes the composite structure made of two sub-halos into a single cluster. While this is a rough approximation of the real dynamics, the description of the outer part of the system is substantially improved as well as the prediction for projected density profiles. Introducing both adaptive smoothing and higher-order corrections to post-collapse PT further improves the results from the visual point of view (Fig.~\ref{fig:merging-cluster_xv_variant}), although the predicted size of the halo tends to be slightly overestimated. As already noticed in \S~\ref{subsec:single_cluster}, the best results are obtained for {\em spl}, which also gives, after adaptive smoothing, comparable projected density profiles to basic post-collapse PT.

The above results demonstrate that post-collapse PT with adaptive smoothing is effective in capturing the main features of phase-space structures. Strictly speaking, it does not give an accurate prescription for the fine structure of high-density regions, but it provides a way of regularizing or mitigating the impact of highly nonlinear dynamics, keeping the location and size of halos reasonably accurate. As we will see next, adaptive smoothing brings a drastic improvement on the prediction of power spectra for systems with random initial conditions. Furthermore, the introduction of adaptive smoothing makes the analytic calculations much less sensitive to the choice of the small-scale cutoff in the initial conditions so that PT predictions become more robust. 
In these respects, the determination of the extension of  multi-streaming regions for a given smoothing scale of initial conditions, as performed in step (iii) of \S~\ref{sec:improvement}, is essential, making the choice of $f_{\rm cross}$ crucial. Concerning the particular merger experiment studied here, $f_{\rm cross}=1$ and $0.5$ seem to be the best choices for tuning post-collapse PT and Zel'dovich solutions, respectively, and we shall adopt these values in subsequent section. We shall however discuss again about the choice of $f_{\rm cross}$ later.

\subsection{Random initial conditions: CDM-like spectrum}
\label{subsec:CDM_initial}

\begin{figure*}
\hspace*{-0.2cm}
\includegraphics[width=8.7cm]{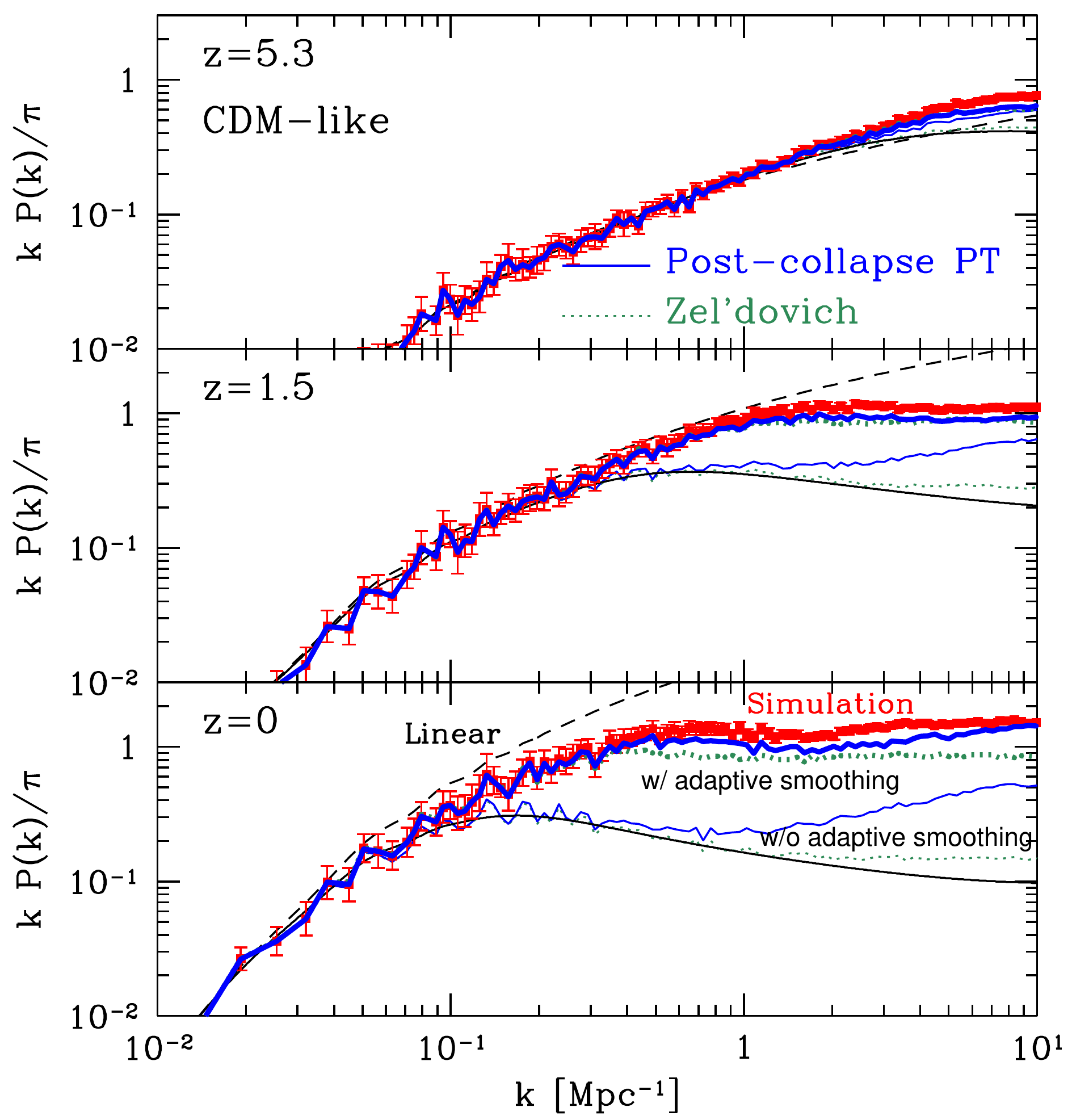}
\hspace*{-0.3cm}
\includegraphics[width=9.2cm]{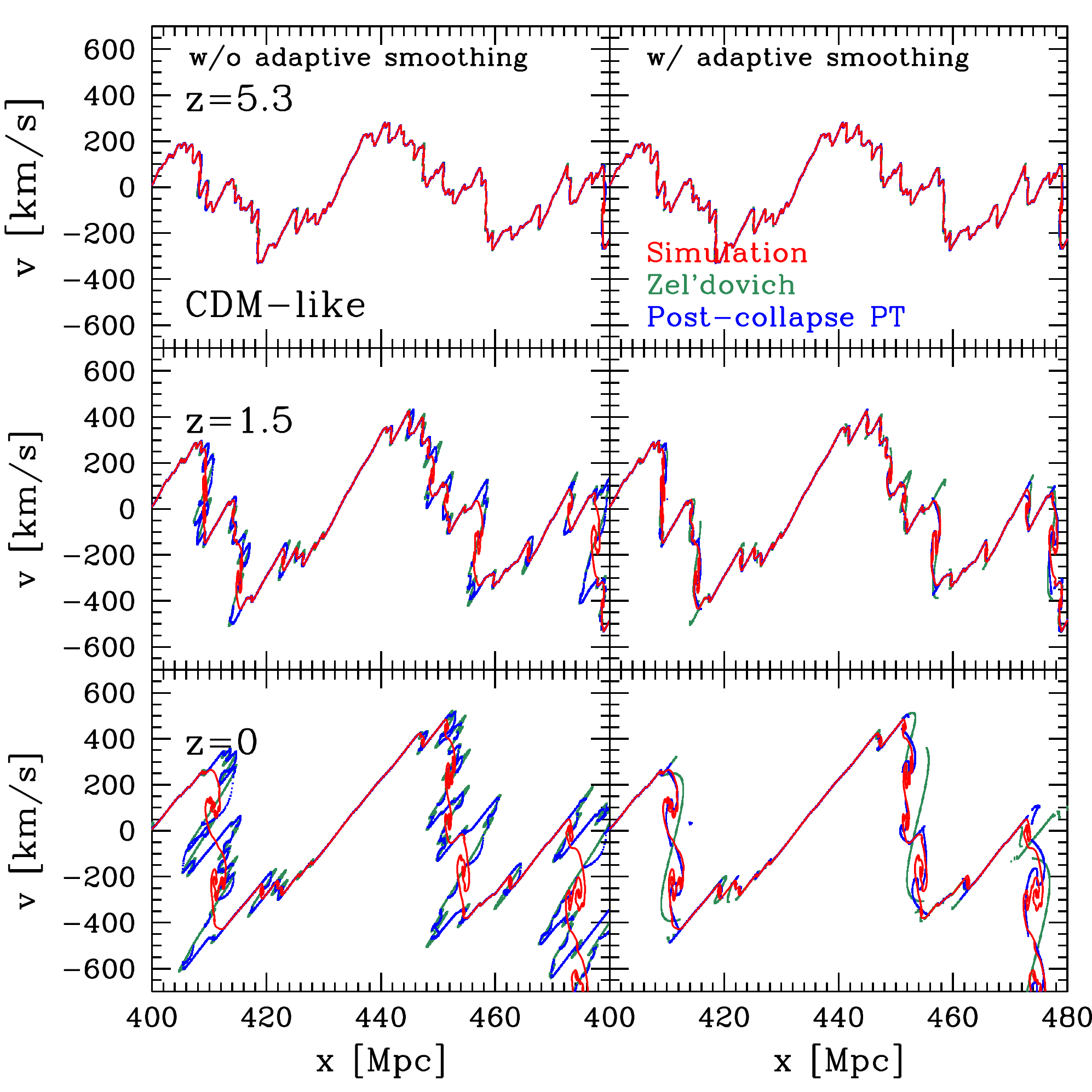}

\vspace*{0.0cm}

\caption{Evolution of the power spectrum (left-panel) and phase-space structure (right panel) in a CDM-like cosmology. From top to bottom, the results are shown at $z=5.3$, $z=1.5$ and $z=0$. On the left panel, symbols with error-bars give the power spectrum measured in the simulations.  The error-bars correspond to the variance over all the modes contained in each bin and sampled from the $50$ realizations. Predictions based on basic post-collapse PT (blue solid) and Zel'dovich solution (green dotted) are plotted, with and without adaptive smoothing respectively for the thick upper curves and thin lower curves. In addition, the analytic result obtained from equation (\ref{eq:pk_ZA_analytic}) is shown as a thin black solid curve and linear theory displayed as a thin black dashed curve. On the right panel, the left and right inserts compare, at various times, the simulated phase-space structure to the prediction of post-collapse PT (blue) and Zel'dovich solution (green),  without and with adaptive smoothing, respectively.  The data are displayed for a particular realization and in a small interval of scales.  
\label{fig:pk_snap_CDM_midres}}
%
%
\hspace*{-0.2cm}
\includegraphics[width=8.7cm]{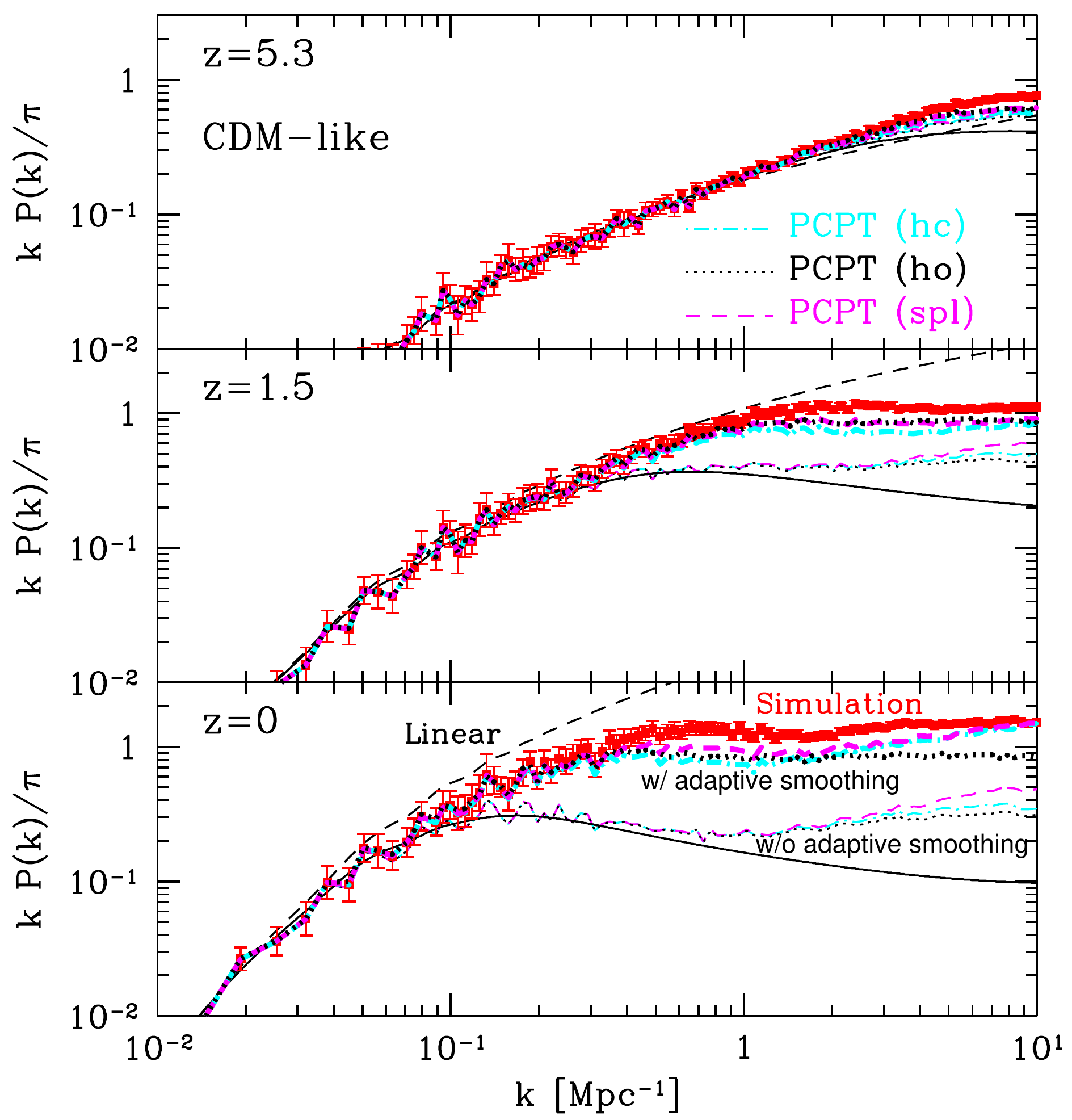}
\hspace*{-0.3cm}
\includegraphics[width=9.2cm]{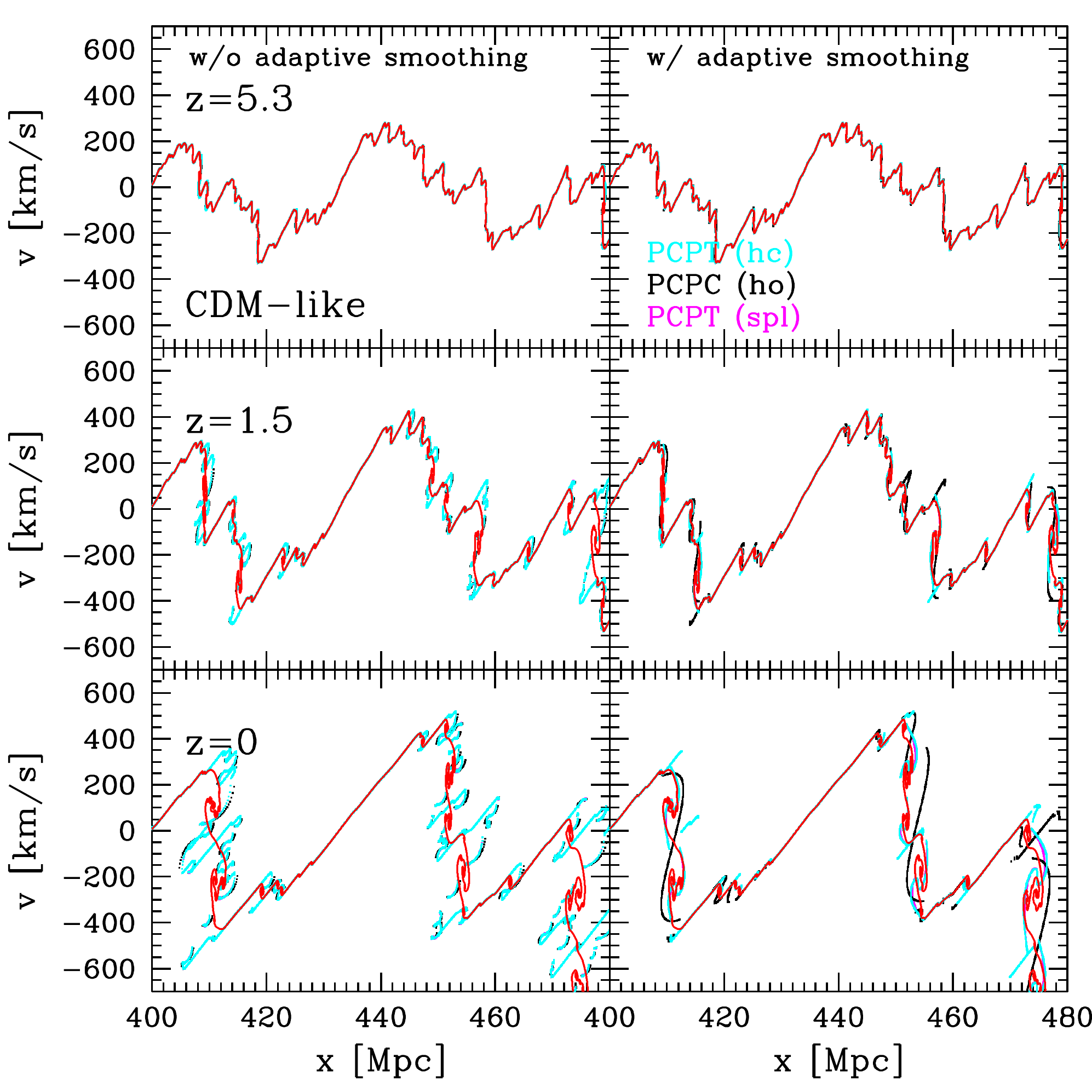}
\caption{Same as in Fig.~\ref{fig:pk_snap_CDM_midres}, but simulations results are now compared to  the variants of post-collapse PT predictions, {\em hc} (cyan dot-dashed), {\em ho} (black dotted) and {\em spl} (dashed magenta).
\label{fig:pk_snap_CDM_midres2}}
\end{figure*}

Let us now consider random CDM initial conditions. Although there is no realistic setup in 1D, a way of mimicking as well as possible the 3D case may be to consider initial conditions given by a random Gaussian field with the following power spectrum \citep[see, e.g.][]{McQuinn:2015tva}:
\begin{align}
P_{\rm 1D}(k)=\frac{k^2}{2\pi}\,P_{\rm 3D}(k),
\end{align}
with $P_{\rm 3D}$ being the 3D matter power spectrum of the initial density fluctuations. Here we use for $P_{\rm 3D}$ the transfer function of \citet{EisensteinHu1998} who wrote the corresponding Fortran module that we included in {\tt Vlafroid}. We set the cosmological parameters to those of the base $\Lambda$CDM model determined by Planck \citep{Planck2015_XIII}: total matter density parameter $\Omega_{\rm m,0}=0.3121$, cosmological constant $\Omega_\Lambda=0.6879$, baryonic matter density parameter $\Omega_{\rm b}=0.04884$, Hubble constant $H_0=67.51\,$km\,s$^{-1}$\,Mpc$^{-1}$, effective slope $n_{\rm s}=0.9653$ and root mean square deviation $\sigma_8=0.815$.\footnote{To be precise, the power spectrum amplitude is normalized by $\sigma_8$ through 
\begin{align}
\sigma_8^2&=\int \frac{{\rm d}^3\bfk}{(2\pi)^3}\,P_{\rm 3D}(k) \{W_{\rm th}(k R_8)\}^2
\\
&=\int_0^\infty \frac{{\rm d}k}{\pi}\,P_{\rm 1D}(k) \{W_{\rm th}(k R_8)\}^2
\label{eq:sigma8_normalization} 
\end{align}
with $R_8=8\,h^{-1}$Mpc, $h=H_0/100$ and $W_{\rm th}(x)=3\{\sin(x)-x\,\cos(x)\}/x^3$. \label{foot:toto}} The simulations were performed with a box size $L=1000\,$Mpc and started at initial redshift $z_{\rm i}=99$. Convergence of simulations results was tested by varying the number of particles $N_{\rm particle}$, the resolution $N_{\rm grid}$ of the grid used to compute the force and the cutoff wavenumber of the initial power spectrum, $k_{\rm max}$, introduced in \S~\ref{subsec:simulation}.\footnote{Namely, we performed three sets of 50 simulations for $(N_{\rm particle},N_{\rm grid}, k_{\rm max} L/2\pi)=(4\times 10^5, 4 \times 10^4, 4\times 10^3)$, $(2 \times 10^5, 2 \times 10^4, 2 \times 10^3)$ and $(10^5, 10^4, 10^3)$. While the highest resolution runs with $(N_{\rm particle},N_{\rm grid}, k_{\rm max} L/2\pi)=(4 \times 10^5, 4 \times 10^4, 4 \times 10^3)$ only allowed us to test the $N$-body results and Zel'dovich solution without adaptive smoothing, lower resolution runs included full PT predictions with and without adaptive smoothing.} Here, we present the results of the analyses for $N_{\rm particle}=200,000$, $N_{\rm grid}=20,000$, and $k_{\rm max}=2000 (2\pi/L)=12.6$ Mpc$^{-1}$. Our convergence study shows, unless specified otherwise below, that the results can be trusted for $k \la 10$ Mpc$^{-1}$, i.e. in the available dynamic range shown in left panels of Figs.~\ref{fig:pk_snap_CDM_midres} and \ref{fig:pk_snap_CDM_midres2}. To have sufficient statistics for measuring power spectra, we ran $50$ simulations with different random seeds.

\begin{figure*}
\hspace*{-0.6cm}
\includegraphics[width=9.5cm]{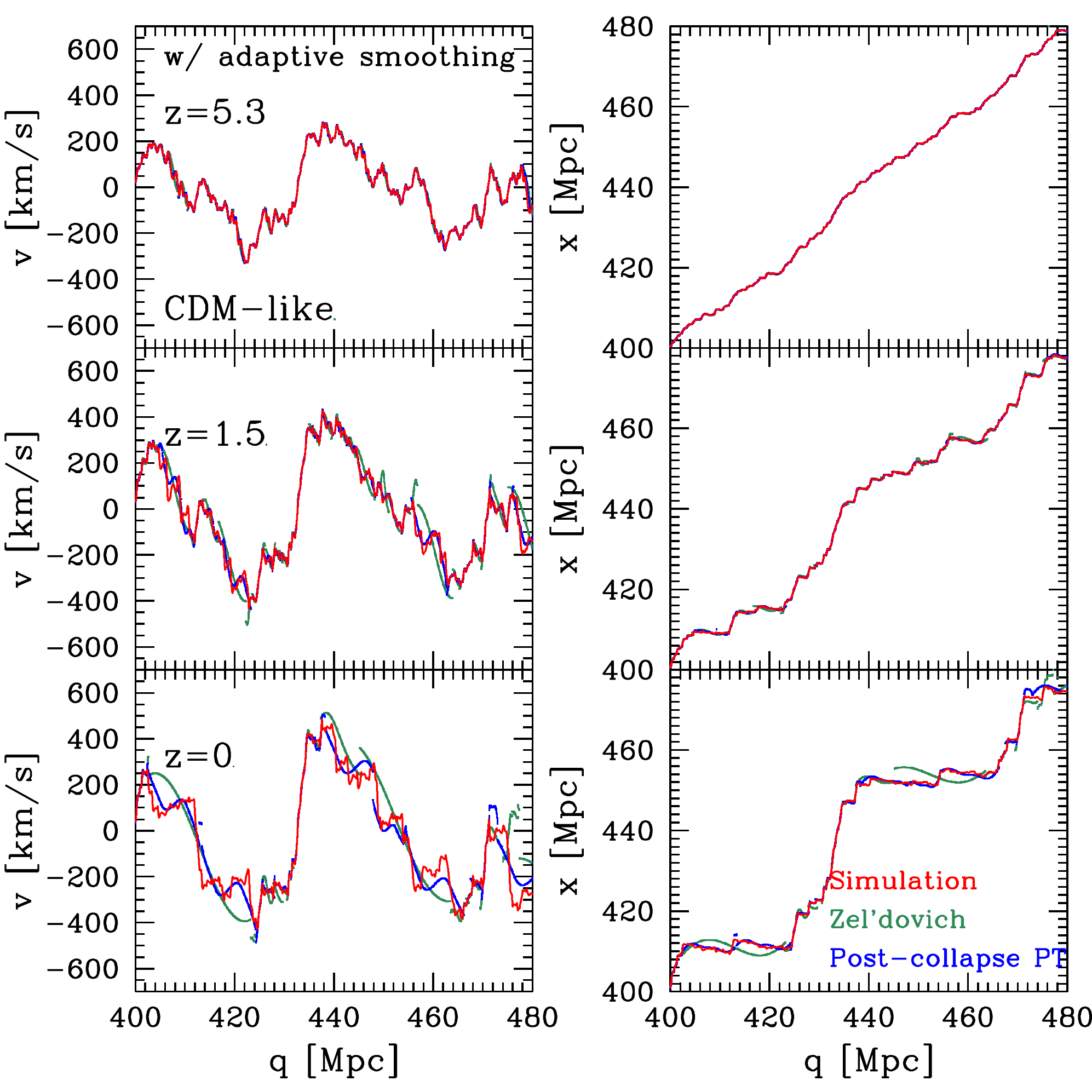}
\caption{Lagrangian/Eulerian correspondence of phase-space structures shown in right panel of Fig.~\ref{fig:pk_snap_CDM_midres}. Left panel shows the velocity as a function of Lagrangian position $q$ at various times, while right panel plots the relation between Lagrangian and Eulerian positions. Predictions presented here correspond to Zel'dovich solution (green) and basic post-collapse PT (blue) with adaptive smoothing. 
\label{fig:snap_qxv_midres}}
\end{figure*}

Figs.~\ref{fig:pk_snap_CDM_midres} and \ref{fig:pk_snap_CDM_midres2} show the measured power spectra at different times (left panel of each figure) as well as snapshots of phase space for a given realization (right panel).  Right panel of Fig.~\ref{fig:pk_snap_CDM_midres}  is supplemented with Fig.~\ref{fig:snap_qxv_midres}, which displays in approximately the same scale interval, Eulerian positions and velocities as functions of Lagrangian coordinates. One can first notice that, in contrast to the 3D case, the amplitude of the power spectrum at small scales is not strongly enhanced in 1D, but damped. For the CDM like initial conditions considered here, function $k P_{1{\rm D}}(k)$ becomes nearly flat in the highly nonlinear regime, which is probably a consequence of stable clustering as discussed further in details in next section, where the case of scale-free initial conditions is considered.

In Fig.~\ref{fig:pk_snap_CDM_midres}, the simulations results are compared to Basic PT predictions (solid blue) and Zel'dovich solution (dotted green), while, in Fig.~\ref{fig:pk_snap_CDM_midres2}, they are compared to various higher-order variants of post-collapse PT.  Note, that except for the solid line that we explain below, all power spectra predicted by post-collapse PT or Zel'dovich solution are computed directly from the particle distribution evolved according to the displacement field predicted by each theoretical model under scrutiny. So it is important to mention that we did not actually compute analytically power spectra in these cases, although it would be in fact possible. The thin black solid lines, on the other hand, provide the analytic results for the Zel'dovich solution without adaptive smoothing, $P_{\rm ZA}(k)$ \citep[e.g.,][]{BondCouchman1986,1995MNRAS.273..475S,McQuinn:2015tva}: 
\begin{align}
&P_{\rm ZA}(k;\,z) = \int_0^\infty {\rm dq}\,\cos(k\,q)\,\Bigl[e^{-k^2\{I(0)-I(q)\}D_+(z)^2}-1\Bigr]\,;
\nonumber\\
&\qquad\qquad\ I(q)=\int_0^\infty \frac{{\rm d}p}{\pi}\cos(p\,q)\frac{P_{\rm 1D}(p)}{p^2},
\label{eq:pk_ZA_analytic}
\end{align}
where $P_{\rm 1D}(p)$ is the {\em un-truncated} power spectrum, i.e. without suppression of power at $k \geq k_{\rm max}$. The black solid curves therefore differ slightly from the green dotted ones on left panel of Fig.~\ref{fig:pk_snap_CDM_midres}, because of the effect of the cut-off at $k_{\rm max}$ (and not because of discreteness effects or finite grid resolution). While the cut-off wavenumber $k_{\rm max}$ has, as expected, a noticeable effect on the un-smoothed Zel'dovich solution as well as the un-smoothed post-collapse PT prediction and its variants (a property that we do not show but that we verified), its influence becomes negligible in the range of values of $k$ under consideration, both for the simulations and the theoretical models with adaptive smoothing. 

As expected, without adaptive smoothing, both post-collapse PT prescriptions and Zel'dovich solution deviate from the simulations as soon as the nonlinear scale becomes larger than the cut-off scale $2 \pi/k_{\rm max}$, which is already the case in upper insert of left panel of Figs.~\ref{fig:pk_snap_CDM_midres} and \ref{fig:pk_snap_CDM_midres2}, i.e. at $z=5.3$. Still, at this redshift, post-collapse PT (or its variants) without adaptive smoothing is able to capture the main features of phase space better than Zel'dovich solution, although this is not obvious to decipher on upper left insert of right panel of Figs.~\ref{fig:pk_snap_CDM_midres} and \ref{fig:pk_snap_CDM_midres2}. At lower redshift, halos relax to a highly nonlinear state and can merge together, which cannot be described well by Zel'dovich solution that gives elongated structures in phase space (middle-left and lower-left inserts of right panels of Fig.~\ref{fig:pk_snap_CDM_midres}. As already argued in previous section, post-collapse PT provides only local corrections to Zel'dovich flow and therefore cannot account for nonlinear dynamics at large scales, hence presents the same defect. This elongation effect obviously leads to a strongly underestimated power spectrum. Note however that this depends also on the choice of the cut-off scale $2\pi/k_{\rm max}$: if this scale would be kept, for the theoretical predictions, variable so that it remains close to the typical transition scale to the nonlinear regime, the results would improve greatly both for Zel'dovich solution and for post-collapse PT, as it is the case for upper insert of left panels of Figs.~\ref{fig:pk_snap_CDM_midres} and \ref{fig:pk_snap_CDM_midres2}. But here, instead of varying the effective smoothing scale $2\pi/k_{\rm max}$ for the PT predictions, we now focus on adaptive smoothing, which should be even better because it provides locally the ``optimal'' softening of initial conditions. 

When adaptive smoothing is employed, agreement between PT predictions and simulation improves strikingly both from the visual and statistical points of view, even for Zel'dovich solution which performs comparably well to post-collapse PT at $z=1.5$, reproducing simulations results in the whole available dynamic range, including the highly nonlinear regime, with a slight underestimation of the power spectrum, though. Of course, superiority of post-collapse PT shows up at $z=0$ but Zel'dovich solution remains quite good. Keep in mind, however, that Zel'dovich solution with adaptive smoothing as presented here uses the calculation of next-crossing time to estimate the local smoothing scale, hence relies as well partly on post-collapse PT. 

These results confirm the conclusions of \S~\ref{subsec:merging_clusters}: adaptive smoothing allows one to summarize composite structures into a single {\cal S} shape like halo with the right size so that two-point statistics matches the true one even in the highly nonlinear regime. In fact, PT predictions reproduce, at least partly, the plateau that can be observed at small scales in middle and lower insert of left panels of Figs.~\ref{fig:pk_snap_CDM_midres} and \ref{fig:pk_snap_CDM_midres2}. In this regime that we will discuss more in details in next section, the system is most probably following stable clustering. Of course, because of the way we implement our adaptive smoothing, continuity of the phase-space sheet is not preserved anymore, as illustrated by right panel of Figs~\ref{fig:pk_snap_CDM_midres} and \ref{fig:pk_snap_CDM_midres2}, but the overall description of phase-space structures is nevertheless improved tremendously. This is illustrated even more clearly by Fig.~\ref{fig:snap_qxv_midres}, which also shows, as expected, that positions predicted by post-collapse PT or Zel'dovich solution match the simulation significantly better than predicted velocities. This is a trivial consequence of the fact that positions correspond to velocities integrated over time: by construction, velocities are better approximated by theoretical models at early than at late times, hence their integrated counterpart, which corresponds to some averaged behavior, will compare better to the simulations. Note thus on right panel of Fig.~\ref{fig:snap_qxv_midres} the excellent agreement with the simulation for the predicted position from basic post-collapse PT, even when significant mergers take place: one just needs to compare bottom right insert of right panel of Fig.~\ref{fig:pk_snap_CDM_midres} to bottom insert of right panel of Fig.~\ref{fig:snap_qxv_midres} to be convinced by this state of fact. As a consequence, we obtain an excellent match between basic post-collapse PT with adaptive smoothing and simulations measurements for the power spectrum, which is a two-point statistics not depending directly on the quality of representation of the velocity. These arguments also apply, to a lesser extent, obviously, to the Zel'dovich solution. 

When focusing again on the power spectrum (left panel of Figs.~\ref{fig:pk_snap_CDM_midres} and \ref{fig:pk_snap_CDM_midres2}), a detailed comparison between various variant of post-collapse PT models reveals that the basic prescription remains the best among all of them, although {\em spl} performs nearly as well. This confirms the results of \S~\ref{subsec:single_cluster} and \ref{subsec:merging_clusters}, where we noticed that higher-order prescriptions for post-collapse PT tended to over-predict the size of halos and mergers (although this is not obvious at first sight when examining phase-space diagrams of right panels of Figs.~\ref{fig:pk_snap_CDM_midres} and \ref{fig:pk_snap_CDM_midres2}), leading to a smaller power spectrum than the basic post-collapse PT prediction at small scales. 

Note that, in agreement with intuition, the results of PT with adaptive smoothing are rather insensitive to the choice of cutoff wavenumber $k_{\rm max}$ (as long as it is kept large enough compared to the wavenumber corresponding to transition towards nonlinearity), but the quality of the agreement with simulations depends, on the other hand, on the value of the parameter $f_{\rm cross}$ introduced in Sec.~\ref{sec:improvement}. Here, we adopt the intuitive choices $f_{\rm cross}=1$ for post-collapse PT and $0.5$ for Zel'dovich solution, but, as illustrated in Appendix \ref{sec:choice_of_f_cross}, it is possible to optimize the value of  $f_{\rm cross}$. For instance at $z=0$, we find for the analyses performed in this section that $f_{\rm cross}=0.6$ and $0.3$ would provide better agreement with simulation measurements. However we also noticed that the optimal value of $f_{\rm cross}$ depends slightly on redshift as well as on initial conditions, this is why we decided for the analyses presented in this article to stick to our fiducial, slightly suboptimal choice.

\begin{figure*}
\hspace*{-0.6cm}
\includegraphics[width=8.7cm]{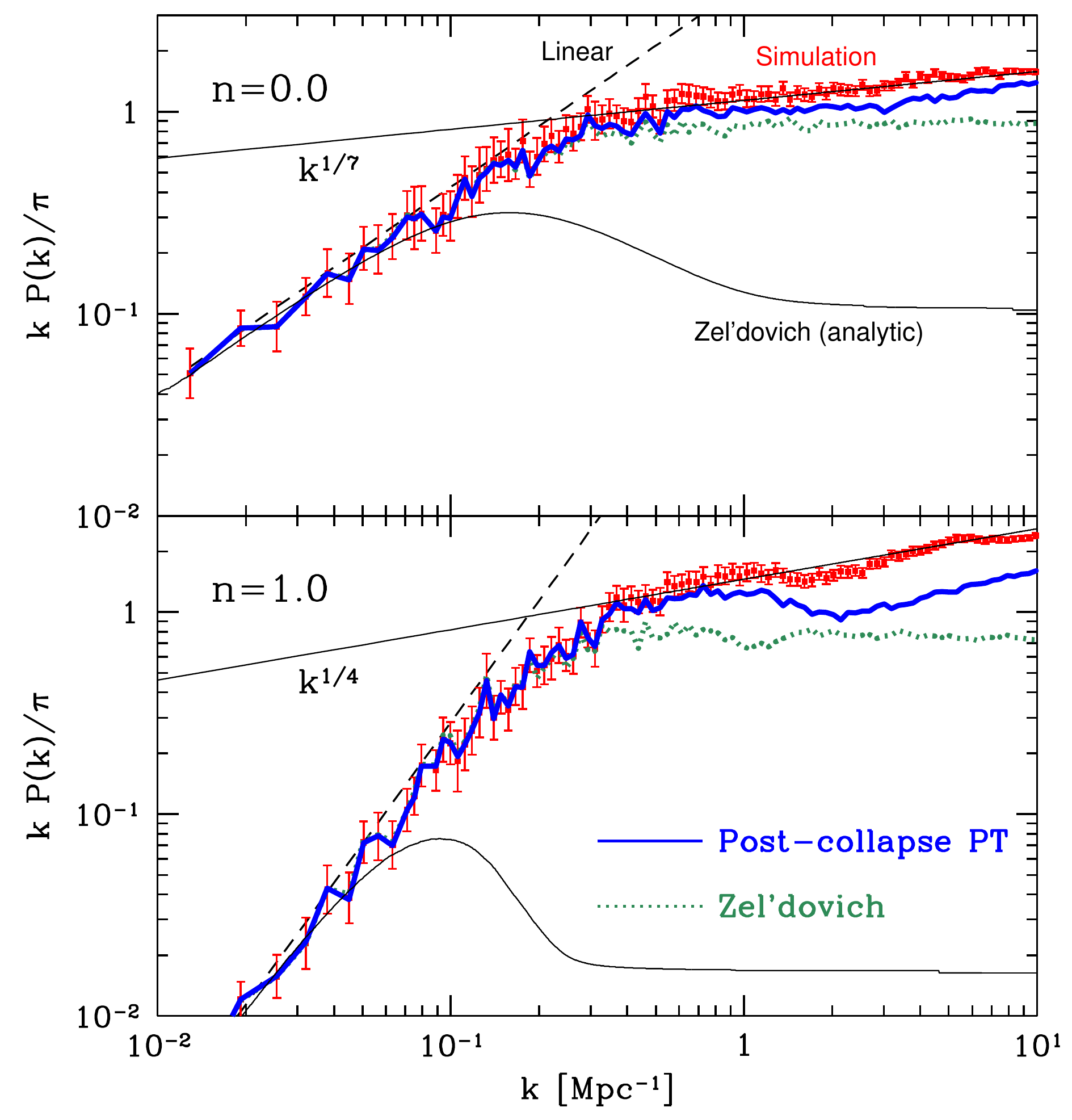}
\includegraphics[width=8.7cm]{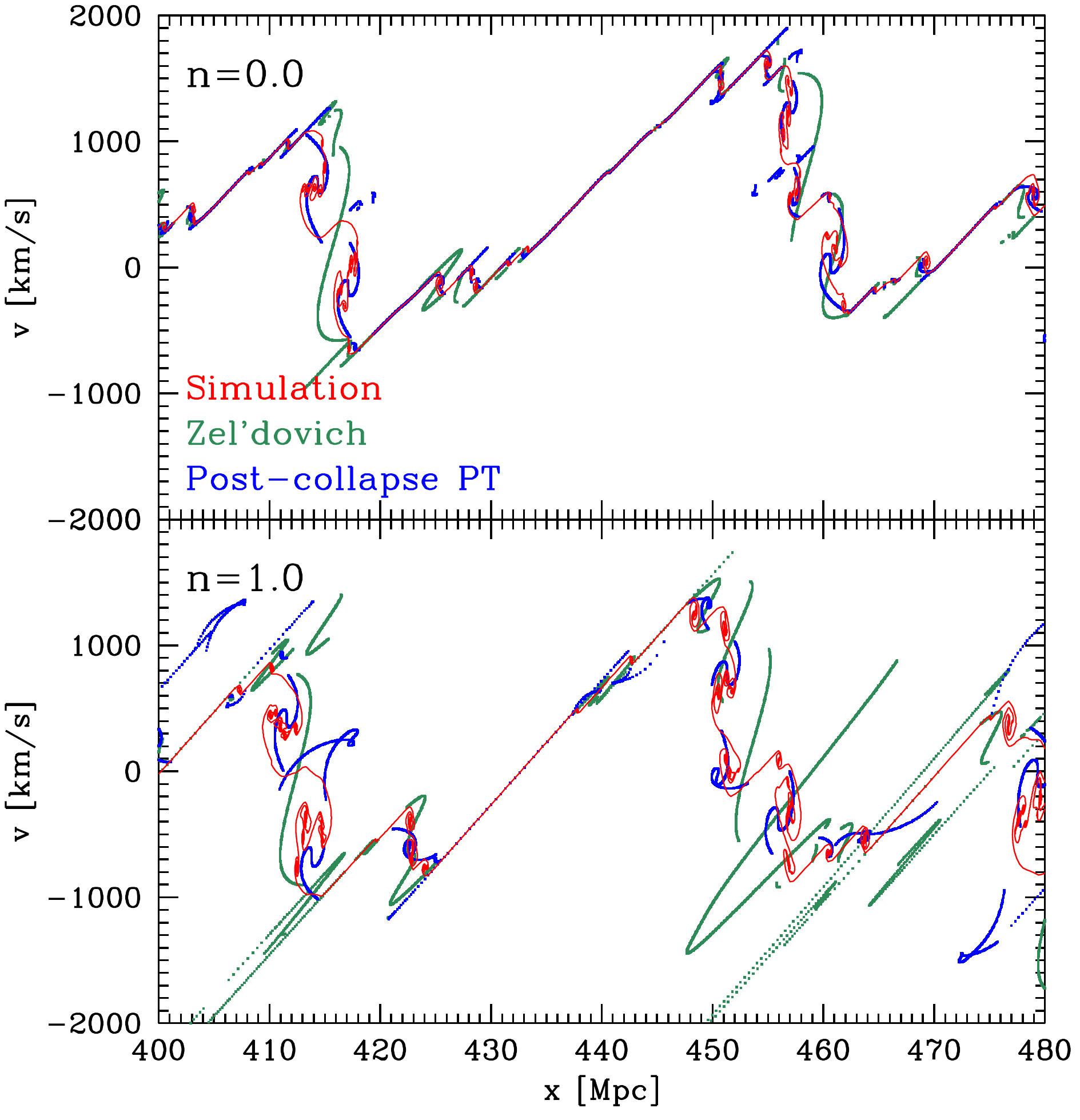}
\caption{Power spectrum (left panel) and phase-space diagram (right panel) from scale free simulations. Left panel shows the matter power spectrum at $z=0$. Upper and lower inserts correspond respectively to $n=0$ and $n=1$. The symbols with errorbars give simulation results, while predictions with adaptive smoothing are plotted as a blue solid curve for post-collapse PT and as a green dotted curve for Zel'dovich solution. For reference, linear theory and analytic Zel'dovich power spectra are also plotted as black dashed and black solid curves as well as the prediction from stable clustering as thin straight lines. Right panel shows, for the same values of $n$, a snapshot of a small region of phase space at $z=0$. Predictions from basic post-collapse PT (blue) and Zel'dovich solutions (green) are plotted together with simulation results (red). Note that the result shown here is generated with the same random seed as in right panels of Figs.~\ref{fig:pk_snap_CDM_midres} and \ref{fig:pk_snap_CDM_midres2}. 
\label{fig:pk_snap_powerlaw}}
\hspace*{-0.6cm}
\includegraphics[width=9.5cm]{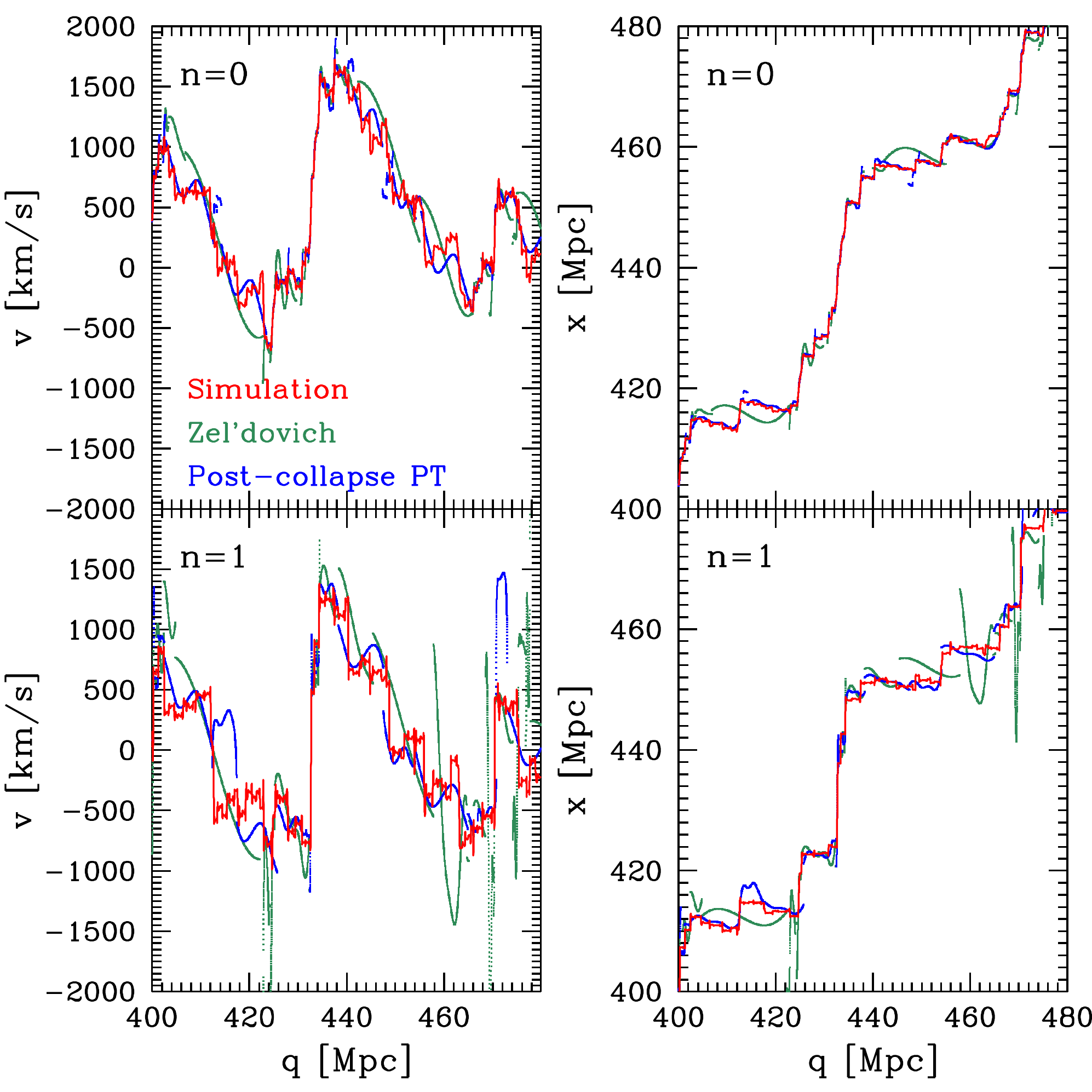}
\caption{Lagrangian/Eulerian correspondence of phase-space structures shown in right panel of Fig.~\ref{fig:pk_snap_powerlaw}.  Upper and lower inserts correspond respectively to $n=0$ and $n=1$, while left and right panels correspond respectively to the velocity and the position as functions of Lagrangian coordinate $q$. Predictions presented here correspond to Zel'dovich solution (green) and basic post-collapse PT (blue) with adaptive smoothing.
\label{fig:snap_qxv_powerlaw}}
\end{figure*}

\subsection{Random initial conditions: power-law power spectrum}
\label{subsec:power-law_initial}
We now turn to random initial conditions with a power-law power spectrum:
\begin{align}
P_{\rm 1D}(k)=A\, k^n.
\end{align}
This allows us to test post-collapse PT for different kinds of initial conditions and to understand more in details small-scale dynamics. Assuming  Einstein-de Sitter universe ($\Omega_{\rm m,0}=1$, $\Omega_{\Lambda}=0$ and $h=H_0/100=1$),
we performed simulations for $n=0$ and $1$, with the same resolution as our main CDM runs, i.e. $N_{\rm particle}=200,000$, $N_{\rm grid}=20,000$ and $k_{\rm max}=12.6$ Mpc$^{-1}$. Although the initial conditions are scale free, we set an actual size for the simulation box, $L=1000$ Mpc, which allows us to normalize the power spectrum using $\sigma_8=1.0$ (see footnote \ref{foot:toto}). With such a normalization, the level of nonlinearity of $P_{\rm 1D}(k)$ at the end of the simulations is roughly the same as in the CDM runs studied in previous section. To measure power spectra with sufficient statistics, we performed sets of 50 simulations with different initial random seed for each value of $n$. 

Fig.~\ref{fig:pk_snap_powerlaw} displays the power spectra (left panel) and phase-space diagrams (right panel) measured at $z=0$ in the scale-free simulations. The numerical results are compared to basic post-collapse PT (blue solid) and to Zel'dovich solution (green dotted) with adaptive smoothing. For reference, linear theory (black dashed) and the analytic prediction given by Eq.~(\ref{eq:pk_ZA_analytic}) for the Zel'dovich solution without smoothing (black solid) are also plotted in left panel.\footnote{While computing $I(q)$ in Eq.~(\ref{eq:pk_ZA_analytic}), we introduced the actual cut-offs in the initial conditions of the simulations, i.e.  below $k_{\rm min}=2\pi/L$ and above $k_{\rm max}=12.6$Mpc$^{-1}$.} In addition, predictions from stable clustering \citep[][]{Joyce2011,Benhaiem2013} are displayed as a thin solid line and will be discussed more in details below. To complete visual inspection of phase space, Fig.~\ref{fig:snap_qxv_powerlaw} plots velocities and positions as functions of Lagrangian coordinate $q$ in the same region of phase space as in right panel of Fig.~\ref{fig:pk_snap_powerlaw}. Note that the simulations considered in right panel of Fig.~\ref{fig:pk_snap_powerlaw} and in Fig.~\ref{fig:snap_qxv_powerlaw} have the same random initial seed as for right panel of Fig.~\ref{fig:pk_snap_CDM_midres} and Fig.~\ref{fig:snap_qxv_midres}, so one can compare directly phase-space features between scale-free and CDM-like simulations. 

Examination of Figs.~\ref{fig:pk_snap_powerlaw} and \ref{fig:snap_qxv_powerlaw} demonstrates again the success of post-collapse PT with adaptive smoothing. While the results are nearly as good for $n=0$ as for the CDM-like case, post-collapse PT does slightly less well for $n=1$. The same trend can be seen for Zel'dovich solution with adaptive smoothing. This result is not surprising, for two reasons. Firstly, the level of nonlinearity in the $n=1$ simulations is slightly larger than for $n=0$ and the CDM-like initial conditions. Indeed, deviation from linear theory or pure Zel'dovich solution happens at smaller values of $k$ for $n=1$ than for other cases. Secondly and more importantly, adaptive smoothing using a sharp low-pass filter is less efficient, from the dynamical point of view, when considering a power spectrum with a high slope. The results would look even worse in the case $n=2$ that we simulated but do not show here. To explain this, one can just take the example of the single $k=k_{\rm single}$ mode system, e.g. initial conditions with a single sine wave as shown in Fig.~\ref{fig:single-cluster_xv_basic}. In this case, with a sharp-$k$ filter,  the fluctuation at $k_{\rm single}$ is either kept, hence post-collapse PT becomes invalid after the system reaches next-crossing time, or it is completely suppressed, which means that the procedure with adaptive smoothing is at the end equivalent to no smoothing at all. Increasing the power spectrum slope $n$ basically makes us approaching the single mode case with $k_{\rm single}=k_{\rm max}$, since we always introduce a cut-off at some value $k_{\rm max}$ to have a sufficiently smooth representation of the phase-space sheet at very small scales.  Of course, this situation is unrealistic in practice, because no such problem occurs when considering CDM-like power spectra, but one has to stay aware of the fact that adaptive smoothing does not always work. A possible way to improve it could consist in changing the filtering window so that all the wavenumbers are affected by the softening procedure, even if it is performed predominantly at a given scale: to this respect, adaptive smoothing with a Gaussian window might represent an interesting alternative to our sharp-$k$ filter. 

Another interesting result of our measurements is the excellent agreement between the simulations and the stable clustering prediction from \cite{Joyce2011} and \cite{Benhaiem2013} in the nonlinear regime\footnote{We also note here that the asymptotic slope seen in the simulations is rather different from the one predicted by the so-called adhesion model based on Burgers' equation, which gives $kP(k)\to k$ at very high-$k$, irrespective of the initial power-law slope \citep[][]{Patrick2009a, Patrick2009b, Valageas_Bernardeau2011}.}, who predict that the slope of the power spectrum should be equal to
\begin{align}
\alpha=\frac{ {\rm d} \log k P_{1{\rm D}}(k)}{{\rm d}\log k}=\frac{n+1}{n+7}.
\end{align}
In fact, even in the CDM-like case, which roughly corresponds, in one dimension, to an asymptotic value of $n$ equal to $-1$ at large $k$, stable clustering prediction works, since it predicts a plateau, $\alpha=0$, which is indeed what we observe on left panel of Fig.~\ref{fig:pk_snap_CDM_midres}. Note that what \cite{Joyce2011}  and \cite{Benhaiem2013} call ``stable clustering'' is very specific. In the 1D case considered here, it is equivalent to assuming that relaxed objects become of constant size in the coordinate $r'=a^{1/3} x$ instead the physical coordinate $r=a x$ as normally supposed in the three-dimensional case \citep[see, e.g.][]{Davis1977,Peebles:1980}. It is important to notice here that post-collapse PT seems to be able to reproduce correctly the measured power spectrum in the beginning of the stable clustering regime. This suggests that stable clustering regime is established only in a few dynamical times, since post-collapse PT is only able to follow the dynamics of a single halo until next-crossing time. 

On the other hand, if one considers a ``halo model'' approach \citep[see, e.g.][and references therein]{Ma2000,Peacock2000,Scoccimarro2001,Cooray2002} and assumes that at large $k$ the power spectrum is dominated by the ``one-halo term'', interesting properties are expected, at least when initial density fluctuations are sufficiently smooth at small scale. Indeed, one dimensional simulations of single cold halos suggest that their central density profile nearly behaves like a power-law
\begin{align}
\rho(x) \propto x^{-\gamma},
\end{align}
with $ 0.4 \la \gamma \la 0.5$. The actual value of $\gamma$ is controversial. While \cite{Binney:2003sn} postulated it to be  equal to $0.5$, detailed $N$-body simulations suggest instead $\gamma \simeq 0.47$ \citep[][]{Schulz2013}. On the other hand, pure Vlasov simulations results using the waterbag method suggest $\gamma=0.4$ \citep[][]{2014MNRAS.441.2414C}. Other values to keep in mind are $\gamma=2/3$ from the singularity occurring exactly at crossing times and $\gamma=1/2$ from the caustics \citep[see, e.g.][]{1989RvMP...61..185S}. These singularities imply, as discussed below, some expected behavior for $P(k)$ at large $k$ \citep[][]{Gouda:1989}. 

A slope $-\gamma$ for the density profile corresponds, after some trivial algebraic calculations, to a power-law slope 
\begin{align}
\alpha=2 \gamma-1
\end{align}
for the power spectrum. For instance, one expects the Zel'dovich approximation to be dominated by the caustics at very small scales, i.e. $k P(k) \rightarrow$ constant at large $k$, which is indeed the case on left panels of Figs.~\ref{fig:pk_snap_CDM_midres} and \ref{fig:pk_snap_powerlaw} in both analytic (black) and numerical (green) cases, irrespective of adaptive smoothing. Turning to post-collapse PT with adaptive smoothing, since, by using $f_{\rm cross}=1$, we push the (smoothed) system exactly to next-crossing time, we are dominated by the singularity with $\gamma=2/3$. Hence, in this case one expects $\alpha$ close to $1/3$ when $k$ becomes very large, as can be observed on all the figures. With a smaller value of $f_{\rm cross}$, the system is dominated, similarly as for Zel'dovich solution, by the caustics at very small scales and then $k P(k)$ reaches a plateau at large $k$ as can be seen in Fig.~\ref{fig:pk_xv_fcross} of appendix \ref{sec:choice_of_f_cross}. Without adaptive smoothing, one also intuitively expect post-collapse PT power spectrum to be dominated by caustics at very small scales, however the objects have a more complex structure than Zel'dovich solution as shown by lower right panel of Fig. 2, hence the plateau at large $k$ is not clearly visible on left panel of Figs.~\ref{fig:pk_snap_CDM_midres}, \ref{fig:pk_snap_CDM_midres2} and \ref{fig:pk_snap_powerlaw} probably because not reached as quickly as for the pure Zel'dovich solution when increasing $k$.

Turning to the actual nonlinear evolution of single objects, assuming as found in the literature $\gamma=0.4$ and $0.5$ for the density profile of a halo  would respectively give $\alpha=-0.2$ and $0$ for the ``one halo'' power spectrum. This of course supposes that statistical averaging over many halos of different masses does not affect the slope of the power spectrum predicted by the single halo. Again, we have to be aware of the fact that at very large $k$, one must be dominated by the caustics, since we introduce a cut-off at large $k$. Such a cut-off enforces some level of smoothness of the curve representing the system in phase space and thus actually imposes the existence of well-defined caustic structures in the nonlinear regime,\footnote{Things would be more difficult to apprehend if the curve representing the phase-space sheet would be un-smooth.} hence a plateau for $k P(k)$ at very large $k$, even for the simulations. However this plateau is not visible in the simulations, because it lies outside the range of $k$ shown in the figures. On the other hand, there must be, according to our ``halo model'' prescription, a regime where $k P(k)$ might decrease ($\gamma=0.4$) or present another plateau ($\gamma=0.5$) or at least some kind of inflection at some intermediate values of $k$. This is what can indeed be observed e.g. on lower insert of left panel of Fig.~\ref{fig:pk_snap_CDM_midres} for $0.6 \la k \la 2$ Mpc$^{-1}$ for the CDM-like case and on lower insert of left panel of Fig.~\ref{fig:pk_snap_powerlaw} for $1 \la k \la 2$ Mpc$^{-1}$ for the $n=1$ simulation. There is no such signature obviously visible in the $n=0$ simulation, although the measurements are too noisy to conclude in this case. Interestingly, when the inflection is visible, its extension in $k$ space is correctly predicted by basic post-collapse PT with adaptive smoothing. This just shows that basic post-collapse PT is able to predict approximately the correct size for the halos, which we already knew. 

Note that here, we did not try to position our single halo term discussion in the context of stable clustering predictions. These latter can be considered as the outcome of some statistical averaging after weighting the one-halo term by the mass function. They are also expected to be valid only in a finite range of values of $k$ if one assumes a cut-off at $k_{\rm max}$: what we mean here is that our ``one-halo term'' discussion is in fact not incompatible with the concept of stable clustering although the link between the halo model and the stable clustering predictions remain to be performed in details in the present case. Clearly, however, the small inflection discussed above marks a small deviation from stable clustering.


\section{Conclusion}
\label{sec:conclusion}

In this paper, we have developed a Lagrangian perturbation theory method for solving, beyond crossing time, Vlasov-Poisson equations in the expanding universe for initially cold systems in the infinite parallel planes geometric set up. The proposed approach captures post-collapse dynamics by computing at leading-order a counter term to the Zel'dovich solution just after first crossing time. It extends earlier work of \citet{Colombi:2014lda} to the cosmological case and, thanks to adaptive smoothing, to random initial conditions instead of a single halo. By performing a local Taylor expansion of the velocity and position of each mass element as functions of Lagrangian coordinate around each initial density peak, we are able to compute the force field in the multi-valued region. A correction to the single-stream flow motion, described by the Zel'dovich solution, is systematically obtained. The results are given as polynomial functions of Lagrangian position and the coefficient of each power is expressed in terms of quantities associated to density peaks. A simple analysis of the formation of a single cluster shows that post-collapse PT reproduces well the phase-space structure up to next-crossing time and can describe the projected density profile quite well even at later time.  

To deal with the bulk properties of phase-space structures, we propose an adaptive filtering scheme. The idea is very close to the peak-patch treatment by \citet{1996ApJS..103....1B}. In particular, the local smoothing scale of initial conditions is determined by the calculation of a dynamical time, but in post-collapse PT this dynamical time is given by next-crossing time instead of collapse time. This allows us to produce a realistic coarse-grained version of the phase-space structure, even in the presence of mergers. After demonstrating that adaptive smoothing indeed works well in a simple setup with coalescing clusters, we applied it to more general, random initial conditions that include CDM-like as well as scale-free power spectra. The results of our analyses show that post-collapse PT predictions with adaptive smoothing reproduce the power spectrum of our numerical simulations remarkably well even at scales sufficiently nonlinear to probe the beginning of the stable clustering regime. 

On important thing to keep in mind for further work is that the critical quantity in post-collapse PT is the time of next crossing, which can in fact be used to also create a spectacularly good pure Zel'dovich prescription with adaptive smoothing. While the calculations of post-collapse PT can be cumbersome, the expression (\ref{eq:tau_cross}) of this time is relatively simple and this can be very interesting when trying to generalize results of post-collapse PT to the three-dimensional case. When analyzing equation (\ref{eq:tau_cross}), one indeed notes that $\tau_{\rm cross}$ depends both on collapse time $\tau_0$  and on the height $\delta_{\rm L}$ of the initial density peak. Since $\tau_0$ depends solely on $\delta_{\rm L}$, we can see that basically the dynamical time we consider for defining the local smoothing scale is a single function of $\delta_{\rm L}$, larger than $\tau_0$. In fact, things are very simple: when performing adaptive smoothing one enforces $\tau_0+\tau_{\rm cross}$ to be equal to present time, or in other words, adaptive smoothing of initial conditions is performed so that the local peak height given by linear theory at present time,  $\delta_{\rm L}$, is a number larger than unity (instead of exactly unity for the collapse time). For instance, in the Einstein-de Sitter case, $\Omega_{{\rm m},0}=1$ and $\Omega_{\Lambda}=0$, we have
\begin{align}
\delta_{\rm L}=\frac{1}{(1- f_{\rm cross}/3)^2},
\end{align}
with $f_{\rm cross}$ chosen in this article to be equal to unity for post-collapse PT and one-half for the Zel'dovich solution. Note that the optimal value of $f_{\rm cross}$ is smaller than these intuitive choices (see Appendix \ref{sec:choice_of_f_cross}).

The present post-collapse PT treatment together with the adaptive smoothing scheme seems a very promising tool. A successful extension to the 3D case should lead to a major breakthrough in the precision PT calculations of large-scale structure statistics beyond the single-stream approximation. Our 1D study is however only a first step. Toward a practical application to the 3D case, there are several issues to be addressed. One is the analytic calculation of statistical quantities such as the power spectrum. Indeed, in the present work, we measured directly the power spectrum from a particle distribution following the dynamics prescribed by the various PT prescriptions under scrutiny. In fact, analytic calculations of the power spectrum are tractable, at least in the presence of  fixed smoothing. Adaptive smoothing itself can in principle be implemented in the analytic framework by inspiring for instance from the theory of excursion sets discussed in e.g. \citet[][]{Bond1991}, although we can foresee that actual calculations will probably be very cumbersome. 

Another issue, when generalizing to the 3D case, is that the Zel'dovich solution no longer provides a sufficiently accurate description of the dynamics before collapse. At least third-order Lagrangian PT is required to estimate correctly collapse times as well as the power spectrum beyond tree-level approximation, which clearly adds some significant level of complexity. Furthermore, the variety and complexity of topological configurations of singularities appearing in the 3D case might in fact represent a nearly impassable obstacle in three dimensions \cite[see, e.g.][]{Hidding2014}. We will tackle these issues in future work.

Finally, while we have focused in this article on the impact of post-collapse dynamics in 1D on the phase-space structure and on the power spectrum of the matter distribution, it would be interesting to see quantitatively how the small-scale modes responsible for the multi-stream flows couple to large-scale fluctuations described by the single-stream flows. This can be presumably clarified by measuring the response of the evolved power spectrum to a small initial perturbation of high-$k$ modes, through the comparison between a pair of particle distributions modeled with post-collapse PT that slightly differ in their initial conditions \citep{Nishimichi:2014rra}. In the 3D case, a strong damping is found for the mode transfer from small to large scales, but the physical origin of it is still unclear. A detailed analysis in the 1D cosmological setup would provide important clues on the nature of mode transfers.

\section*{Acknowledgments}
AT thanks Patrick Valageas for references and useful comments on the adhesion model. SC greatly acknowledges hospitality of Yukawa Institute for Theoretical Physics (YITP), where this work was initiated. We also acknowledge the support of YITP in organizing the workshop ``Vlasov-Poisson: towards numerical methods without particles'' in Kyoto, funded by grant YITP-T-15-02, ANR grant ANR-13-MONU-0003 and by Institut Lagrange de Paris (ANR-10-LABX-63 and ANR-11-IDEX-0004-02). This work was also supported in part by MEXT/JSPS KAKENHI Grant Numbers JP15H05889 and JP16H03977 (AT) as well as ANR grant ANR-13-MONU-0003 (SC).




\bibliographystyle{mnras}

%
\input{ms_pcpt.bbl}



\appendix
\section{Derivation of integral expression for the potential $\Phi$}
\label{sec:Green_func}

In this appendix,  we derive the expression in Eq.~(\ref{eq:Phi_Poisson}). To do this, we first consider the 1D Green function which satisfies the following equation:
\begin{align}
\frac{\partial^2}{\partial x^2}\,G(x,x')=-\delta_{\rm D}(x-x').
\label{eq:Green_func}
\end{align}
With this Green function, the solution of the Poisson equation is formally expressed as
\begin{align}
\Phi(x)=-\frac{3}{2}\,H_0^2\omegam \,a\,\int_0^L{\rm d}x'G(x,x') \delta(x').
\label{eq:integral}
\end{align}
To solve Eq.~(\ref{eq:Green_func}) under periodic boundaries condition, we move to  harmonic space and express the Green function as:
\begin{align}
G(x,x')=\sum_{n=-\infty}^{+\infty}
G_n\,\,e^{i\,2n\,\pi\,x/L}.
\label{eq:G_Fourier}
\end{align}
Recalling that the Fourier transform of Dirac's delta function reads
$\delta_{\rm D}(y)=(1/L)\sum_n\,\exp[i\,2n\,\pi\, y/L]$,\footnote{Here we assume $0\leq y\leq L$.}  substitution of Eq.~(\ref{eq:G_Fourier}) into Eq.~(\ref{eq:Green_func}) gives
\begin{align}
G_n=\frac{L}{(2n\,\pi)^2}\,e^{-i\,2n\,\pi\,x'/L}\,\,;\quad n\ne0.
\end{align}
Thus, we obtain
\begin{align}
G(x,x')&=\sum_{n=-\infty}^{+\infty}
\frac{L}{(2n\,\pi)^2}\,e^{-i\,2n\,\pi(x- x')/L}
\nonumber\\
&=\sum_{n=1}^{+\infty}
\frac{2L}{(2n\,\pi)^2}\,\cos\left\{\frac{2n\,\pi(x- x')}{L}\right\}.
\end{align}
Note that there is a subtlety in the above equation. In deriving it, we ignore the contribution from $n=0$ mode in $G_n$, while the $n=0$ mode does not vanish in Dirac's delta function. The result remains however correct as long as the density field does not contain a $n=0$ mode, or equivalently, $\int_0^L{\rm d}x\,\delta(x)=0$.  

Keeping this point in mind, we can further simplify the expression. We use the following formula \citep[see, e.g. Chap.1.44 of][]{Gradshteyn1980book,Iwanami}:
\begin{align}
\sum_{n=1}^\infty \frac{\cos(n\,y)}{n^2}=\frac{1}{4}(y-\pi)^2-\frac{\pi^2}{12};\quad 0\leq y\leq 2\pi.
\end{align}
The final expression for the Green function is then given by
\begin{align}
G(x,x')=\frac{L}{2}\left[\left\{\frac{|x-x'|}{L}-\frac{1}{2}\right\}^2-\frac{1}{12}\right],\qquad |x-x'|\leq L.
\end{align}
Substituting this into Eq.~(\ref{eq:integral}), we obtain the expression in 
Eq.~(\ref{eq:Phi_Poisson}).

\section{Derivation of the expression for the force}
\label{sec:derivation}

In this Appendix, we derive the expression for the force in the multi-stream region, given in Eqs.~(\ref{eq:force_inner}), (\ref{eq:force_inner_Q}) and (\ref{eq:force_inner_const}).

To start with, we divide the integral in Eq.~(\ref{eq:Force_Poisson}) into several pieces according to Eq.~(\ref{eq:part_by_part_integrals}): 
\begin{align}
&F(x)=-\frac{3}{2}\,\omegam H_0^2\,a\,
\nonumber
\\
&\quad \times\Biggl[\,\int_{x(-\Qchat)}^{x(\Qchat)}{\rm d}x'\,
 \frac{\delta(x')}{2}
\left\{\Theta\left(x(Q)-x'\right)-\Theta\left(x'-x(Q)\right)\right\}
\nonumber
\\
&\qquad +\frac{1}{L}\int_{x(-\Qchat)}^{x(\Qchat)}{\rm d}x'\,\,x'\,\delta(x') 
\nonumber
\\
&\qquad+\left(\int_0^{x(-\Qchat)}+\int_{x(\Qchat)}^L\right)\,
{\rm d}x'\,\frac{\delta(x')}{2}
\nonumber \\
& \quad \quad \quad \quad \quad \quad \quad \quad \quad \quad \times \left\{\Theta(x-x')-\Theta(x'-x)\right\}
\nonumber
\\
&\qquad+\frac{1}{L}\left(\int_0^{x(-\Qchat)}+\int_{x(\Qchat)}^L\right)\,
{\rm d}x'\,x'\,\delta(x')\,\Biggr].
\label{eq:force_pieces}
\end{align}
We shall below evaluate these integrals one by one.

Consider the first integral in right-hand side of Eq.~(\ref{eq:force_pieces}). We evaluate it at position $x_*=x(Q_*)=x(Q_\pm)$ in Fig.~\ref{fig:shellcrossing}. Using Eq.~(\ref{eq:jacobian_delta}), the integral is rewritten 
\begin{align}
&\int_{x(-\Qchat)}^{x(\Qchat)}{\rm d}x'\,
\frac{\delta(x')}{2}
\left\{\Theta(x_*-x')-\Theta(x'-x_*)\right\}
\nonumber\\
&=
\int_{-\Qchat}^{\Qchat} \frac{{\rm d}Q'}{2}\,
\left\{1-\left(\frac{\partial x}{\partial q}\right)_{Q'}\right\}
\nonumber\\
&\quad\times
\Bigl[\Theta\Bigl(x(Q_*;\tau)-x(Q';\tau)\Bigr)-
\Theta\Bigl(x(Q';\tau)-x(Q_*;\tau)\Bigr)\Bigr].
\end{align}
Exploiting the geometric properties of the configuration in Fig.~\ref{fig:shellcrossing}, 
substitution of the explicit form of $x(Q)$ (Eq.~\ref{eq:ballistic_shellcrossing}) into the above leads to 
\begin{align}
&\Bigl(\int_{-\Qchat}^{Q_-}-\int_{Q_-}^{Q_*}+\int_{Q_*}^{Q_+}-\int_{Q_+}^{\Qchat}\Bigr)\,
\nonumber\\
&\qquad\qquad\qquad\times
\frac{{\rm d}Q'}{2}\Bigl\{1+B(q_0;\tau)-3C(q_0;\tau)\,Q'^2\Bigr\}
\nonumber\\
&\qquad=\Bigl\{1+B(q_0;\tau)\Bigr\}\,(Q_-+Q_+-Q_*)
\nonumber\\
&\qquad\qquad\qquad\qquad\qquad -C(q_0;\tau)\,(Q_-^3+Q_+^3-Q_*^3).
\label{eq:integral_inner1_mid}
\end{align}
Here we used the fact that $Q_-<Q_*<Q_+$. Recalling that $Q_\pm$ and $Q_*$ are related with each other through Eqs.~(\ref{eq:relation_three-values_1}) and (\ref{eq:relation_three-values_2}), Eq.~(\ref{eq:integral_inner1_mid}) can be recast as a single function of either $Q_\pm$ or $Q_*$, which respectively provides the general expression valid for $\Qc<|Q|<\Qchat$ and $|Q|<\Qc$. Omitting the subscript $\pm$ or $*$ in either cases, we finally obtain
\begin{align}
&\int_{x(-\Qchat)}^{x(\Qchat)}{\rm d}x'\,
\frac{\delta(x')}{2}
\left\{\Theta(x-x')-\Theta(x'-x)\right\}
\nonumber\\
&\quad
=\left\{
\begin{array}{l}
\Bigl\{1+B(q_0;\tau)\Bigr\}\,Q-C(q_0;\tau)\,Q^3\\
\quad\quad\quad\quad -\mbox{sgn}(Q)\sqrt{3(\Qchat^2-Q^2)};\,\, Q_{\rm c}<|Q|<\Qchat,
\\
\\
 \Bigl\{-2+B(q_0;\tau)\Bigr\}\,Q-C(q_0;\tau)\,Q^3\,\,;\,\,  |Q|<\Qc. \\
\end{array}
\right.
\label{eq:integral_inner1}
\end{align}

Next, we consider the second integral in Eq.~(\ref{eq:force_pieces}). Rewriting the integrand  
in terms of the variable $Q$ with the help of Eqs.~(\ref{eq:jacobian_delta}) and (\ref{eq:ballistic_shellcrossing}), it is straightforward to obtain
\begin{align}
&\frac{1}{L}\int_{x(-\Qchat)}^{x(\Qchat)}{\rm d}x'\,\,x'\,\delta(x') 
\nonumber\\
&= \frac{1}{L}\int_{-\Qchat)}^{\Qchat}{\rm d}x'\,\,\,
\Bigl\{A(q_0;\tau)-B(q_0;\tau)\,Q'+C(q_0;\tau)\,Q'^3\Bigr\}
\nonumber\\
&\qquad\qquad\qquad\times\Bigl\{1+B(q_0;\tau)-3C(q_0;\tau)\,Q'^2\Bigr\}
\nonumber\\
&
=\frac{2}{L} A(q_0;\,\tau)\,\left\{1+B(q_0;\,\tau)-C(q_0;\,\tau)\,\Qchat^2\,\right\}\Qchat.
\label{eq:integral_inner2}
\end{align}

The remaining integrals to be computed, i.e., the third and fourth terms in Eq.~(\ref{eq:force_pieces}), are performed over intervals outside the multi-valued region, which allows us to use Zel'dovich solution to evaluate the integrands. For the third integral in Eq.~(\ref{eq:force_pieces}), we obtain
\begin{align}
&\Bigl(\int_0^{x(-\Qchat)}+\int_{x(\Qchat)}^L\Bigr)\,
{\rm d}x'\,\frac{\delta(x')}{2}
\left\{\Theta(x-x')-\Theta(x'-x)\right\}
\nonumber\\
&=\Bigl(\int_{Q(x=0)}^{-\Qchat}-\int_{\Qchat}^{Q(x=L)}\Bigr)\,{\rm d}Q'\,
\frac{1}{2}\left(-\frac{\partial \psi}{\partial Q'}\right)D_+(\tau)
\nonumber\\
&=\frac{1}{2}\Bigl\{\psi(Q(x=0))+\psi(Q(x=L)) - \psi(-\Qchat)-\psi(\Qchat)\Bigr\} \nonumber \\
 & \quad \quad \quad \quad \quad \quad \quad \quad \quad \times D_+(\tau).
\label{eq:integral_outer1_mid}  
\end{align}
Although this expression is exact, our perturbative treatment implicitly assumes that the multi-stream region is small and the quantities inside this region are written in polynomial forms of $Q$. We may expand the last two terms of Eq.~(\ref{eq:integral_outer1_mid}) as
\begin{align}
\psi(-\hat{Q}_{\rm c})+\psi(\hat{Q}_{\rm c})
&\simeq 2\psi(q_0)+\mathcal{O}(\hat{Q}_{\rm c}^4),
\end{align}
where we used the fact that $\psi''(q_0)=0$ (see Eq.~\ref{eq:shellcrossing_peak}). Additionally, periodic boundaries imply $\psi(Q(x=0))=\psi(Q(x=L))$. Therefore, Eq.~(\ref{eq:integral_outer1_mid}) can be recast as 
\begin{align}
&\left\{\int_0^{x(-\Qchat)}+\int_{x(\Qchat)}^L\right\}\,
{\rm d}x'\,\frac{\delta(x')}{2}
\left\{\Theta(x-x')-\Theta(x'-x)\right\}
\nonumber\\
&\quad \quad \quad =\Bigl[ \psi(q) \Bigr]_{q_0}^{q(x=L)} D_+(\tau).
\label{eq:integral_outer1}
\end{align}

In similar manner, we evaluate the fourth integral in Eq.~(\ref{eq:force_pieces}) as follows:
\begin{align}
&\frac{1}{L}\Bigl(\int_0^{x(-\Qchat)}+\int_{x(\Qchat)}^L\Bigr)\,
{\rm d}x'\,x'\,\delta(x')
\nonumber\\
&
=\frac{1}{L}\Bigl(\int_{q(x=0)}^{q(x=L)}-\int_{q_0-\Qchat}^{q_0+\Qchat}\Bigr)\,{\rm d}q'\,\left\{q'+\psi(q')\,D_+(\tau)\right\}
\nonumber\\
&\quad\qquad\times\left(-\frac{\partial \psi(q')}{\partial q'}\right)\,D_+(\tau).
\label{eq:outer2_step1}
\end{align}
The first of these two integrals is performed exactly to give 
\begin{align}
&\frac{1}{L}\int_{q(x=0)}^{q(x=L)}
{\rm d}q'\,\left\{q'+\psi(q')\,D_+(\tau)\right\}\left(-\frac{\partial \psi(q')}{\partial q'}\right)\,D_+(\tau)
\nonumber\\
&\quad
=\frac{1}{L}\Biggl\{
-\Bigl[q'\psi(q')\Bigr]_{q(x=0)}^{q(x=L)} -\frac{1}{2}
\Bigl[\psi(q')^2\Bigr]_{q(x=0)}^{q(x=L)}\,
\nonumber\\
&\quad\quad
+\int_{q(x=0)}^{q(x=L)} {\rm d}q'\,\psi(q')\Biggr\}\,D_+(\tau)
\nonumber\\
&\quad
=-\psi(q(x=0))\,D_+(\tau), 
\label{eq:outer2_part1}
\end{align}
where we exploited properties of periodic boundaries,
$q(x=L)-q(x=0)=L$ and $\psi(q(x=0))=\psi(q(x=L))$, as well as momentum conservation,  which implies
$\int_{q(x=0)}^{q(x=L)}{\rm d}q'\,\psi(q')=0$. Turning to the second integral in Eq.~(\ref{eq:outer2_step1}), we use again the fact that the domain of integration is assumed to be narrow to Taylor expand the integrand using the formula
\[
\int_{-\Delta s}^{\Delta s} {\rm d}s' \,f(s)\,\,\simeq\,\, 2\,f(0)\,\Delta s +\frac{1}{3}f''(0)(\Delta s)^3,
\]
to obtain the following expression, valid at third-order in $\Qchat$:
\begin{align}
&\frac{1}{L}\int_{-\Qchat+q_0}^{\Qchat+q_0}
{\rm d}q'\,\left\{q'+\psi(q')\,D_+(\tau)\right\}\left(-\frac{\partial \psi(q')}{\partial q'}\right)\,D_+(\tau)
\nonumber\\
&
\simeq
\frac{2}{L}\,\Bigl\{q_0+\psi(q_0)\,D_+(\tau)\Bigr\}\, \nonumber \\
&
\quad \quad \times \left\{\,\delta_{\rm L}(q_0)\,\,\Qchat + 
\frac{1}{6}\,\delta_{\rm L}''(q_0)\,\,\Qchat^3\right\}\,D_+(\tau),
\nonumber\\
&
=\,\frac{2}{L}\,A(q_0;\tau)\left\{\,1+B(q_0;\tau)\,-\,C(q_0;\tau)\,
\Qchat^2\,\right\}\,\Qchat.
\label{eq:outer2_part2}
\end{align}
Here, Eqs.~(\ref{eq:def_A})--(\ref{eq:def_C}) are used to rewrite the expression in the last line. Summing up the results in Eqs.~(\ref{eq:outer2_part1}) and (\ref{eq:outer2_part2}), we obtain, at the end,
\begin{align}
&\frac{1}{L}\left(\int_0^{x(-\Qchat)}+\int_{x(\Qchat)}^L\right)\,
{\rm d}x'\,x'\,\delta(x')
\nonumber\\
&
\qquad=-
\frac{2}{L}\,A(q_0;\tau)\left\{\,1+B(q_0;\tau)\,-\,C(q_0;\tau)\,
\Qchat^2\,\right\}\,\Qchat
\nonumber\\
&\qquad\qquad-\psi(q(x=0))\,D_+(\tau).
\label{eq:integral_outer2}
\end{align}

Plugging Eqs.~(\ref{eq:integral_inner1}), (\ref{eq:integral_inner2}), (\ref{eq:integral_outer1}) and (\ref{eq:integral_outer2}) into Eq.~(\ref{eq:force_pieces}), we finally obtain the expression of the force in the multi-valued region (Eqs.~\ref{eq:force_inner}, \ref{eq:force_inner_Q} and \ref{eq:force_inner_const}).

\section{Derivation of basic post-collapse PT results}
\label{sec:pcpt_basic}

In this Appendix, we detail the calculations of basic post-collapse PT predictions summarized in Eqs.~(\ref{eq:p_for_tau_c>tau>hattau_c}), (\ref{eq:x_for_tau_c>tau>hattau_c}), (\ref{eq:p_for_tau>tau_c}) and (\ref{eq:x_for_tau>tau_c}), together with the coefficients in Table~\ref{tab:xv_pcpt1} and \ref{tab:xv_pcpt2}.

\subsection{Derivation of $\Delta\rmv_{\rm out}$}
\label{subsec:Delta_v_out}

Our starting point is Eq.~(\ref{eq:v_ii}), which involves two integrals. To evaluate the first integral, the approximate formula given by Eq.~(\ref{eq:integ_formula_J}) is applied. Recalling that we are considering the outer part of the multi-valued region, $\Qc<|Q|<\Qchat$, we have 
\begin{align}
&\int_{\widehat{\tau}_{\rm c}(Q)}^\tau {\rm d}\tau'\,a(\tau')\,\mathcal{J}(Q;\,q_0,\tau') 
\nonumber\\
&\quad \simeq a(\tau_0)\Biggl[\,
T\,Q +
\left\{-\frac{\kappa}{8}+\frac{1}{6}\delta_{\rm L}''(q_0)D_+(\tau_0)\,T\right\}Q^3
\nonumber\\
&\qquad
-\mbox{sgn}(Q)\frac{\kappa}{4\sqrt{3}}\left(\Qchat(\tau)^2-Q^2\right)^{3/2}
\nonumber \\
&\qquad -\frac{\kappa}{48}\delta_{\rm L}''(q_0)D_+(\tau_0)\,\,Q^5
\Biggr],
\label{eq:integ_J1}
\end{align}
with $T\equiv \tau-\tau_0$. Here, we used the leading-order expression for critical time $\widehat{\tau}_{\rm c}$,  $\widehat{\tau}_{\rm c}-\tau_0\simeq (\kappa/8) \,Q^2$, and ignore in Eq.~(\ref{eq:integ_formula_J}) the higher-order terms involving the quantities $g(\tau_0)$ or $H(\tau_0)$.

On the other hand, the second integral in Eq.~(\ref{eq:v_ii}) is analytically performed. Using the evolution equation in Eq.~(\ref{eq:evolv_lin}), we obtain 
\begin{align}
&-\frac{3}{2}H_0^2\Omega_{\rm m,0} \,
\int_{\widehat{\tau}_{\rm c}(Q)}^\tau {\rm d}\tau'\,a(\tau')\,\mathcal{F}(q_0,\tau') 
\,\,
\nonumber\\
&\quad= \,\,\psi(q_0)\,\int_{\widehat{\tau}_{\rm c}(Q)}^{\tau} {\rm d}\tau'\,\,
\frac{{\rm d}^2\,D_+(\tau')}{{\rm d}\tau'^2}
=\,\psi(q_0)\,\Bigl[\frac{{\rm d}D_+(\tau')}{{\rm d}\tau'}\Bigr]_{\widehat{\tau}_{\rm c}(Q)}^\tau.
\label{eq:integ_F}
\end{align}

Summing up the above two contributions, we obtain the expression for $\Delta\rmv_{\rm out}$ summarized in Eq.~(\ref{eq:p_for_tau_c>tau>hattau_c}).

\subsection{Derivation of $\Delta x_{\rm out}$}
\label{subsec:Delta_x_out}

To derive the expression for $\Delta x_{\rm out}$, we just need to integrate over time Eq.~(\ref{eq:p_for_tau_c>tau>hattau_c}), as written in Eq.~(\ref{eq:x_ii}):
\begin{align}
&\Delta x_{\rm out}(Q;\tau,\,\widehat{\tau}_{\rm c})=\int_{\widehat{\tau}_{\rm c}(Q)}^\tau {\rm d}\tau'
\,\Delta\rmv_{\rm out}(Q,\tau')
\nonumber\\
&~
=-\frac{3}{2}H_0^2\,\Omega_{\rm m,0}\,a(\tau_0)
\int_{\widehat{\tau}_{\rm c}(Q)}^\tau {\rm d}\tau'\,\Bigl[\,\widetilde{\alpha}_1(\tau')\,Q+
\widetilde{\beta}_1(\tau')\,Q^3
\nonumber\\
&~
\quad +\widetilde{\gamma}_1(\tau_0)\,
\Bigl\{\Qchat^2(\tau)-Q^2\Bigr\}^{3/2}
+\widetilde{\delta}_1(\tau_0)\,Q^5\,\Bigr]
\nonumber \\
& ~\quad +\int_{\widehat{\tau}_{\rm c}(Q)}^\tau {\rm d}\tau'\,\,\widehat{\epsilon}_1
\,(\tau',\widehat{\tau}_{\rm c}).
\label{eq:x_ii_prime}
\end{align}

Making use of the expressions summarized in Table \ref{tab:xv_pcpt1}, we evaluate the first integral. A straightforward calculation leads to 
\begin{align}
&\int_{\widehat{\tau}_{\rm c}(Q)}^\tau {\rm d}\tau'\,\Biggl[\,\widetilde{\alpha}_1(\tau')\,Q+
\widetilde{\beta}_1(\tau')\,Q^3+\widetilde{\gamma}_1(\tau_0)\,
\Bigl\{\Qchat^2(\tau)-Q^2\Bigr\}^{3/2}
\nonumber\\
&\qquad\qquad+\widetilde{\delta}_1(\tau_0)\,Q^5\,\Biggr]
\nonumber\\
&\simeq 
\frac{T^2}{2}\,Q+
\Bigl\{-\frac{\kappa}{8}\,T+\frac{\delta_{\rm L}''(q_0)}{6}D_+(\tau_0) \,
\Bigr\}Q^3
\nonumber\\
&\quad
-\mbox{sgn}(Q)\,\frac{\kappa^2}{80\sqrt{3}}\,
\left\{\Qchat^2(\tau)-Q^2\right\}^{5/2}
\nonumber\\
&\quad
+\Biggl[\frac{1}{2}\left(\frac{\kappa}{8}\right)^2-\left(\frac{\kappa}{8}\right)\,\frac{\delta_{\rm L}''(q_0)}{6}D_+(\tau_0)\,T\,\Biggr]Q^5
\nonumber\\
&\quad+
\frac{1}{2}\left(\frac{\kappa}{8}\right)^2 \frac{\delta_{\rm L}''(q_0)}{6}D_+(\tau_0)Q^7.
\label{eq:tilde_x_i}
\end{align}
Note again that we used the leading-order expression for $\widehat{\tau}_{\rm c}$ (see Eq.~\ref{eq:hattau_c}).

The second integral in Eq.~(\ref{eq:x_ii_prime}) is analytically performed to give
\begin{align}
&\int_{\widehat{\tau}_{\rm c}(Q)}^\tau {\rm d}\tau'\,\widetilde{\epsilon}_1(\tau,\widehat{\tau}_{\rm c})
=\psi(q_0)\int_{\tauchat(Q)}^\tau {\rm d}\tau'\,\left[\frac{{\rm d}D_+(\tau'')}{{\rm d}\tau''}\right]^{\tau'}_{\tauchat(Q)}
\nonumber\\
&\quad
=\psi(q_0)\,\Bigg\{ D_+(\tau)-D_+(\tauchat(Q))  \nonumber\\
& \quad \quad \quad \quad -\left.\left.\frac{{\rm d}D_+}{{\rm d}\tau}\right|_{\tauchat(Q)}(\tau-\tauchat(Q))\right\}.
\label{eq:integ_eps_1}
\end{align}

Collecting Eqs.~(\ref{eq:tilde_x_i}) and (\ref{eq:integ_eps_1}), we obtain Eq.~(\ref{eq:x_for_tau_c>tau>hattau_c}).

\subsection{Derivation of $\Delta\rmv_{\rm in}$}
\label{subsec:Delta_v_in}

For $\Delta\rmv_{\rm in}$, the expression to be evaluated is given by Eq.~(\ref{eq:Delta_p_2}), which involves three integrals. The first integral is obtained from Eq.~(\ref{eq:integ_J1}) by setting $\tau-\tau_0\equiv \tau_{\rm c}(Q)-\tau_0\simeq (\kappa/2)\,Q^2$ (Eq.~\ref{eq:tau_c}).  Recalling the fact that $\Qchat(\tau_{\rm c})=2Q$ (Eq.~\ref{eq:Qchat_Qc}), we have
\begin{align}
& \int_{\widehat{\tau}_{\rm c}(Q)}^{\tau_{\rm c}(Q)} {\rm d}\tau'\,a(\tau')\,
\mathcal{J}(Q;\,q_0,\tau')  
\nonumber\\
&\qquad\qquad\simeq \,a(\tau_0)\,\Biggl[
-\frac{3}{8}\,\kappa\,Q^3
+\frac{\kappa}{16}\,\delta_{\rm L}''(q_0)\,D_+(\tau_0)\,Q^5\,\Biggr].
\end{align}

For the second integral in Eq.~(\ref{eq:Delta_p_2}), we use the approximate formula given by Eq.~(\ref{eq:integ_formula_J}) in the regime $|Q|<Q_{\rm c}$.  Ignoring higher-order terms involving the quantities $g(\tau_0)$ or $H(\tau_0)$, we obtain 
\begin{align}
&\int_{\tau_{\rm c}(Q)}^{\tau} {\rm d}\tau'\,a(\tau')\,\mathcal{J}(Q;\,q_0,\tau') 
\nonumber\\
&\qquad\qquad
\simeq a(\tau_0)\,\Biggl[
-2T\,Q+
\left\{\kappa+\frac{T}{6}\delta_{\rm L}''(q_0)\,D_+(\tau_0)\right\}Q^3
\nonumber\\
&\qquad\qquad\quad
-\frac{\kappa}{12}\,\delta_{\rm L}''(q_0)\,D_+(\tau_0)\,Q^5 \Biggr].
\end{align}

Finally, the last integral in Eq.~(\ref{eq:Delta_p_2}) is evaluated using the formula given in Eq.~(\ref{eq:integ_F}). 
Collecting the three contributions above, we obtain the expression for  $\Delta\rmv_{\rm in}$ (Eq.~\ref{eq:p_for_tau>tau_c}).

\subsection{Derivation of $\Delta x_{\rm in}$}
\label{subsec:Delta_x_in}

The starting expression for $\Delta x_{\rm in}$ is given in Eq.~(\ref{eq:x_iii}), which involves two integrals. To evaluate the first integral in Eq.~(\ref{eq:x_iii}), we use Eq.~(\ref{eq:tilde_x_i}). Setting $T=\tau_{\rm c}-\tau_0\simeq (\kappa/2)\,Q^2$, an expression at seventh-order in $Q$ is obtained: 
\begin{align}
&\int_{\widehat{\tau}_{\rm c}(Q)}^{\tau_{\rm c}(Q)} {\rm d}\tau'
\,\Delta\widetilde{\rmv}(Q;q_0,\tau') \simeq 
-\frac{3}{2}H_0^2\,\Omega_{\rm m,0}\,a(\tau_0)
\nonumber\\
&\qquad\times
\Biggl(
\,-\frac{27}{640}\,\kappa^2\,Q^5 +
\frac{3}{4}\left(\frac{\kappa}{8}\right)^2\delta_{\rm L}''(q_0)\,
D_+(\tau_0)\,Q^7\Biggr).
\label{eq:integ_x_ii_part1}
\end{align}

To compute the second integral in Eq.~(\ref{eq:x_iii}), we substitute Eq.~(\ref{eq:p_for_tau>tau_c}) into the integrand, but without the term $\tilde{\epsilon}_1$. Using the explicit expression for the coefficients in Table \ref{tab:xv_pcpt2}, a straightforward calculation gives
\begin{align}
&\int_{\tau_{\rm c}(Q)}^{\tau} {\rm d}\tau'\,\Delta\widetilde{\rmv}(Q;q_0,\tau')
=-\frac{3}{2}H_0^2\,\Omega_{\rm m,0}\,a(\tau_0)\,
\nonumber\\
&~
\times\int_{\tau_{\rm c}(Q)}^{\tau} {\rm d}\tau'\,\Biggl[\,\widetilde{\alpha}_3(\tau')\,Q+
\widetilde{\beta}_3(\tau')\,Q^3+
\widetilde{\delta}_3(\tau_0)\,Q^5+\widetilde{\zeta}_3(\tau_0)\,Q^7
\Bigr]
\nonumber\\
&~=-\frac{3}{2}H_0^2\,\Omega_{\rm m,0}\,a(\tau_0)\,
\nonumber\\
&\quad
\times\Biggl(\,-T^2\,Q+
\left\{\frac{5\kappa}{8}\,T+\frac{\delta_{\rm L}''(q_0)}{12}\,D_+(\tau_0)
T^2\right\}\,Q^3
\nonumber\\
&\qquad\quad-
\Bigl\{\left(\frac{\kappa}{4}\right)^2\,
+\left(\frac{\kappa}{8}\right)\frac{\delta_{\rm D}''(q_0)}{6}\,D_+(\tau_0)\,T\,
\Bigr\}\,Q^5
\nonumber\\
&\qquad\quad
-4\left(\frac{\kappa}{8}\right)^2\frac{\delta_{\rm L}''(q_0)}{6}\,D_+(\tau_0)\,Q^7 \Biggr).
\label{eq:integ_x_ii_part2}
\end{align}

Combining the above two results, we obtain Eq.~(\ref{eq:x_for_tau>tau_c}), which gives the final expression for $\Delta x_{\rm in}$ along with the coefficients in Table \ref{tab:xv_pcpt2}.

\section{Higher-order corrections to post-collapse PT}
\label{sec:pcpt_higher-order}

In this Appendix, we detail calculations of the various higher-order corrections to post-collapse PT we propose in  \S~\ref{subsec:higher-order}. 

One possible way to improve on basic post-collapse PT is to Taylor expand at next-to-leading order the expression for $\Qchat$ (as well as  $\Qc$) that determines the boundary of the post-collapse region (Eqs.~\ref{eq:Qchat_Qc} and \ref{eq:Qchat_approx}):   
\begin{align}
\Qchat&=2Q_{\rm c}=\sqrt{\frac{4B}{3C}}
\nonumber\\
&\simeq \left\{\frac{8}{\kappa(q_0,\tau_0)}\right\}^{1/2}
T^{1/2}\Biggl\{1-\frac{\eta(\tau_0)}{2}\,T+\cdots\Biggr\}, 
\label{eq:higher_order_Qhat}
\end{align}
with 
\begin{align}
&\eta(\tau_0)\equiv 
\frac{1}{D_+(\tau_0)}\frac{{\rm d}D_+(\tau_0)}{{\rm d}\tau_0}
-\frac{1}{2}\frac{{\rm d}^2D_+(\tau_0)}{{\rm d}\tau_0^2}\bigg/\frac{{\rm d}D_+(\tau_0)}{{\rm d}\tau_0}.
\label{eq:def_eta}
\end{align}
Inverting the relation $Q=\Qchat(\widehat{\tau}_{\rm c})$ [$Q=\Qc(\tau_{\rm c})$], we obtain the corresponding critical times at next-to-leading order, 
\begin{align}
&\widehat{\tau}_{\rm c}(Q)-\tau_0 \quad
\simeq \frac{\kappa(q_0,\tau_0)}{8}\,Q^2 
+\left\{\frac{\kappa(q_0,\tau_0)}{8}\right\}^2\eta(\tau_0)\,Q^4,
\label{eq:higher_order_hattau_c}
\end{align}
and
\begin{align}
\tau_{\rm c}(Q)-\tau_0 &= \widehat{\tau}_{\rm c}(2Q)-\tau_0
\nonumber\\
&\simeq \frac{\kappa(q_0,\tau_0)}{2}\,Q^2 
+\left\{\frac{\kappa(q_0,\tau_0)}{2}\right\}^2\eta(\tau_0)\,Q^4.
\label{eq:higher_order_tau_c}
\end{align}
Using these expressions, we repeat below the same calculations as in Sec.~\ref{subsec:corrections_pcpt}. Higher-order expressions for the correction to Zel'dovich flow are derived on top of basic post-collapse PT results and are denoted by $\Delta x^{\rm (hc)}$ and $\Delta \rmv^{\rm (hc)}$. In Sec.~\ref{subsec:higher-order}, a semi-analytic treatment designed by {\em spl} is also discussed. In this case,  we adopt the basic post-collapse PT results in the inner part of the multi-stream region, but the extension of this latter is calculated using the higher-order expression (\ref{eq:higher_order_Qhat}) for $\Qc$. Then, a straightforward third-order spline interpolation, that we do not detail here, is used to connect the inner part to the Zel'dovich solution, assuming again that the Lagrangian boundary $\Qchat$ of the multi-stream region is given by the higher-order expansion (\ref{eq:higher_order_Qhat}). 
 
Another improvement may come from the calculation of the integrals in Eqs.~(\ref{eq:Delta_p_formal}) and (\ref{eq:Delta_x_formal}). In Sec.~\ref{subsec:corrections_pcpt}, a part of the integrands is Taylor-expanded in time and the integration is performed for the leading-order terms. The resultant expressions include time-dependent terms up to $\mathcal{O}(T^1)$ for the velocity and $\mathcal{O}(T^2)$ for the position. Here, using the approximate integral formulae for the Taylor-expanded integrands in Appendix \ref{sec:integral}, we keep terms up to $\mathcal{O}(T^2)$ for the velocity, $\mathcal{O}(T^3)$ for the position, and derive the corresponding higher-order corrections to the Zel'dovich flow, $\Delta x^{\rm (ho)}$ and $\Delta \rmv^{\rm (ho)}$. These corrections assume, of course, higher-order expansions for critical times $\widehat{\tau}_{\rm c}$ and $\tau_{\rm c}$. 

Below, we present the expressions for $\Delta x^{\rm (ho)}$ and $\Delta \rmv^{\rm (ho)}$. The expressions for $\Delta x^{\rm (hc)}$ and $\Delta \rmv^{\rm (hc)}$ are obtained by simply setting $g(\tau_0)$ and $H(\tau_0)$ to zero.

\subsection{Velocity and position in the outer part of the multi-stream region: $\Qc<|Q|\leq\Qchat$}

The expression for the higher-order correction to the velocity, $\Delta\rmv^{\rm(ho)}$, is
\begin{align}
& \Delta \rmv_{\rm out}^{\rm (ho)}(Q;\tau)=
-\frac{3}{2}H_0^2\,\Omega_{\rm m,0}\,a(\tau_0)
\Bigl[\,\widetilde{\alpha}_1(\tau)\,Q
\nonumber\\
&\qquad\qquad
+\widetilde{\beta}_1(\tau)\,Q^3
+\widetilde{\delta}_1(\tau_0)\,Q^5
+\widetilde{\zeta}_1(\tau_0)\,Q^7
\Bigr],
\label{eq:p_higher_tau_c>tau>hattau_c_higher}
\end{align}
with 
\begin{align}
& \widetilde{\alpha}_1(\tau)=\frac{g(\tau_0)}{2}\,T^2,
\label{eq:def_delta_alpha_1_higher}
\\
& \widetilde{\beta}_1(\tau)=
\frac{g(\tau_0)}{2}\,T^2\,\frac{\delta_{\rm L}''(q_0)}{6}\,D_+(\tau_0),
\label{eq:def_delta_beta_1_higher}
\\
&\widetilde{\delta}_1(\tau_0)=-\left(\frac{\kappa}{8}\right)^2\,\left\{\frac{g(\tau_0)}{2}+\eta\right\},
\label{eq:def_delta_delta_1_higher}
\\
& \widetilde{\zeta}_1(\tau_0)=-\left(\frac{\kappa}{8}\right)^2\left[\frac{\delta_{\rm L}''(q_0)}{6}D_+(\tau_0)\,
\left\{\frac{g(\tau_0)}{2}+\eta\right\}+\frac{\kappa}{8}\,g(\tau_0)\eta
\right].
\label{eq:def_delta_zeta_1_higher}
\end{align}
Here, we have introduced the following function:
\begin{align}
&g(\tau_0)\equiv H(\tau_0)\left\{a(\tau_0)\right\}^2+\frac{{\rm d}\ln D_+(\tau_0)}{{\rm d}\tau_0}, 
\label{eq:def_T_g}
\end{align}
where $H(\tau_0)$ is the value of the Hubble parameter at time $\tau_0$. 
For the higher-order correction to the position, $\Delta x_{\rm out}^{\rm(ho)}$, we have
\begin{align}
&\Delta x_{\rm out}^{\rm(ho)}(Q;\tau)=
-\frac{3}{2}H_0^2\,\Omega_{\rm m,0}\,a(\tau_0)
\Bigl[\, \widetilde{\alpha}_2(\tau)\,Q+\widetilde{\beta}_2(\tau)\,Q^3
\nonumber\\
&\qquad\qquad
+ \widetilde{\delta}_2(\tau)\,Q^5\,+
 \widetilde{\zeta}_2(\tau)\,Q^7\,
+ \widetilde{\mu}_2(\tau)\,Q^9\,+
 \widetilde{\nu}_2(\tau)\,Q^{11}\,
\Bigr],
\label{eq:x_higher_tau_c>tau>hattau_c_higher}
\end{align}
with 
\begin{align}
& \widetilde{\alpha}_2(\tau)=\frac{g(\tau_0)}{6}T^3,
\label{eq:def_delta_alpha_2_higher}
\\
&\widetilde{\beta}_2(\tau)=\frac{\delta_{\rm L}''(q_0)}{6}D_+(\tau_0) \,\left\{\frac{T^2}{2}+\frac{g(\tau_0)}{6}T^3\right\},
\label{eq:def_delta_beta_2_higher}
\\
& \widetilde{\delta}_2(\tau)=-
\left(\frac{\kappa}{8}\right)^2\,\left(\frac{g(\tau_0)}{2}+\eta\right)\,T,
\label{eq:def_delta_delta_2_higher}
\\
&\widetilde{\zeta}_2(\tau)=-T\,\left\{\frac{g(\tau_0)}{2}+\eta\right\} \left(\frac{\kappa}{8}\right)^2 \frac{\delta_{\rm L}''(q_0)}{6}D_+(\tau_0) 
\nonumber\\
&\qquad\qquad+ 
\left(\frac{\kappa}{8}\right)^3\left\{\frac{g(\tau_0)}{3}+ \eta\Bigl(1-g(\tau_0)\,T\Bigr)\right\},
\label{eq:def_delta_zeta_2_higher}
\\
&\widetilde{\mu}_2(\tau)=\left(\frac{\kappa}{8}\right)^3 \frac{\delta_{\rm L}''(q_0)}{6}D_+(\tau_0)  \left\{\frac{g(\tau_0)}{3}+\eta\,\Bigl(1-g(\tau_0)\,T\Bigr)\,\right\}
\nonumber\\
&\qquad\qquad+
\left(\frac{\kappa}{8}\right)^4 \,\eta\,\left\{g(\tau_0)+\frac{\eta}{2}\,\Bigl(1-g(\tau_0)\,T\Bigr)\,\right\},
\label{eq:def_delta_mu_2_higher}
\\
& \widetilde{\nu}_2(\tau)=\frac{1}{49152}\,\kappa^4\eta^2\delta_{\rm L}''(q_0)\,D_+(\tau_0).
\label{eq:def_delta_nu_2_higher}
\end{align}
Note the term proportional to $Q^{11}$, which is necessary to enforce continuity of the solution.

\subsection{velocity and position in the inner part of the multi-stream region: $|Q|\leq\Qc$}

The expression for the higher-order corrections to the velocity, $\Delta\rmv^{\rm (ho)}_{\rm in}$, becomes
\begin{align}
&\Delta \rmv_{\rm in}^{\rm(ho)}(Q;\tau)=
-\frac{3}{2}H_0^2\,\Omega_{\rm m,0}\,a(\tau_0)
\Biggl[\,\widetilde{\alpha}_3(\tau)\,Q
\nonumber\\
&\qquad\qquad+ \widetilde{\beta}_3(\tau)\,Q^3+
 \widetilde{\delta}_3(\tau_0)\,Q^5
+\widetilde{\zeta}_3(\tau_0)\,Q^7
\Biggr],
\label{eq:p_higher_tau>tau_c}
\end{align}
with 
\begin{align}
& \widetilde{\alpha}_3(\tau)=
\frac{T^2}{2}\,\left[\,g(\tau_0)-3\,\{a(\tau_0)\}^2\,H(\tau_0)\right],
\\
&\widetilde{\beta}_3(\tau)=
\frac{g(\tau_0)}{2}T^2\,\frac{\delta_{\rm L}''(q_0)}{6}\,D_+(\tau_0),
\\
&
 \widetilde{\delta}_3(\tau_0)=\frac{47}{64}\kappa^2\,\eta 
-\frac{\kappa^2}{128} 
\Bigl[\,g(\tau_0)-48\{a(\tau_0)\}^2\,H(\tau_0)\Bigr],
\\
&
\widetilde{\zeta}_3(\tau_0)=-\left(\frac{\kappa}{8}\right)^2\,
\left\{\,\frac{g(\tau_0)}{2}+\eta\,\right\}
\frac{\delta_{\rm L}''(q_0)}{6}\,D_+(\tau_0)
\nonumber\\
&\qquad\qquad-\left(\frac{\kappa}{8}\right)^3\,\eta\
\Bigl[\,g(\tau_0)-192\{a(\tau_0)\}^2\,H(\tau_0)\Bigr].
\end{align}

Finally, the higher-order corrections to the position, $\Delta x_{\rm in}^{\rm (ho)}$, are given by 
\begin{align}
&\Delta x_{\rm in}^{\rm (ho)}(Q;\tau)=
-\frac{3}{2}H_0^2\,\Omega_{\rm m,0}\,a(\tau_0)
\Biggl[\,\widetilde{\alpha}_4(\tau)\,Q
+  \widetilde{\beta}_4(\tau)\,Q^3
\nonumber\\
&\qquad\qquad
+\widetilde{\delta}_4(\tau)\,Q^5+
 \widetilde{\zeta}_4(\tau)\,Q^7\,
+
 \widetilde{\mu}_4(\tau)\,Q^9\,
+\widetilde{\nu}_4(\tau)\,Q^{11}\,
\Biggr],
\label{eq:x_higher-order_tau>tau_c}
\end{align}
with 
\begin{align}
&\widetilde{\alpha}_4(\tau)=\frac{T^3}{6}\Bigl[\,g(\tau_0)-3\,\{a(\tau_0)\}^2H(\tau_0)\,\Bigr],
\label{eq:def_delta_alpha_4}
\\
&\widetilde{\beta}_4(\tau)=
\frac{\delta_{\rm L}''(q_0)}{6}\,D_+(\tau_0)
\frac{g(\tau_0)}{6}\,T^3, 
\label{eq:def_delta_beta_4}
\\
&\widetilde{\delta}_4(\tau)=
-\left(\frac{\kappa}{4}\right)^2\,\left(\frac{g(\tau_0)}{8}\,T-6\{a(\tau_0)\}^2H(\tau_0)\,T\,\right)
\nonumber\\
&\qquad\quad+\,47\left(\frac{\kappa}{8}\right)^2\,\eta\,T,
\label{eq:def_delta_delta_4}
\\
&\widetilde{\zeta}_4(\tau)=
-\left(\frac{\kappa}{8}\right)^2 T
\left(\frac{g(\tau_0)}{2}+\eta\right)\frac{\delta_{\rm L}''(q_0)}{6}\,D_+(\tau_0)
\nonumber\\
&\qquad\quad
+\left(\frac{\kappa}{8}\right)^3\Biggl\{\,\frac{g(\tau_0)}{3}-64\,\{a(\tau_0)\}^2H(\tau_0)
\nonumber\\
&\qquad\quad
-\eta\,\Bigl(95+g(\tau_0)\,T-192\,\{a(\tau_0)\}^2H(\tau_0)\,T\Bigr)\,\Biggr\},
\label{eq:def_delta_zeta_4}
\\
&\widetilde{\mu}_4(\tau)=\kappa^3\,\delta_{\rm L}''(q_0)\,D_+(\tau_0)
\left(\frac{g(\tau_0)}{9216}+\frac{1}{3072}\eta-\frac{g(\tau_0)\,\eta\,T}{3072}\right)
\nonumber\\
&\qquad\quad
+\kappa^4\,\eta\,\Biggl\{\frac{g(\tau_0)}{4096}-\frac{3}{16}\,\{a(\tau_0)\}^2H(\tau_0)-\frac{767\eta}{8092}
\nonumber\\
&\qquad\quad
-\frac{g(\tau_0)\,\eta\,T}{8192}+\frac{3}{32}\eta\,\{a(\tau_0)\}^2H(\tau_0)\,T\Biggr\},
\nonumber\\
& \widetilde{\nu}_4(\tau)=\frac{1}{49152}\,\kappa^4\eta^2\delta_{\rm L}''(q_0)\,D_+(\tau_0).
\label{eq:def_delta_nu_3}
\end{align}

\section{Approximate formulae for integrals}
\label{sec:integral}

\begin{figure*}
\includegraphics[width=8.0cm]{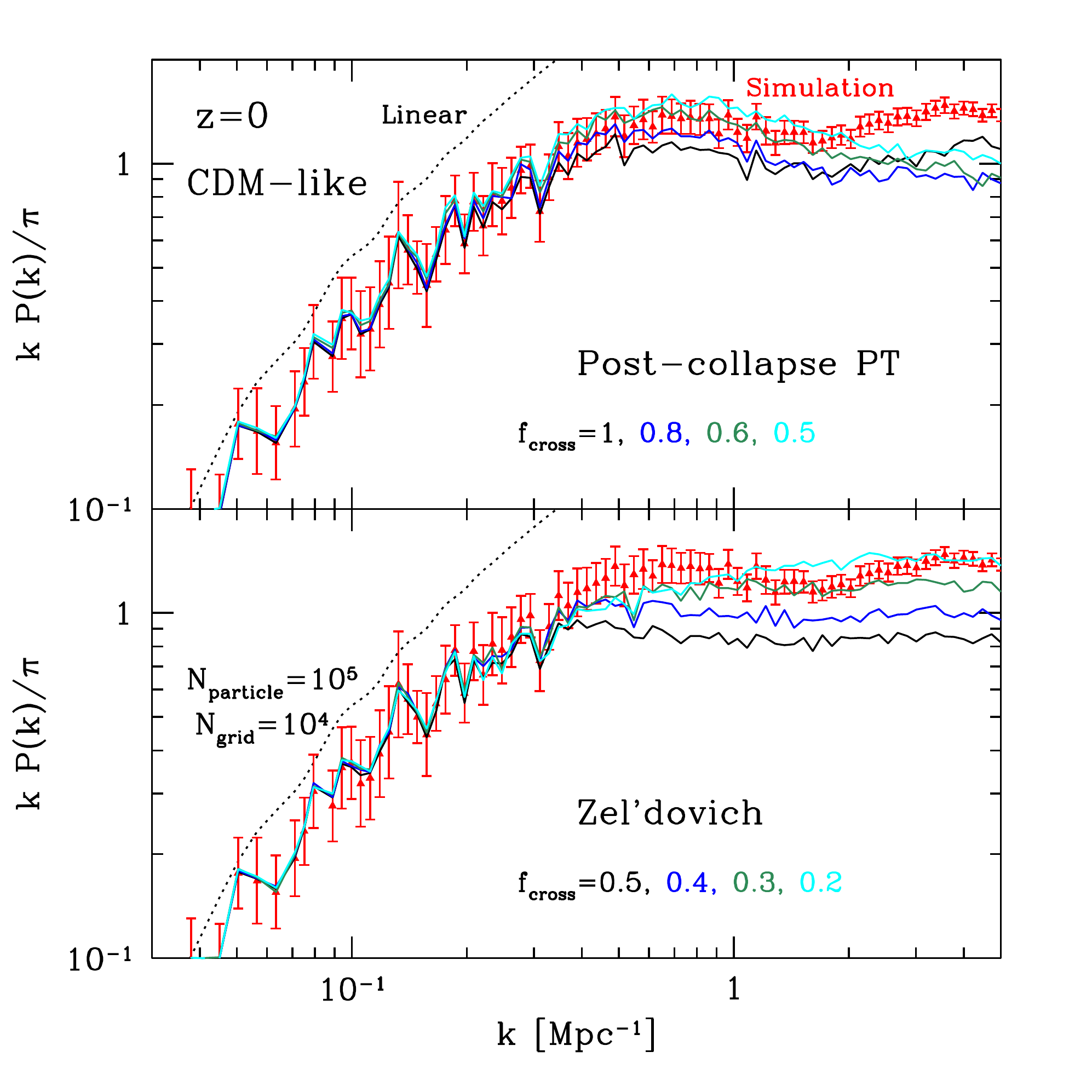}
\includegraphics[width=8.0cm]{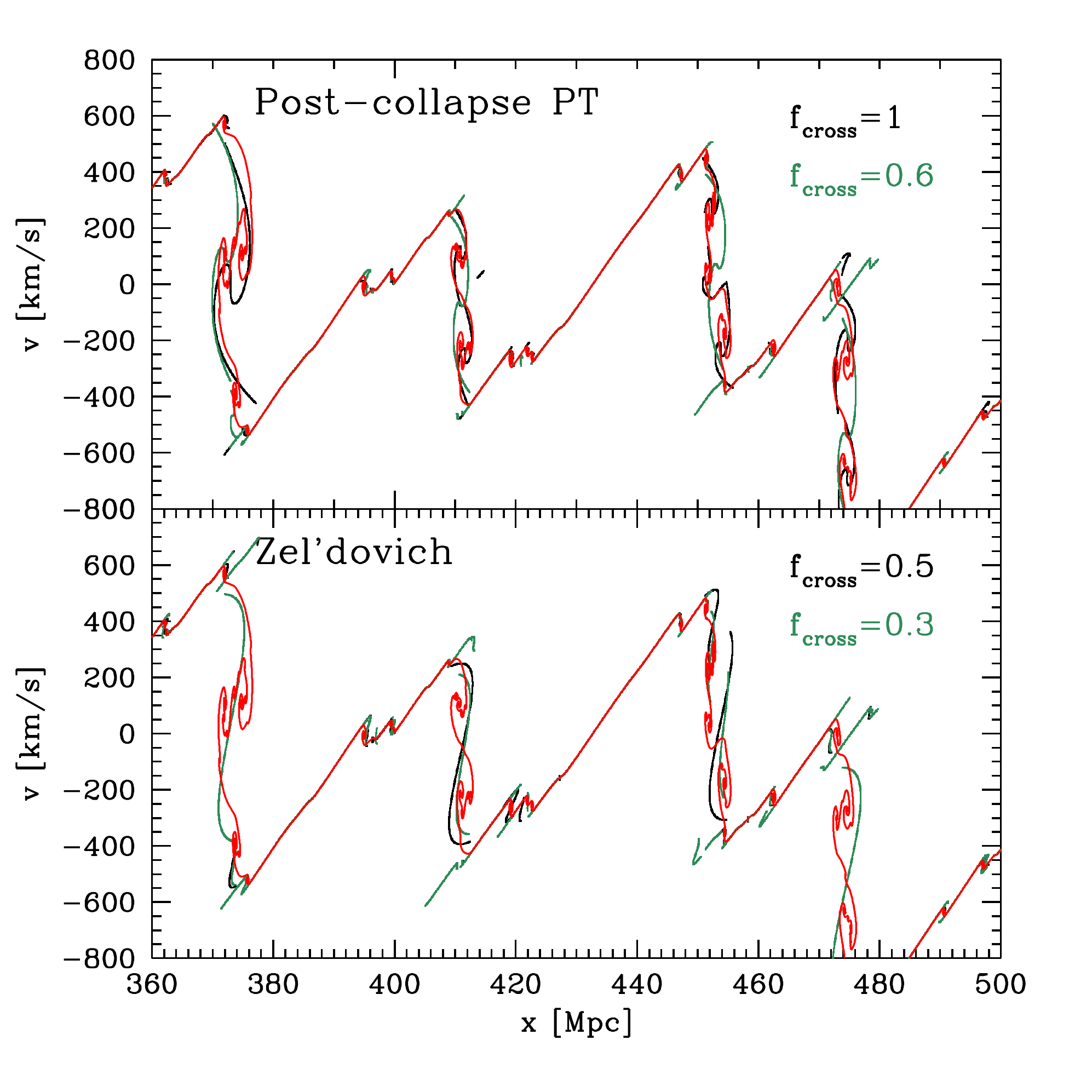}
\caption{Dependence of the PT predictions on the choice of parameter $f_{\rm cross}$ controlling adaptive smoothing in the case of a CDM-like power spectrum. Left panel shows the power spectrum while right panel plots the  phase-space portraits. In each panel, the predictions of post-collapse PT and Zel'dovich solution are respectively shown in upper and lower inserts, with various values of $f_{\rm cross}$ (indicated by different colors). 
\label{fig:pk_xv_fcross} }
\end{figure*}

In this appendix, we provide useful formulae for primitive integrals over time, which are used to derive the post-collapse PT results in Appendices \ref{sec:pcpt_basic} and \ref{sec:pcpt_higher-order}. 

Our aim here is to evaluate the time-integral of the force in the multi-valued region, given in Eq.~(\ref{eq:force_inner}). The time-integral of the second term, involving function $\mathcal{F}(q_0,\tau)$, is performed exactly (see Eq.~\ref{eq:integ_F}). On the other hand, the integral of the first term needs an approximate treatment which will be valid as long as the multi-valued region is sufficiently small, or equivalently, shortly after collapse time $\tau_0$. Below, we evaluate the integral by Taylor-expanding the integrand. We first give the results term by term: 
\begin{align}
&\int {\rm d}\tau' \,a(\tau')\simeq a(\tau_0)(\tau-\tau_0) +\frac{1}{2}\,H(\tau_0)a(\tau_0)^3\,(\tau-\tau_0)^2,
\\
&\int {\rm d}\tau' \,a(\tau')\,B(q_0;\tau')\simeq\frac{\delta_{\rm L}(q_0)}{2}\,
\nonumber\\
&\qquad\times 
a(\tau_0)D_+(\tau_0)\Bigl\{g(\tau_0)-H(\tau_0)\{a(\tau_0)\}^2\Bigr\}\,\,(\tau-\tau_0)^2,
\\
&\int {\rm d}\tau' \,a(\tau')\,C(q_0;\tau')\simeq
\left\{-\frac{\delta_{\rm L}''(q_0)}{6}\right\}
\nonumber
\\
&\qquad\times a(\tau_0)D_+(\tau_0)\Bigl\{
\,(\tau-\tau_0) + \frac{1}{2}g(\tau_0)(\tau-\tau_0)^2+\cdots\Bigr\},
\end{align}
where $g(\tau_0)$ is given by Eq.~(\ref{eq:def_T_g}). For the term with fractional power of time, we obtain
\begin{align}
&\int {\rm d}\tau' \,a(\tau')\,\sqrt{\Qchat^2(\tau')-Q^2}\simeq
a(\tau_0)\,\frac{\kappa(q_0,\tau_0)}{12}\,
\left\{\Qchat^2-Q^2\right\}^{3/2},
\nonumber\\
\end{align}
where we used the leading-order approximation for $\Qchat$, i.e., $\Qchat\simeq(8/\kappa)^{1/2}(\tau-\tau_0)^{1/2}$.

Summing up the above formulas,  we obtain the primitive of the time integrals of the force involving function $\mathcal{J}(Q;\,q_0,\tau)$. The result including polynomials of time up to $(\tau-\tau_0)^2$ is
\begin{align}
&\int_{\tau_{\rm i}}^{\tau_{\rm f}} {\rm d}\tau'\,a(\tau')\,\mathcal{J}(Q;\,q_0,\tau') 
\nonumber\\
&\quad\simeq \left\{
\begin{array}{l} 
a(\tau_0)\,\Biggl\{\,\Bigl[\,(\tau-\tau_0)+
\frac{1}{2}g(\tau_0)\,(\tau-\tau_0)^2 \Bigr]_{\tau_{\rm i}}^{\tau_{\rm f}}\,Q 
\\
\\
 \qquad+\frac{1}{6}\delta_{\rm L}''(q_0)\,D_+(\tau_0)\,
\\
\\
\qquad\times\,
\Bigl[\,
 (\tau-\tau_0)+ \frac{1}{2}g(\tau_0)\,(\tau-\tau_0)^2
\Bigr]_{\tau_{\rm i}}^{\tau_{\rm f}}\,Q^3
\\
\qquad -
\mbox{sgn}(Q)\,{\displaystyle \frac{\kappa}{4\sqrt{3}}}\,\Bigl[\,
\left(\Qchat(\tau)^2-Q^2\right)^{3/2}
\Bigr]_{\tau_{\rm i}}^{\tau_{\rm f}} \,\Biggr\}
\\
\qquad\qquad\qquad\qquad\qquad\qquad;\quad Q_{\rm c}<|Q|<\Qchat,
\\
\\
a(\tau_0)\,\Biggl\{\,
\Bigl[\,-2 (\tau-\tau_0)+
\frac{1}{2}\Bigl\{-3\,H(\tau_0)\,a(\tau_0)^2
\\
\qquad
+g(\tau_0)\Bigr\}(\tau-\tau_0)^2\,
\Bigr]_{\tau_{\rm i}}^{\tau_{\rm f}}\,Q 
 +  \frac{1}{6}\delta_{\rm L}''(q_0)\,D_+(\tau_0)
\\
\qquad
\times
\Bigl[\,(\tau-\tau_0)+ \frac{1}{2}g(\tau_0)\,(\tau-\tau_0)^2\,
\Bigr]_{\tau_{\rm i}}^{\tau_{\rm f}}\,Q^3
\Biggr\}
\\
\qquad\qquad\qquad\qquad\qquad\qquad ; \quad |Q|<Q_{\rm c}.
\end{array}
\right. 
\label{eq:integ_formula_J}
\end{align}

\section{On the choice of parameter in adaptive smoothing}
\label{sec:choice_of_f_cross}

In this Appendix, we study the effect of varying the parameter $f_{\rm cross}$ on the performances of PT predictions with adaptive smoothing (see Sec.~\ref{sec:improvement} for definition). In the main text, we adopt the intuitive setup $f_{\rm cross}=1$ for post-collapse PT and $0.5$ for Zel'dovich solution. However, PT predictions are prone, after shell-crossing, to get worse over time, so we can naively expect that choosing a smaller value of $f_{\rm cross}$ will provide a better agreement with simulations.

Fig.~\ref{fig:pk_xv_fcross} shows the results obtained in the case of the CDM-like power spectrum studied in \S~\ref{subsec:CDM_initial} when varying the value of $f_{\rm cross}$. Here, the calculations are performed using $N_{\rm particle}=10^5$, $N_{\rm grid}=10^4$ and $k_{\rm cut}=6.3$\,Mpc$^{-1}$ for a  boxsize $L=1000$ Mpc, which is a relatively low resolution set up, but this will not have any consequence on the discussion that follows. Left panel shows the power spectrum at $z=0$ and the resultant predictions for different values of $f_{\rm cross}$ are plotted as solid lines with different colors. As anticipated, decreasing $f_{\rm cross}$ results in an enhancement of small-scale power and a better agreement between post-collapse PT and the simulation is obtained. But a too small value of $f_{\rm cross}$ overshoots the simulation and an optimal choice of $f_{\rm cross}$ roughly corresponds to  $f_{\rm cross}\sim0.6$. The same trend can be also seen for the Zel'dovich solution, but the improvement of the power spectrum amplitude is rather mild: while $f_{\rm cross}=0.3$ seems to provide the best overall behavior for the Zel'dovich solution, the result is clearly still not as good as what post-collapse PT can provide, in particular in the regime $0.5 \la k \la 1$ Mpc$^{-1}$.

Right panel of Fig.~\ref{fig:pk_xv_fcross} shows the phase-space portraits and compares the simulation results (red) to predictions obtained with our ``standard'' value of $f_{\rm cross}=1$ for post-collapse PT (upper insert) and $f_{\rm cross}=0.5$ for Zel'dovich solution (lower insert) and to predictions obtained with respective nearly optimal values of $f_{\rm cross}=0.6$ and $0.3$. It seems difficult to judge by eye whether the prediction with a smaller value of $f_{\rm cross}$ really improves the description at small scales, although more structures seems to be captured for post-collapse PT with $f_{\rm cross}=1$, which contradict the results obtained for the power spectrum. However the simple examination of this figure corresponding to a single realization of the random initial conditions is not conclusive, obviously: it does not preclude the fact that a smaller value of $f_{\rm cross}$ can give a better result for the power spectrum obtained from averaging over many realizations. 

Finally, to conclude this section, although we do not show the results here, we noticed as well that the best choice of $f_{\rm cross}$ could vary slightly according to redshift or initial condition: for example, for the CDM cosmology considered here, a better choice of $f_{\rm cross}$ at redshift $z=1.5$ is $f_{\rm cross}=0.7$ and $0.4$ respectively for post-collapse PT and Zel'dovich solution. This finally explains why we decided to keep, for simplicity, the generic values $f_{\rm cross}=1$ and $f_{\rm cross}=0.5$ respectively for post-collapse PT and Zel'dovich solution, even if they are sub-optimal. 

\bsp	
\label{lastpage}
\end{document}